\documentclass[11pt]{article}

\usepackage{amssymb,amsmath,latexsym,bm,amsfonts}

\usepackage{longtable}
\usepackage{color,xcolor}
\usepackage{indentfirst}
\usepackage[margin=1in]{geometry}
\usepackage{mathtools,tensor,graphicx}
\usepackage{cite}

\usepackage{stmaryrd}
\usepackage{breqn}
\usepackage[colorlinks=true,filecolor=red,citecolor=blue ]{hyperref}
\usepackage{cleveref} 
\usepackage{eufrak}
 \usepackage{mathrsfs}

\usepackage{tikz,colortbl}
\usetikzlibrary{calc}
\usepackage{zref-savepos}
\usepackage{array}
\PassOptionsToPackage{linktocpage}{hyperref}

\newcounter{NoTableEntry}
\renewcommand*{\theNoTableEntry}{NTE-\the\value{NoTableEntry}}

\newcommand*{\notableentry}{%
  \multicolumn{1}{@{}c@{}|}{%
    \stepcounter{NoTableEntry}%
    \vadjust pre{\zsavepos{\theNoTableEntry t}}
    \vadjust{\zsavepos{\theNoTableEntry b}}
    \zsavepos{\theNoTableEntry l}
    \hspace{0pt plus 1filll}%
    \zsavepos{\theNoTableEntry r}
    \tikz[overlay]{%
      \draw[black]
        let
          \n{llx}={\zposx{\theNoTableEntry l}sp-\zposx{\theNoTableEntry r}sp},
          \n{urx}={0},
          \n{lly}={\zposy{\theNoTableEntry b}sp-\zposy{\theNoTableEntry r}sp},
          \n{ury}={\zposy{\theNoTableEntry t}sp-\zposy{\theNoTableEntry r}sp}
        in
        (\n{llx}, \n{lly}) -- (\n{urx}, \n{ury})
        (\n{llx}, \n{ury}) -- (\n{urx}, \n{lly})
      ;
    }%
  }%
}
 
\DeclareMathOperator*{\limitarrow}{\xrightarrow{\hspace*{1cm}}}

\def\simlt{\mathrel{\lower2.5pt\vbox{\lineskip=0pt\baselineskip=0pt
           \hbox{$<$}\hbox{$\sim$}}}}
\def\simgt{\mathrel{\lower2.5pt\vbox{\lineskip=0pt\baselineskip=0pt
           \hbox{$>$}\hbox{$\sim$}}}}

\newcommand{\be}{\begin{equation}}
	\newcommand{\ee}{\end{equation}}
\newcommand{\ba}{\begin{eqnarray}}
	\newcommand{\ea}{\end{eqnarray}}

\newcommand{\RV}{}

 \numberwithin{equation}{section}
 \allowdisplaybreaks

\begin{document}
\renewcommand{\baselinestretch}{1.2}

\begin{titlepage}

\begin{center} 
\vphantom{htr}

\vspace{3cm}

\huge{\bf Hybrid inflation and waterfall field in string theory from D7-branes}
    \vspace{1.5 cm}
    
\large{Ignatios Antoniadis\footnote{e-mail: antoniad@lpthe.jussieu.fr}$^{a}$,  \, \, Osmin Lacombe\footnote{e-mail: osmin@lpthe.jussieu.fr}$^{a}$, \, \, George K. Leontaris\footnote{{e-mail: leonta@uoi.gr}}$^{b}$
}

    \vspace{0.8cm}

\normalsize{\it $^a$ Laboratoire de Physique Th\'eorique et Hautes Energies - LPTHE\\ Sorbonne Universit\'e, 
CNRS, 4 Place Jussieu, 75005 Paris, France 
\vspace{0.2cm}


$^b$ Physics Department, University of Ioannina, 

45110, Ioannina, Greece}

    \vspace{1.2cm}
    
\begin{abstract}
\normalsize
We present an explicit string realisation of a cosmological inflationary scenario we proposed recently within the framework of type IIB flux compactifications in the presence of three magnetised $D7$-brane stacks. Inflation takes place around a metastable de Sitter vacuum. The inflaton is identified with the volume modulus and has a potential with a very shallow minimum near the maximum. Inflation ends due to the presence of ``waterfall" fields that drive the evolution of the Universe from a nearby saddle point towards a global minimum with tuneable vacuum energy describing the present state of our Universe.

\end{abstract}
\end{center} 

\end{titlepage}

\newpage

\pagestyle{plain}
\pagenumbering{arabic}
\vspace{3cm}

 {\hypersetup{linkcolor=black}
\large \tableofcontents
}

\section{Introduction}

During the last years, there  has been intense activity on the implications of quantum corrections to the moduli stabilisation problem in string compactifications, in relation to the possible existence of de Sitter (dS) vacua and realisations of inflationary models. Type IIB  string theory and more generally its geometric F-theory variant, compactified on a Calabi-Yau (CY) threefold, is of particular interest since it provides a framework for addressing these issues~\cite{Kachru:2003aw, Conlon:2005ki}. 
 	
Recently, within this framework, an economical  scenario has been proposed for stabilising  the K\"ahler moduli and  ensuring a positive cosmological constant~\cite{Antoniadis:2018hqy, Antoniadis:2019rkh}. The proposed mechanism only relies on perturbative in $\alpha'$ and string  loop contributions without resorting to non-perturbative corrections. A  representative geometric set-up consists of (a minimum of) three magnetised $D7$ brane stacks mutually orthogonal in the internal six-dimensional (6d) space. The corresponding magnetic fields are turned on along $U(1)$ directions on their internal worldvolumes.  

The framework  takes advantage  of an induced four-dimensional (4d) Einstein-Hilbert (EH) term, localised in the internal space and proportional to its Euler characteristic~\cite{Antoniadis:2002tr}. This term emanates from the ${\cal R}^4$ couplings present in the  ten-dimensional (10d) string effective action, when three Riemann tensors multiplied with the wedge product are integrated over the CY space. Graviton emission from the localised EH term into the bulk towards the distinct $D7$-brane sources leads to local tadpoles where gravitons propagate through a co-dimension-two bulk, giving rise to logarithmic dependent corrections on the size of the bulk, in the large (transverse to the $D7$-brane) volume limit~\cite{Antoniadis:1998ax, Antoniadis:2019rkh}.

These logarithmic contributions break the tree-level no scale structure of the K\"ahler potential~\cite{Cremmer:1983bf, Ellis:1983ei, Becker:2002nn} and an F-term dependent on the K\"ahler moduli is induced in the scalar potential. On the other hand, magnetic fluxes of $U(1)$ gauge symmetries associated with the $D7$ brane stacks provide (positive) D-term  contributions to the scalar potential~\cite{Burgess:2003ic}. Both, logarithmic corrections and D-terms are sufficient to stabilise the K\"ahler moduli and  support a positive cosmological constant~\cite{Antoniadis:2018hqy}. 

A meticulous examination of the resulting scalar potential shows that cosmological inflation can be implemented with the internal volume modulus acting as the inflaton field~\cite{Antoniadis:2020stf}. It has been found that the horizon exit is just above the inflection point and the  accumulation of the 60 e-folds required to realise inflation happens when the inflaton field approaches the minimum of  the potential. However, this dS minimum generated by radiative corrections is a false vacuum with a value of the cosmological constant much larger than the one observed  today. A plausible solution to this issue is through hybrid inflation~\cite{Linde:1993cn} where a second (waterfall) field ends the inflation phase and  settles to a  lower (true) minimum with the right value of the cosmological constant. 
 	
In the above geometric setup,  possible available candidates for the role of a waterfall field, are charged matter fields from the $D7$-branes; they correspond to excitations of open strings ending on the $D7$-brane stacks or their intersections. In this paper, we investigate this possibility and work out an explicit realisation. In general, a charged open string scalar gets two types of contributions to its mass:
\begin{itemize}
\item A positive supersymmetric contribution corresponding to turning on a Wilson line or introducing a brane separation, which are equivalent by T-duality and described by an appropriate superpotential.
\item A non-supersymmetric contribution due to the presence of the worldvolume magnetic fields that can be negative depending on the spin-magnetic field interaction along the internal wolrdvolume directions. This contribution is described by an appropriate Fayet-Iliopoulos (FI) term entering the D-auxiliary component of the $U(1)$ with internal magnetic field. 
\end{itemize}
Thus, the above contributions can differ in their sign  and their dependence on the internal volumes. As a result, tachyonic fields may appear in particular regions of values for the internal volumes. Selecting appropriate magnetic fluxes and making a judicious choice of (quantised) Wilson lines and brane positions, we construct first a model with the following properties:
\begin{enumerate}
\item The ratios of the internal worldvolumes along the three $D7$-brane stacks are fixed in terms of the (ratios of the) FI parameters depending on the quantised magnetic fluxes. On the other hand, all complex structure moduli and the string dilaton are assumed to be fixed at weak coupling by appropriate 3-form fluxes in a supersymmetric way (i.e. with vanishing F-auxiliary components), leading to a constant flux-dependent superpotential~\cite{Frey:2002hf,Kachru:2002he}.
\item The total 6d internal volume (which is the only leftover K\"ahler modulus) can then be stabilised in the large volume regime by minimising the full scalar potential of the theory containing the logarithmic corrections in the K\"ahler potential through the F-term contribution. All charged open string states have positive squared-masses in the large volume limit, larger than a certain value near the minimum ensuring sufficient inflation (around 55 e-folds).
\item There is only one charged open string scalar that becomes tachyonic when the internal volume becomes less than the above critical value and can thus be identified with the waterfall field. The condition that there is no other tachyon for all values of the volume implies within our framework that this state comes necessarily from the same magnetised $D7$-brane identified with its image under the corresponding $O7$-orientifold but being separated in the transverse plane.
\end{enumerate}

It turns out that the above model, although it provides an explicit string construction that implements the waterfall proposal for ending inflation within our perturbative framework of moduli stabilisation, it does not lead to a sufficiently deep vacuum that can accommodate the present dark energy. We therefore explose generalisations with more than one waterfall fields that become tachyonic at nearby points successively (as a rollercoaster), ending up to a vacuum of an infinitesimal (tuneable) energy.

The paper is organised as follows. In Section~2, we give a short review of the mechanism of moduli stabilisation (subsection 2.1) and the corresponding model for inflation (subsection 2.2). In Section~3, we perform a general analysis of the open string spectrum and the tachyons' appearance in a $\mathbb{Z}_2\times \mathbb{Z}_2$ orientifold of type IIB string with three mutual orthogonal sets of $D7$-branes. We first review the model (subsection 3.1) and then analyse the possible tachyonic states upon turning on worldvolume magnetic fields, first in the simplest case of along one internal torus in one of the $D7$-brane stacks (subsection 3.2), then in all three stacks (subsection 3.3) and finally along both tori on the worldvolume of all stacks (subsection 3.4). Next we solve the constraint of having only one possible tachyon that can play the role of waterfall field in all cases. In Section~4, we analyse the dynamics of the waterfall field on the metastable dS vacuum and inflation by computing first the scalar potential (D-term part in subsection 4.1 and F-term part in subsection 4.2) and then the new vacuum where the waterfall field develops a non-vanishing vacuum expectation value (VEV) at a saddle point, breaking the corresponding $U(1)$ symmetry in a (global) minimum at a lower energy than the scale of inflation but for a value of the volume modulus of the same order as the original one (subsection 4.3). However, the vacuum energy  cannot be made sufficiently small to accommodate the present dark energy.
In Section~5, we study the global minimum and generalise the model by adding more tachyons at nearby points around the saddle point, that allow tuning the vacuum energy at an infinitesimally small value.
Finally, Section~6 contains our conclusions and outlook. There are also three Appendices containing the expressions of Riemann theta-functions and $SO(2)$ characters (Appendix~A), the lattice of momenta and winding modes that we use in Section~4 (Appendix~B) and the study of tachyons in the case of magnetised $D7$-branes on their entire worldvolume (Appendix~C).

\section{A short review of the framework 
}

In this section we  review the salient features of previous work~\cite{Antoniadis:2019rkh,Antoniadis:2020stf}, focusing mainly on the mechanism of   K\"ahler moduli stabilisation with  perturbative radiative corrections  and the  implementation of the hybrid inflationary scenario.

 In the framework of type IIB superstring theory, we consider a configuration of three mutually orthogonal $D7$-brane stacks and 
 three (four-cycle) K\"ahler moduli ${\cal T}_k=e^{-\RV{\phi_{10}}}\tau_k+ib_k$, with $ k=1,2,3$ and $\RV{\phi_{10}}$ the dilaton.  The total 6d internal volume can be  expressed \RV{in string units} in terms of $\mathcal{T}_k$~\cite{Becker:2002nn}  and here is simply given  by 
 \be 
 {\cal V} =\sqrt{\tau_1\tau_2\tau_3}\,, \qquad \hat{\cal V}=e^{-\frac 32\RV{\phi_{10}}}{\cal V}= \sqrt{\prod_i \frac 12 (\mathcal{T}_i+\bar{\mathcal{T}}_i)}\, \, , \quad \label{6vol}
 \ee
 where $\tau_k$ correspond to the internal worldvolumes of the three $D7$-brane stacks \RV{in string units}. 
The two basic ingredients  we are interested in the following analysis are the superpotential of the moduli fields and the K\"ahler potential. 
 The  tree-level superpotential induced by 3-form fluxes~\cite{Gukov:1999ya} is $ {{\cal W}_{\rm flux}}=\int\, G_3\wedge { \Omega}( z_a)$, where $ G_3$ is defined in terms of the field strengths of the two 2-form gauge potentials ($C_2$ and $B_2$) $F_3= d\,C_2,\,H_3= d\, B_2$ and of the axion-dilaton $S=e^{-\RV{\phi_{10}}}+iC_0$, through  $ G_3=   F_3-iS\,  H_3$. $ \Omega(z_a)$ is the holomorphic 3-form of the CY internal manifold which depends on the complex structure moduli $z_a$.  

Supersymmetric minimisation conditions fix the  moduli $S, z_a$ (thus the string coupling $g_s=\langle e^\phi\rangle$ and the complex structure of the internal manifold), but  the K\"ahler moduli ${\cal T}_k$ remain undetermined. Indeed, the classical K\"ahler potential is of no-scale type for ${\cal T}_k$ and is expressed as
\begin{eqnarray} 
	{\cal K} &=&-2\ln{\left[\hat{\cal V}(\mathcal{T}_k)\right]}-\ln\left[-i\int[{\Omega}\wedge {\bar{\Omega}}](z_a,{\bar z}_a)\right]
	-\ln\left[S+{\bar S}\right]\,,
	\label{ClassKahler}
\end{eqnarray}
while the superpotential is reduced to a flux-dependent constant ${\cal W}_0$. Note that once the complex structure moduli and the dilaton are stabilised, $\mathcal{V}$ and $\hat{\cal V}$ are interchangeable in the K\"ahler potential, as well as $\tau_k$ and ${\cal T}_k$. 
Due to the no-scale structure and the supersymmetry conditions for $z_a$ and $S$, the induced scalar potential is identically zero 
\begin{equation} { V}_{\rm no-scale }\,=\, e^{\cal K}\left(\sum_{I,J} {\cal D}_I{\cal W}_0{\cal K}^{I\bar J}{\cal D}_{\bar J}{\cal W}_0-3 |{\cal W}_0|^2\right) \equiv 0, \label{noscaleV}
\end{equation} 
hence it is not possible to stabilise the K\"ahler moduli at the classical level. 
 As it is well known, a crucial role in the resolution of this issue is played by the  perturbative  and non-perturbative   corrections. In the particular geometric 
 configuration considered in the present work, it has been shown that
 perturbative  quantum corrections which depend logarithmically on
 the internal volume suffice to stabilise all K\"ahler moduli in a dS vacuum~\cite{Antoniadis:2018hqy, Antoniadis:2019rkh}.
 Below, we give a brief summary of the main points of the  derivation  of these  corrections. 

\subsection{Logarithmic corrections and scalar potential}  \label{sectionintromodel}

 The 10d effective action of type IIB superstring theory, in addition
to  the Einstein-Hilbert (EH) term linear in the scalar curvature $\cal R$,  includes  also the leading order gravitational term which depends on the fourth power of the Riemann tensor $\cal R$. Such ${\cal R}^4$-terms are induced from graviton scattering 
and do not receive any other perturbative corrections beyond one-loop~\cite{Grisaru:1986kw, Antoniadis:1997eg}. 

The low energy limit of type IIB theory is described by its effective action obtained 
upon compactification to four dimensions ${\cal M}_{10}\to  {\cal X}_6\times{\cal M}_4$, where ${\cal X}_n$ is a $n$-dimensional compact manifold
and ${\cal M}_d$ the $d$-dimensional Minkowski spacetime. Under reduction of the 10-dimensional
action, the ${\cal R}^4$ couplings induce a novel EH term localised in the bulk, denoted in the following with ${\cal R}_{(4)}$. The 4d effective action relevant to our discussion takes the form~\cite{Antoniadis:2003sw, Antoniadis:2002tr}: 
 \ba
 	{\cal S }_{\rm grav}&=& \frac{1}{(2\pi)^7 \alpha^{\prime 4}} \int\limits_{{\cal X}_6 \times {\mathcal M_{4}}} e^{-2\RV{\phi_{10}}} {\cal R} + \frac{\chi}{(2\pi)^4 \alpha'} \int\limits_{\mathcal M_{4}} \left(2\zeta(3) e^{-2\RV{\phi_{10}}}  + 4\zeta(2) \right) {\cal R}_{(4)}\,,  
 	\label{4d_IIBAct} 
 \ea
 where $\alpha'$ is the string Regge slope and $\zeta(2)=\pi^2/6$.
 The tree-level term proportional to $\zeta(3)$ is vanishing for orbifolds.
The proportionality factor of the ${\cal R}_{(4)}$ term depends on the Euler characteristic $\chi$ of the compactification manifold given by 
\ba	
\chi&=&	\frac{3}{4\pi^3}\int\limits_{{\cal X}_6} {\cal R}\wedge {\cal R}\wedge {\cal R}~\cdot 
\label{EulerX}
\ea
From (\ref{4d_IIBAct}) and (\ref{EulerX}) it is readily inferred that the ${\cal R}_{(4)}$ term exists only
in four dimensions and is localised at points in the internal space where the Euler number is concentrated in the large volume (decompactification) limit.

It follows that from these points of 4d localised gravity kinetic terms, 10d gravitons represented by closed strings can be emitted in the bulk towards distinct $D$-brane and orientifold sources, leading to local tadpoles~\cite{Antoniadis:1998ax}.
In a geometric configuration with $D7$-brane stacks (as well as $O7$-orientifold planes) spanning four out of the six internal dimensions, a novel type of radiative corrections emerges.  
More concretely, by momentum conservation, gravitons  
emitted from the localised ${\cal R}_{(4)}$-vertices and ending on 7-brane sources propagate effectively in the two dimensions transverse to the 7 branes, acquiring a logarithmic propagator as a function of the distance. As a result, for generic $D7$-brane distribution at the boundaries of the compactified space, `far' away from the localised EH-term, 
they give rise to corrections depending logarithmically on the size of the bulk, of the  form~\cite{Antoniadis:2019rkh}
\begin{eqnarray}
	\frac{4\zeta(2)}{(2\pi)^3}\chi \int_{M_4} \left(1-\sum_{k=1}^{3} e^{2\RV{\phi_{10}}}T_k \ln(R^k_{\bot}/\mathtt{w})\right)\,{\cal R}_{(4)}~,\label{allcor}
\end{eqnarray}
where we considered the case of orbifolds, where computations can be done explicitly.
Here, $T_k$ is the (effective) tension of the $k^{th}$ 7-brane stack, $R^k_{\bot}$ stands for the size of the two-dimensional space transverse to the corresponding brane stack, and $\mathtt{w}$ is the width of the ${\cal R}_{(4)}$ localisation, playing the role of an effective ultraviolet cutoff for the graviton propagator in the bulk~\cite{Antoniadis:2002tr}.

Incorporating  the above  corrections into the K\"ahler potential (\ref{ClassKahler})  we obtain the following K\"ahler moduli dependence
\begin{equation}
	{ \cal  K}(\tau_k)=
	-2\ln\left(\sqrt{{\tau}_1{\tau}_2{\tau}_3}+{\xi}+\sum_k \gamma_k \ln{ {\tau}_k}\right)= 
	-2\ln\left({\cal V}+{\xi}+{\gamma} \ln {\cal V}\right)~, 
\label{CorrectedKahler}
\end{equation}
where in the last equality we assume for simplicity the same tension $T_k\equiv T=e^{-\RV{\phi_{10}}}T_0$ for all the brane stacks, which amounts to identical $\gamma_k\equiv \gamma/2$. The parameters $\xi$ and $\gamma$ are given by~\cite{Antoniadis:2002tr, Antoniadis:2019rkh}
\be 
\gamma\equiv -\frac{1}{2} g_sT_0\xi\;, \qquad {\rm with }\;\; \xi=-\frac{\chi}{4} \times 
\begin{cases}
	\frac{\pi^2}{3}g_s^2\quad {\rm for\ orbifolds}
	\\[3pt]
	\zeta(3)\;\quad {\rm for\ smooth\ CY} 
\end{cases} .\label{gammaxi}
\ee 
These corrections induce a non-zero F-term effective potential $V_F$. In addition, 
the effective potential receives contributions from D-terms associated with (magnetised) 
$U(1)$ factors of the $D7$-brane stacks. The D-term effective potential $V_D$ can be minimised to fix the ratios $\tau_i/\tau_j$. \RV{These ratios are related to moduli orthogonal to the total internal volume modulus. When the masses of these moduli are large compared to the mass of the total volume, one can indeed study the resulting effective potential of the total volume after minimisation over the ratios. We check explicitly this assumption for the model studied hereafter, in \cref{secnewcacuum}. } 

The sum of F- and D-term  contributions constitutes the effective scalar potential $V_{\rm eff}$ which 
after minimising the ratios in the large volume limit, can be cast in the form
\begin{align}
	V_{\rm eff} (\mathcal{V})=V_F+V_D&\simeq \frac{3 \mathcal{W}_0^2}{2\kappa^4 \mathcal{V}^3}\left(2\gamma(\ln \mathcal{V}-4)+\xi \right) +\frac d{\kappa^4\mathcal{V}^2}\equiv \frac{C}{\kappa^4} \left( -\frac{\ln \mathcal{V}-4+q}{{\mathcal V}^3}-\frac{3\sigma}{2\mathcal{V}^2}\right), \label{Vlargelimit}
\end{align}
where the term in the right-hand side  proportional to $\gamma$ is $V_F$, while the term proportional to $d$ is $V_D$. 
The constant $d$ is related to D-terms, $\kappa=\sqrt{8\pi G_N}$ is the reduced Planck length, and  we have defined 
\begin{equation}
	q\equiv \frac \xi{2 \gamma}=-{1\over g_sT_0}, \quad C\equiv -3{\mathcal{W}_0}^2\gamma 
\,,  \quad \sigma \equiv \frac{2d}{9 {\mathcal{W}_0}^2\gamma}=-\frac{2d}{3C}. \label{modelparameters}
\end{equation}
It can be readily  shown that within the above procedure, positive square masses are provided to all the K\"ahler moduli fields and at the same time a local de Sitter vacuum is obtained in a narrow region of $\sigma$, at weak coupling and large volume for $\gamma$ and $q$ negative, implying positive tension and negative Euler number, $T_0>0$ and $\chi<0$. 

\subsection{Implementation of  Hybrid Inflation} \label{sectionhybridinflation}

In~\cite{Antoniadis:2020stf}   slow-roll inflation was successfully  implemented  with the internal volume $\mathcal{V}$ 
playing the role of the inflaton field. Introducing  the  canonically normalised inflaton
 \begin{equation}
	\phi/\kappa\equiv{\sqrt{6}}/({3\kappa})\ln(\mathcal{V}), \label{inflaton}
\end{equation}
the potential \eqref{Vlargelimit}  reads
 \begin{equation}\label{potwitht}
 	V(\phi)\simeq -\frac{C}{\kappa^4}e^{-3 \sqrt{\frac 32} \phi} \left(\sqrt{\frac 32} \phi - 4 + q + \frac 32 \sigma e^{\sqrt{\frac 32} \phi}\right).
 \end{equation}
The extrema of (\ref{potwitht}) are found to be 
 \begin{align}
 	\phi_{-/+}=-\sqrt{\frac 23}\left(q-\frac{13}3+W_{0/-1}\left(-e^{-x-1}\right)\right), \label{max}
 \end{align}
 where $\phi_-\,(\phi_+) $ is the local minimum (maximum)  with $\phi_-<\phi_+$,  and $W_{0/-1}$ are the two branches of the Lambert-W function, whilst  $x$ is a convenient parameter defined through the relation 
 \begin{equation}
 	x\equiv q-\frac{16}3-\ln(-\sigma) \quad \leftrightarrow \quad \sigma=-e^{q-\frac{16}3-x}. \label{xdefinition}
 \end{equation}
From \eqref{max} we observe  that variation of the parameter $q$, while keeping  $x$  constant, implies only a common shift  of the local extrema. Moreover, a simple inspection of the form of the potential \eqref{potwitht} shows that $x$ is the only real parameter of the model, while $q$ shifts the origin of the field and $C$ rescales the potential. 

Note that the value of the volume at the minimum is given by\footnote{Please note that there is a typo regarding the sign in front of $W_0$ in equation (56) of \cite{Antoniadis:2020stf}. }
\begin{equation}
\mathcal{V}_-=\exp\left(\sqrt{\frac 32}\phi_-\right)=e^{-q} \times \exp\left(\frac{13}{3}-W_0\left(-e^{-x-1}\right)\right).\label{Vmin}
\end{equation}
Thus, for a given value of $x$, one obtains a large volume for a  large (negative) $q=-1/(g_sT_0)$, which is reached exponentially fast as long as $g_s$ is small. Hence the weak coupling and large volume limits are related naturally in a simple way.  \RV{For simplicity, in the following we take $q=0$ 
emphasising that the parameter $q$ does not change the properties and the analysis of the inflationary phase, but can be used to reach parametrically large volumes. }

It turns out~\cite{Antoniadis:2020stf}  that the critical value $x_{c} \simeq 0.072$ 
gives a Minkowski minimum, $i.e.$ $V(\phi_-)=0$, the region $0<x<x_c$ ensures de Sitter minima, the values  $x>x_c$ yield  anti-de Sitter (AdS) vacua, and the region $x<0$ corresponds to the case where the two branches of the Lambert function join and the potential loses its local extrema. 
Slow-roll inflation compatible with observations can be realised  for $x\simeq3.3\times10^{-4}$, 
while the field separation between the two extrema is given by $\phi_+-\phi_-=0.042$.
The inflaton starts rolling near the maximum with no initial speed, these initial conditions being motivated if one considers that this maximum is related to a symmetry restoration point. The inflationary phase corresponds to the inflaton rolling down its potential. An analysis of the slow roll parameters $\epsilon=(V'/V)^2/2$ and $\eta=V''/V$ shows that $\epsilon \ll |\eta|$ holds in the whole region of the field space $[\phi_-,\phi_+]$ and thus the spectral index of primordial density fluctuation  $n_s\simeq 1+2\eta$ is fixed by $\eta$ which has to be around $-0.02$ at the horizon exit $\phi\equiv\phi^*$ to agree with the data.

As the inflaton $\phi$ goes down from the maximum to the minimum, the second derivative $V''(\phi)$ changes sign and as the slow roll parameter $\eta(\phi_+)<-0.02$, it passes through the value $\eta(\phi_*)=-0.02$ before the inflection point. The $x$ parameter of the model is chosen so that at least $60$ e-folds are obtained from this point to the end of inflation.  The required  number of $N_*\simeq60$ e-folds is 
computed from the horizon exit $\phi_*\simeq 
\phi_-+0.02$ at which $\eta(\phi_*)=-0.02$, to the minimum $\phi_-$. The modes exit the horizon just before the inflection point is reached and most of the e-folds are obtained around the minimum. Furthermore, it should be emphasised that the corresponding inflaton field displacement is $	\Delta\phi\simeq 0.02$, 
which is much less than one in Planck units, corresponding to small field inflation compatible with the validity of the effective field theory.
 
In the model described above, the dS vacuum energy is constrained by the choice of the value of the parameter $x$ and for its value of interest for inflation, the potential at the minimum is practically of the same order given by the inflation scale, $V(\phi_-)\simeq V(\phi^*)$. This amount of vacuum energy is way much greater than the observed value today, hence it  could not be the true vacuum of the theory. Indeed, with such a big value, the Universe would continue expanding and never reach the standard cosmology with radiation and matter domination eras. 
 As suggested in~\cite{Antoniadis:2020stf}, the introduction of new physics near the minimum of the potential brings in a natural scenario for the end of the inflation epoch. This relates the model to the hybrid inflation proposal \cite{Linde:1993cn}, where a second field $Y$ is added to the model. This ``waterfall'' field $Y$ adds another direction to the scalar potential. If falling towards this direction becomes favorable at a certain point of the inflaton trajectory, this immediately ends the inflation era and the theory reaches another minimum at a different energy scale which should coincide with the true vacuum today dominated by the observed dark energy.
 
The main features of the hybrid scenario adapted to our model  are described by the following potential
 \begin{equation}
 	V_Y(\phi,Y)=V(\phi)+\frac 12m_Y^2(\phi) Y^2+\frac \lambda 4 \,Y^4~,\label{Vsfs}
 \end{equation}
where  $V(\phi)$ is the inflaton potential (\ref{potwitht}) and the extra terms contain the dependence in $Y$ together with its coupling to the inflaton $\phi$. Depending on the sign of its effective squared mass $m_Y^2(\phi)$, the  waterfall field $Y$ stays in two separate phases. When $m_Y^2>0$, the minimum in the $Y$-field direction is at the origin 
 \begin{equation}
 	\langle Y \rangle=0, \quad {\rm when} \quad m_Y^2(\phi)>0\,,
 \end{equation}
 and the extra contribution to the scalar potential vanishes
 \begin{equation}
 	V_Y(\phi,0)=V(\phi)\,.
 \end{equation}
 When the mass of $Y$ becomes tachyonic, a phase transition occurs and the new vacuum is obtained at a non-vanishing VEV for $Y$: 
 \begin{equation}
 	\label{newmin}
 	\langle Y \rangle=\pm \frac{\lvert{m_Y}\rvert}{\sqrt{\lambda}}\equiv\pm{v}, \quad {\rm when} \quad m_Y^2(\phi)<0.
 \end{equation}
 The value of the potential $V_Y$ at the minimum of this broken phase is
 \begin{equation}
 	V_Y(\phi,v)= V(\phi)-\frac{m_Y^4(\phi)}{4\lambda}\,. 
\label{vacuumbrokenphase}
 \end{equation}
 
For suitable $m_Y(\phi)$  during the inflationary phase when the field $\phi$ rolls down the potential, the system is in the symmetric phase and the $Y$ field is stabilised with a vanishing VEV and a large mass. The inflationary phase is then equivalent to the one field inflation model. Subsequently, if  $m_Y^2$ turns negative near the minimum, a phase transition occurs and the $Y$ field attains its value given in~\eqref{newmin} at the new minimum. This amounts to a change of the potential $V(\phi)$ near the minimum, by a negative constant  $V_{down}=-m_Y^4/(4\lambda)<0$. The effect of such a downlift is double: it decreases the value of the cosmological constant and if the waterfall direction is steep enough, it gives a natural criterion to stop inflation ($\epsilon>1$). 
In the next sections we will  propose a possible implementation of the scenario of hybrid inflation in a string theory framework by demonstrating   how the waterfall field can be identified with  an open string state on $D7$-branes stacks.
~\\
 
\section{Toroidal model of a matter waterfall field} \label{section3}

In this section we implement a toy model of toroidal compactification with magnetic fluxes giving rise to a matter waterfall field, located at an intersection of the $D7$-branes stacks. We will consider a $\mathbb{Z}_2\times \mathbb{Z}_2$ orbifold on a factorised 6-torus $T^6=T^2\times T^2\times T^2$, for which the associated Euler characteristic\footnote{For toroidal orbifolds, the Euler characteristic is defined as  $ \chi= 1/|P| \sum_{g,h \in P} \chi(g,h)$ where $P$ is the point group of the orbifold and $\chi(g,h)$ the number of fixed points under both twists $g$ and $h$, taken zero when there is a common fixed torus. In the $\mathbb{Z}_2\times \mathbb{Z}_2$ example generated by the basis $\alpha=(-,-,+)$ and $\beta=(+,-,-)$ twists, a non-trivial $(g,h)$ pair is either $(\alpha, \beta)$, $(\alpha,\alpha \beta)$ or $(\beta,\alpha \beta)$, and have $\chi(h,g)=\chi(g,h)=4^3=64$. Hence the Euler characteristic is  $\chi= 1/4 \times 2 \times 64 \times 3 = 96$, with the factor of  $2$ coming from the interchange of $g$ and $h$ in the sum.}  is $\chi=96$. As explained in \cite{Antoniadis:2002tr, Antoniadis:2019rkh}, a large Euler characteristic is necessary to control the approximations in the computation of the localisation width of the induced 4d graviton kinetic terms and the logarithmic K\"ahler quantum corrections $\gamma_k$ and $\gamma$ of \eqref{CorrectedKahler} and \eqref{gammaxi}. The $\mathbb{Z}_2\times \mathbb{Z}_2$ orbifold is therefore a valid and simple candidate for a specific model.

In the following, we first review toroidal string compactifications in the presence of magnetised branes and then show how to obtain a waterfall field. The idea is to generate a tachyonic field, whose mass-squared depends non-trivially on the total internal volume and becomes negative around the minimum of the scalar potential \eqref{potwitht}.
\subsection{ $T^6/\mathbb{Z}_2\times \mathbb{Z}_2$ with magnetic fields : setup and notations} \label{sectiontoroidal}

\paragraph{Toroidal orbifold}
We consider for simplicity a factorised 6-torus $T^6=T^2_1\times T^2_2 \times T^2_3$ with $i=1,2,3$ indices denoting  the (45), (67) and (89) internal directions respectively. 
To fix notations we define the $i$-th torus $T^2_i$ as
\begin{align}
T^2_i\equiv\mathbb{R}^2/2\pi\Lambda_i, \quad \Lambda_i= \left\{ q \, \bold{R}_{ix} + r \, \bold{R}_{iy} ; q,r \in \mathbb{Z} \right\}, \label{torusdefinition}
\end{align}
with $\bold{R}_{ix}$,  $\bold{R}_{iy}$ two linearly independent vectors of norm $R_{ix}$, $R_{iy}$ and relative angle $\alpha_i$. The dual lattice $\Lambda^*_i$ is generated by the dual vectors $\bold{R}_i^{*x},\bold{R}_i^{*y}$ satisfying $\bold{R}_{ik}\cdot \bold{R}_i^{*l}=\delta_k^l$.  
The torus metric reads
\begin{equation}
g^{(i)}_{kl}=\bold{R}_{ik}\cdot \bold{R}_{il}=\frac{\mathcal{A}_i}{\text{Re}(U_i)}
 \begin{pmatrix}
1 & \text{Im}(U_i) \\
\text{Im}(U_i) & |U_i|^2
\end{pmatrix}, \label{torusmetric}
\end{equation}
and its inverse can be used to raise the indices and express the dual vectors $\bold{R}_i^{*k}=g^{{\scriptscriptstyle (i)}kl}\bold{R}_{il}$.
In the above metric we have defined by $\mathcal{A}_i$ the  
unit cell area of the torus $T_i^2$
\begin{equation}
\mathcal{A}_i \equiv \sqrt{\det g^{\scriptscriptstyle{(i)}}}= \frac{\text{vol}(2\pi\Lambda_i)}{(2\pi)^2}= R_{ix}R_{iy} \sin \alpha_i, \quad \text{with} \quad \bold{R}_{ix} \cdot  \bold{R}_{iy}= R_{ix}R_{iy} \cos \alpha_i, \label{Kahlertorus}
\end{equation}
and by $U_i$, the torus complex structure modulus
\begin{equation}
U_i\equiv i\frac{R_{iy}}{R_{ix}}e^{-i\alpha_i}=\frac 1{{R_{ix}}^2}(\mathcal{A}_i + i \bold{R}_{ix} \cdot  \bold{R}_{iy}). \label{Complextorus}
\end{equation}

We now consider the following $D7$ branes configuration, dual to the configuration containing $D9$ and $D5$ branes as in the toroidal orbifold model on $T^6/\mathbb{Z}_2\times \mathbb{Z}_2$ described in \cite{Larosa:2003mz,Aldazabal:1998mr}: 

\vspace{0.5cm}
\begin{minipage}[t]{0.13\linewidth}
\begin{tabular}{c|ccc}
& (45) & (67) & (89) \\
  \hline
 $ D7_1$ & $\cdot$ & $ \times$  & $\times$  \\
$D7_2$ & $\times$ & $ \times$  & $\cdot$  \\
$D7_3 $ &$\times$ & $\cdot$ &$ \times $ \\
\end{tabular}\end{minipage}\hfill
\begin{minipage}[t]{0.1\linewidth}
$\underset{\text{T-duality along (45)}}{ \xleftrightarrow{\hspace*{4cm}} }$
\end{minipage}\hfill
\begin{minipage}[t]{0.3\linewidth}
\begin{tabular}{c|ccc}
& (45) & (67) & (89) \\
  \hline
 $ D9_1$ & $\times$ & $ \times$  & $\times$  \\
$D5_2$ & $\cdot$ & $ \times$  & $\cdot$  \\
$D5_3 $ &$\cdot$ & $\cdot$ & $ \times $ \\
\end{tabular}\end{minipage}

\vspace{0.5cm}
\noindent
In the above tables,  a cross $\times$ represents the D7 worldvolume spanning the corresponding torus, while a dot $\cdot$ indicates the transverse directions where the D7 brane is  localised.  In the following we will introduce magnetic fields and circled crosses $\otimes$ will represent directions of a magnetic flux for the worldvolume $U(1)$ gauge fields.

The torus, Klein bottle, annulus and M\"obius amplitudes are computed using standard methods \cite{Bianchi:1991eu,Bianchi:1990yu,Angelantonj:2002ct} and the specific ones for the $T^6/\mathbb{Z}_2\times\mathbb{Z}_2$ model can be found in $e.g.$ \cite{Larosa:2003mz}. The torus amplitude (without discrete torsion) reads
\begin{align}
4\mathsf{T}=&\left|T_{oo}\right|^2\Lambda_1\Lambda_2\Lambda_3 +16\left(\left|T_{og}\right|^2\Lambda_1+\left|T_{of}\right|^2\Lambda_2+\left|T_{oh}\right|^2\Lambda_3\right) \left|\frac{\eta^2}{\vartheta_2^2}\right|^2 \nonumber\\
&+16 \left(\left|T_{go}\right|^2\Lambda_1+\left|T_{fo}\right|^2\Lambda_2+\left|T_{ho}\right|^2\Lambda_3\right)\left|\frac{\eta^2}{\vartheta_4^2}\right|^2 + 16 \left(\left|T_{gg}\right|^2\Lambda_1+\left|T_{ff}\right|^2\Lambda_2+\left|T_{hh}\right|^2\Lambda_3\right)\left|\frac{\eta^2}{\vartheta_3^2}\right|^2\nonumber\\
&+64\left(\left|T_{gh}\right|^2+\left|T_{gf}\right|^2+\left|T_{fg}\right|^2+\left|T_{fh}\right|^2+\left|T_{hg}\right|^2+\left|T_{hf}\right|^2\right)\left|\frac{\eta^3}{\vartheta_2\vartheta_3\vartheta_4}\right|^2  \label{torusamp},
\end{align}
where $\vartheta_i$'s are the Riemann theta-functions and $\eta$ the Dedekind function, depending on the world-sheet torus modular parameter $\tau$, given in \cref{AppendixA}.
The tori lattice sums $\Lambda_i$ are given in \cref{Appendixsums}. There is a clash on notations with the defining torus lattice of \eqref{torusdefinition}, but this should not cause any problem. The characters $T_{kj}$ are expressed in terms of the $16$ $\mathbb Z_2\times \mathbb Z_2$ characters $\tau_{kl}$ constructed from quadruple products of the four level-one $SO(2)$ characters, see Appendix \ref{AppendixA}. The $T_{kj}$ characters used for the $T^6/\mathbb{Z}_2\times \mathbb{Z}_2$ model are defined in \cite{Angelantonj:2002ct,Larosa:2003mz}
\begin{align}
T_{ko}=\tau_{ko}+\tau_{kg}+\tau_{kh}+\tau_{kf}, \quad T_{kg}=\tau_{ko}+\tau_{kg}-\tau_{kh}-\tau_{kf}, \nonumber \\
T_{kh}=\tau_{ko}-\tau_{kg}+\tau_{kh}-\tau_{kf}, \quad T_{kf}=\tau_{ko}-\tau_{kg}-\tau_{kh}+\tau_{kf},  \label{Tkjfirst}
\end{align}
for $k=o,f,h,g$. 

In our $D7$-branes setup, the world-sheet involution $\Omega$ projection is implemented by adding the Klein-bottle amplitude $\mathsf{K}$ to the half torus $\frac 12 \mathsf{T}$ of equation \eqref{torusamp}, following the conventions of \cite{Angelantonj:2002ct}. The Klein-bottle amplitude reads
\begin{align}
8 \mathsf{K}=&\left(\vphantom{1^{2^2}}W_1P_2P_3+P_1W_2P_3+P_1P_2W_3\right)T_{oo} + 2 \times 16 \left[\vphantom{1^{2^2}}P_1T_{go}+P_2T_{fo}+P_3T_{ho}\right]\left(\frac{\eta}{\vartheta_4}\right)^2\,,
\end{align}
where the Klein-bottle lattice sums $P_i, W_i$ are given in \Cref{Appendixsums}. 
The open string spectrum can be obtained through the annulus and M\"obius amplitudes; we describe them in the following subsections, in the presence of magnetic fields. We then use $(x,y,z)$ arguments for the $T_{kj}$ open-string characters referring to the internal oscillator shifts. For instance, the first character of \eqref{Z2characters} reads   
\begin{align}
\tau_{oo}(x,y,z)= V_2(0) &O_2(x) O_2(y)O_2(z)+O_2(0)V_2(x)V_2(y)V_2(z)\nonumber\\
-S_2(0)&S_2(x)S_2(y)S_2(z)-C_2(0)C_2(x)C_2(y)C_2(z), \label{dependencetau}
\end{align}
and the $T_{kj}(x,y,z)$ follow the same logic.
Of course in  the torus amplitude \eqref{torusamp}, $T_{kj}$ stand for $T_{kj}(0,0,0)$.
 
 \paragraph{Introducing magnetic fields}
 We  now give a few elements of toroidal compactifications in the presence of worldvolume magnetic fields, that we use in the following subsections. We generically  denote by $H_a^{\scriptscriptstyle(i)}$ a magnetic field introduced on the $D7_a$ stack, in the $i$-th internal plane, with $i=1,2 $ and $3$ for $(45), (67)$ and $(89)$ respectively. 
 
 Magnetic fields modify the world-sheet action by introducing  boundary terms \cite{Abouelsaood:1986gd, Bachas:1995ik}. The solution of the wave equations depends on the charge of the open string. Neutral strings have standard oscillators while charged ones see their modes shifted by the magnetic field through the theta function argument 
 \begin{equation}
\zeta_{a}^{\scriptscriptstyle(i)}=\frac 1\pi \text{Arctan}(2\pi\alpha' q_a H^{\scriptscriptstyle(i)}_a)~. \label{defzeta}
\end{equation} 
In the following we choose a normalisation for the $U(1)$ charges at the endpoints of an open string 
$q=\pm1,0$.  

For NN boundary conditions (with N standing for Neumann), this argument appears through a factor ${\eta}/{\vartheta_1\left(\zeta^{\scriptscriptstyle(i)}_a\tau\right)}$, replacing the standard ${P}/{\eta^2}$ bosonic oscillators contribution of a complex (compact) coordinate. The argument in $\vartheta_1$ contains in particular the field-theory Landau levels, replacing the lattice momenta sums.
For ND or DN boundary conditions (with D standing for Dirichlet), it gives an argument to the $\vartheta_4$ function appearing in the ${\eta}/{\vartheta_4\left(\zeta^{\scriptscriptstyle(i)}_a\tau\right)}$ factors.
The dipole strings (with ends of opposite charges, $i.e.$ attached to the same $D$-brane) have special quantised zero-modes inducing  ``boosted''   string momenta ${m^{\scriptscriptstyle(i)}_a}/{\sqrt{1+\left(2\pi\alpha' H^{\scriptscriptstyle(i)}_a\right)^2}}$\cite{Abouelsaood:1986gd}.

The magnetic fields $H^{\scriptscriptstyle(i)}_a$ are quantised through the standard Dirac quantisation on fluxes 
\begin{equation}
m \int_{T^2} H = 2\pi n,
\end{equation}
where $m$ is the wrapping number and $n$ the flux quantum. This leads to the magnetic field quantisation 
\begin{equation}
2 \pi H^{\scriptscriptstyle(i)}_a \mathcal{A}_i=k^{\scriptscriptstyle(i)}_a,  \qquad k^{\scriptscriptstyle(i)}_a=\frac{n^{\scriptscriptstyle(i)}_a}{m^{\scriptscriptstyle(i)}_a} \in \mathbb{Q}~, \label{defH}
\end{equation}
with $k^{\scriptscriptstyle(i)}_a$ the ratio of the flux number $n^{\scriptscriptstyle(i)}_a$ over the wrapping number $m^{\scriptscriptstyle(i)}_a$ of the D7$_a$ brane on the $i$-th torus $T^2_i$.  We recall that the $T^2_i$ area is $4 \pi^2 \mathcal{A}_i$, see \eqref{Kahlertorus}. Note that due to the $\mathbb{Z}_2$ quotient, $n^{\scriptscriptstyle(i)}_a$ can take half-integer values. This does not change the allowed values for $k^{\scriptscriptstyle(i)}_a$.

In the next sections we will extract the open string mass spectrum from the annulus amplitude. The masses can also be extracted by looking at the different spins of the internal components of the massless states (without magnetic fields), through the mass shift formula given in  \cite{Bachas:1995ik}
\begin{equation}
\Delta m^2=\frac{1}{2 \alpha'} \sum_{i}\left[(2n_i+1)\left|{\zeta_{L}^{\scriptscriptstyle(i)}}+{\zeta_{R}^{\scriptscriptstyle(i)}}\right| + 2 \Sigma_i\left({\zeta_{L}^{\scriptscriptstyle(i)}}+{\zeta_{R}^{\scriptscriptstyle(i)}}\right)\right]. \label{mass_shift}
\end{equation}
The $L,R$ subscripts indicate the string endpoints and have to be replaced by the corresponding brane in the oscillator shift defined in \eqref{defzeta}. The first term in the sum corresponds to the Landau levels, while the second one corresponds to the magnetic moments for the internal $\Sigma_i$ helicities. Landau levels appear only for NN boundary conditions. This formula, which can be derived from the annulus amplitude, can be understood using the field theoretical description of magnetised branes.

\subsection{One magnetised stack} \label{section1field}

We first consider a toy model with only one magnetic field. We turn this magnetic field on the $D7_2$ stack and align it with the common $U(1)$. Hence, the whole stack is magnetised and there is no neutral $D7_2$ brane.  We choose the magnetisation to be on the third torus $T^2_3$ ($i.e.$ in the (89)  direction). Then, according to the notation introduced in equation \eqref{defH}, we denote the magnetic field by $H_2^{\scriptscriptstyle(3)}$, and the associated oscillator shift by $\zeta_2^{\scriptscriptstyle(3)}$.  The configuration is summarised in the following table:
\vspace{5pt}
\begin{center}
\centering
\begin{tabular}{c|ccc}
& (45) & (67) & (89) \\
  \hline
 $ D7_1$ & $\cdot$ & $ \times$  & $\times$  \\
$D7_2 $ &$\times$ & $\cdot$ &$ \otimes $ \\
$D7_3$ & $\times$ & $ \times$  & $\cdot$  \\
\end{tabular}
\end{center}
\vspace{5pt}

The annulus amplitude is computed using the techniques and conventions of \cite{Angelantonj:2000hi,Angelantonj:2002ct,Larosa:2003mz}. For our $D7$-branes model, the different contributions to the annulus amplitude read
\begin{flalign}
8 \mathsf{A}_{0}= & \left({N_1}^2 W_1P_2 P_3+ {N_3}^2 P_1P_2W_3 + 2 {N_2}\bar{N}_2 P_1 W_2 \tilde{P}_3\right) T_{oo}(0,0,0) \nonumber &\\
& \vphantom{ \tilde{P}_3}+2  N_1 N_3 P_2  \,T_{fo}(0,0,0) \left(\frac{\eta}{{\vartheta}_4(0)}\right)^2, \hspace{-10pt} \label{annulus1}
\end{flalign}
\begin{flalign}
8 \mathsf{A}_{1}=  &-2iN_1 {N_2}  T_{ho}(0,0,\zeta_2^{\scriptscriptstyle(3)}\tau) \frac{k_2^{\scriptscriptstyle(3)}\eta}{\vartheta_1(\zeta_2^{\scriptscriptstyle(3)}\tau)} \left(\frac{\eta}{\vartheta_4(0)}\right)^2  +2i N_1 \bar{N}_2  T_{ho}(0,0,-\zeta_2^{\scriptscriptstyle(3)}\tau) \frac{k_2^{\scriptscriptstyle(3)}\eta}{\vartheta_1(-\zeta_2^{\scriptscriptstyle(3)}\tau)} \left(\frac{\eta}{\vartheta_4(0)}\right)^2\nonumber &\\  
&+ \left(2N_3 {N_2}  P_1 T_{go}(0,0,\zeta_2^{\scriptscriptstyle(3)}\tau) +2 N_3 \bar{N}_2 P_1  T_{go}(0,0,-\zeta_2^{\scriptscriptstyle(3)}\tau)\right) \frac{\eta}{\vartheta_4(0)}  \frac{\eta}{\vartheta_4(\zeta_2^{\scriptscriptstyle(3)} \tau)}\,, \hspace{-8pt} \label{annulus2}& 
\end{flalign}
\begin{flalign}
8 \mathsf{A}_{2}= & -i{N_2}^2 P_1W_2 T_{oo}(0,0,2\zeta_2^{\scriptscriptstyle(3)}\tau)  \frac{2k_2^{\scriptscriptstyle(3)}\eta}{\vartheta_1(2\zeta_2^{\scriptscriptstyle(3)}\tau)} +i \bar{{N}_2}^2 P_1W_2 T_{oo}(0,0,-2\zeta_2^{\scriptscriptstyle(3)}\tau)  \frac{2k_2^{\scriptscriptstyle(3)}\eta}{\vartheta_1(-2\zeta_2^{\scriptscriptstyle(3)}\tau)}\,,&  \label{annulus4}
\end{flalign}
where $\mathsf{A}_{0}, \mathsf{A}_{1},\mathsf{A}_{2}$ correspond respectively to the neutral, charged $\pm 1$ and charged $\pm 2$ strings with respect to the magnetised $U(1)$.
In the above expressions, $P_i, W_i$ are the standard momentum and winding sums defined in \Cref{Appendixsums}, while $\tilde{P}_3$ is the sum over boosted momenta $m_3/$\hspace{-2pt}{\small$\sqrt{1+(2\pi\alpha' H_2^{\scriptscriptstyle(3)})^2}$} coming along with dipole strings. Note also that for notational simplicity, the parameter $\tau$ is used instead of the direct channel annulus parameter $\frac i2\, {\rm Im}\tau$.

The $T_{ko}$ characters were introduced in \cref{Tkjfirst} and Appendix \ref{AppendixA}. Their dependence in the magnetic fields is explained around \cref{dependencetau}.
As explained shortly in \cref{sectiontoroidal}, it is easy to trace back the different state contributions to the amplitude: each of ND or DN mixed boundary conditions contributes by a ${\eta}/{\vartheta_4}$ factor (instead of ${1}/{\eta^2}$ for standard bosonic coordinates), with oscillator shift $\zeta_i\tau$ when a magnetic field is present on the N boundary, and each NN boundary condition with magnetic fields introduces Landau levels through a ${\eta}/{\vartheta_1}$ factor with oscillator shifts.

We present hereafter the M\"obius contributions $\mathsf{M}_0$ and $\mathsf{M}_2$, corresponding to neutral and doubly charged strings.  In the M\"obius amplitude both endpoints have to be identical hence there is no simply charged contribution:
\begin{flalign}
8\mathsf{M}_{0}=& - \left({N_1} W_1P_2 P_3+ {N_3} P_1P_2W_3 \right) \hat{T}_{oo}(0,0,0) +( N_1 W_1+N_3P_1)  \,\hat{T}_{og}(0,0,0) \left(\frac{2\hat{\eta}}{{\hat{\vartheta}}_2(0)}\right)^2\nonumber &\\
& + ( N_1 P_2 + N_3 P_2)  \,\hat{T}_{of}(0,0,0) \left(\frac{2\hat{\eta}}{{\hat{\vartheta}}_2(0)}\right)^2+(N_1 P_3+N_3W_3) \,\hat{T}_{oh}(0,0,0) \left(\frac{2\hat{\eta}}{{\hat{\vartheta}}_2(0)}\right)^2,\hspace{-10pt}\label{moebius1}
\end{flalign}
\begin{flalign}
8 \mathsf{M}_{2}=&\,\, i {N_2} P_1W_2 \hat{T}_{oo}(0,0,2\zeta_2^{\scriptscriptstyle(3)}\tau)  \frac{2k_2^{\scriptscriptstyle(3)}\hat{\eta}}{\hat{\vartheta}_1(2\zeta_2^{\scriptscriptstyle(3)}\tau)} -i {\bar{N}_2} P_1W_2 \hat{T}_{oo}(0,0,-2\zeta_2^{\scriptscriptstyle(3)}\tau)  \frac{2k_2^{\scriptscriptstyle(3)}\hat{\eta}}{\hat{\vartheta}_1(-2\zeta_2^{\scriptscriptstyle(3)}\tau)}\,,\label{moebius2}& \\
&- N_2 P_1  \,\hat{T}_{og}(0,0,2\zeta_2^{\scriptscriptstyle(3)}\tau) \frac{2\hat{\eta}}{\hat{\vartheta}_2(0)}\frac{2k_2^{\scriptscriptstyle(3)}\hat{\eta}}{{\hat{\vartheta}}_2(2\zeta_2^{\scriptscriptstyle(3)}\tau)}-\bar{N}_2 P_1 \,\hat{T}_{og}(0,0,-2\zeta_2^{\scriptscriptstyle(3)}\tau) \frac{2\hat{\eta}}{\hat{\vartheta}_2(0)}\frac{2k_2^{\scriptscriptstyle(3)}\hat{\eta}}{\hat{\vartheta}_2(-2\zeta_2^{\scriptscriptstyle(3)}\tau)}\nonumber\\
&-N_2W_2\,\hat{T}_{of}(0,0,2\zeta_2^{\scriptscriptstyle(3)}\tau) \frac{2\hat{\eta}}{\hat{\vartheta}_2(0)}\frac{2k_2^{\scriptscriptstyle(3)}\hat{\eta}}{\hat{\vartheta}_2(2\zeta_2^{\scriptscriptstyle(3)}\tau)}-\bar{N}_2W_2\,\hat{T}_{of}(0,0,-2\zeta_2^{\scriptscriptstyle(3)}\tau) \frac{2\hat{\eta}}{\hat{\vartheta}_2(0)}\frac{2k_2^{\scriptscriptstyle(3)}\hat{\eta}}{\hat{\vartheta}_2(-2\zeta_2^{\scriptscriptstyle(3)}\tau)}\nonumber &\\
&+iN_2\,\hat{T}_{oh}(0,0,2\zeta_2^{\scriptscriptstyle(3)}\tau) \left(\frac{2\hat{\eta}}{\hat{\vartheta}_2(0)}\right)^2\frac{2k_2^{\scriptscriptstyle(3)}\hat{\eta}}{\hat{\vartheta}_1(2\zeta_2^{\scriptscriptstyle(3)}\tau)}-i\bar{N}_2\,\hat{T}_{oh}(0,0,-2\zeta_2^{\scriptscriptstyle(3)}\tau) \left(\frac{2\hat{\eta}}{\hat{\vartheta}_2(0)}\right)^2\frac{2k_2^{\scriptscriptstyle(3)}\hat{\eta}}{\hat{\vartheta}_1(-2\zeta_2^{\scriptscriptstyle(3)}\tau)}.\nonumber
\end{flalign}
The M\"obius amplitude modifies the unitary groups of the unmagnetised branes to orthogonal groups (branes on top of orientifolds). On the other hand, it acts on the magnetised branes by forming states in the antisymmetric representation. The hatted $\hat{T}_{ij}$ characters and $\hat{\vartheta}$ functions are related as usual to the choice of a real basis of characters \cite{Angelantonj:2002ct}. 

The different Chan-Paton multiplicities are as follows: $N_1,N_3$ for the string endpoints aligned with the $D7_1,D7_3$ branes and $N_2,\bar{N}_2$ for the $D7_2$ string endpoints aligned with the $U(1)$ magnetic field, with charge $\pm1$. $N_1$ and $N_3$ are real because they index orthogonal groups. These Chan-Paton multiplicities include the wrapping numbers factors so that they are in fact written as
\begin{equation}
N_a=N_a'm_a^{\scriptscriptstyle{(j)}}m_a^{\scriptscriptstyle{(k)}}, \qquad a\neq j \neq k\neq a, \label{ChanPatonwrappings}
\end{equation}
where $N_a'$ is the true number of branes in the $a$-th stack. Replacing \cref{ChanPatonwrappings} in the amplitudes of \cref{annulus1,annulus2,annulus4,moebius1,moebius2}, one can read the chiral fermion multiplicities through the "intersection number" defined for each magnetised torus $T^2_j$ as
\begin{equation}
I^{\scriptscriptstyle{(j)}}_{ab}=n_a^{\scriptscriptstyle{(j)}}m_b^{\scriptscriptstyle{(j)}}-m_a^{\scriptscriptstyle{(j)}}n_b^{\scriptscriptstyle{(j)}}. \label{intersectionnumber}
\end{equation}
This intersection number  $I_{ab}=\prod_j I^{\scriptscriptstyle{(j)}}_{ab} $ is the index of the Dirac operator of the charged fermions. Taking a specific example, the total multiplicity of the doubly charged states between the $D7_2$ brane and its orientifold image, described by  the amplitude $\mathsf{A}_2$ given in \cref{annulus4}, can be written as 
\begin{equation}
{N_2}^2 \, k_2^{\scriptscriptstyle{(3)}}-{\bar{N}_2}^2 \, k_2^{\scriptscriptstyle{(3)}}=2 {N_2'}^2 (m_2^{\scriptscriptstyle{(1)}}m_2^{\scriptscriptstyle{(3)}})^2k_2^{\scriptscriptstyle{(3)}}=2 {N_2'}^2 {m_2^{\scriptscriptstyle{(1)}}}^2 m_2^{\scriptscriptstyle{(3)}}n_2^{\scriptscriptstyle{(3)}} ={N_2'}^2 {m_2^{\scriptscriptstyle{(1)}}}^2 I_{22'}.
\end{equation}
The ${N_2'}^2 {m_2^{\scriptscriptstyle{(1)}}}^2$ is just the Chan-Paton multiplicity for the unmagnetised torus while $I_{22'}$ is the chiral fermions multiplicity (which is then modified by the orientifold projection). This multiplicity can be understood from the field theoretical point of view as the degeneracy of each Landau level \cite{Bachas:1995ik}.

As usual, the various multiplicities are subject to tadpole cancellation conditions, modified in general by the presence of 3-form fluxes needed for complex structure moduli stabilisation.

The massless states of the original orbifold model are modified by the magnetic field. The charged states receive different contributions (according to the internal spins) resulting to the mass shift \eqref{mass_shift}. We show in the following table the smallest mass shifts for each state ($i.e.$ the new lowest-lying states after magnetic deformation). In the table, the lines and columns entries represent the two possible string endpoints of each state.
\vspace{0.3cm}
\begin{center}
\begin{tabular}{| > {\centering\arraybackslash}m{30pt} | >{\centering\arraybackslash}m{65pt} | >{\centering\arraybackslash}m{80pt}  |  >{\centering\arraybackslash}m{83pt} | }
\hline
& $D7_1 $ & ${D7_2}$ &$ {D7_3}$ \\
  \hline
\rule{0pt}{12pt}$D7_1$ & $\alpha'm^2 =0$  &$\hspace{-24pt} \alpha'm^2 =0$  & \hspace{-22pt} $\alpha'm^2 = 0$ \\
 \hline
  \rule{0pt}{12pt} $D7_2$ & \notableentry & $  \hspace{5pt} \alpha'm^2=-2|\zeta_2^{\scriptscriptstyle(3)}|$  & \hspace{2pt}  $\alpha'm^2 = - |\zeta_2^{\scriptscriptstyle(3)}|$ \\
  \hline
\rule{0pt}{12pt} $D7_3$ &\notableentry  &\notableentry  &$\hspace{-18pt} \alpha'm^2=0$ \\
  \hline
\end{tabular}
\end{center}
\vspace{0.3cm}

We see that tachyonic states can appear in the spectrum \cite{Bachas:1995ik,Angelantonj:2000hi}. 
In order to eliminate them, we introduce appropriate brane separations and/or Wilson lines. In the annulus amplitudes of equations \eqref{annulus1} to \eqref{annulus4}, Wilson lines amount to shifting the momentum numbers in $P_i$ according to the endpoint charges of the strings. Similarly, brane separations shift in $W_i$ the windings numbers.
We then introduce Wilson lines and separations as follows:
\vspace{5pt}
\begin{center}
\begin{tabular}{c|ccc}
& (45) & (67) & (89) \\
  \hline
 $ D7_1$ & $\cdot$ & $ \times$  & $\times$  \\
$D7_2 $ &$\times$ & $\cdot$ &$ \otimes $ \\
$D7_3$ & $\times$ & $ \times$  & $\cdot$  \\
\end{tabular}
\hspace{20pt}$\underset{\text{}}{ \xrightarrow{\hspace*{2cm}} }$\hspace{20pt}
\begin{tabular}{c|lll}
& (45) & (67) & (89) \\
  \hline
 $ D7_1$ & \hspace{4pt} $\cdot$ &   \hspace{6pt}$ \times$  & \hspace{6pt}$\times$  \\
 $D7_2 $ & \hspace{5pt}$\times$ & \hspace{5pt} $\cdot_{\,\, \pm {x}_2 }$ &  \hspace{3pt} $  \otimes $ \\
$D7_3$ &  \hspace{5pt}$\times_{\,A_3}$ &  \hspace{6pt}$ \times$  &   \hspace{5pt} $\cdot$  \\
\end{tabular}
\end{center}
\vspace{5pt}
The index $A_3$ indicates a $U(1)$ Wilson line gauge field, that we take again along the diagonal abelian factor of the $D7_3$ stack and turned on within the torus $T^2_1$ in the (45) internal plan. The $ {x}_2$ index represents the brane positions of the  $D7_2$ brane stack (and $-{x}_2$ for its orientifold image). 

If the Wilson line modulus is projected out by the orbifolding procedure, the model would generally only allow for discrete Wilson lines that can be expressed in the dual lattice as
 \begin{equation}
A_{3}= {a_{3x}}{\bold{R}_1^{*x}}+{a_{3y}}{\bold{R}_{1}^{*y}}, \quad \text{with} \quad a_{3x}, a_{3y} \in \mathbb{Q}\,.\label{wilsonline}
\end{equation}
For $\mathbb{Z}_2$ orbifolds we typically get $a_{3x/y}= \frac12$ (if non-vanishing). This Wilson line gives a mass for the charged fields of the form
\begin{align} 
\alpha' m^2 &=  \alpha'  {A_3} \cdot {A_3}= \alpha' a_{3k}a_{3l}g^{{\scriptscriptstyle{(1)}}kl}=\frac {\alpha'}{\mathcal{A}_1\text{Re}(U_1)}\left| a_{3y}+i U_1 a_{3x}\right|^2 \equiv \frac{\alpha' a_{3}^2(U_1)}{\mathcal{A}_1}\label{Wilson}\,,
\end{align}
where the dimensionful area $\mathcal{A}_1$, the dimensionless complex structure $U_1$,  and the torus metric $g^{\scriptscriptstyle(1)}$ were defined in \eqref{Kahlertorus}, \eqref{Complextorus} and \eqref{torusmetric}. In the last equality we separated the complex structure and K\"ahler modulus $(\mathcal{A}_1)$ dependences. 

Similarly, the $D7_2$ brane position ${x}_2$ can be expressed as 
\begin{equation}
{x}_2 \equiv x_2^x \, \bold{R}_{2x} + x_2^y \, \bold{R}_{2y} \quad \text{with} \quad x_2^x,x_2^y \in \mathbb{Q},\label{positionbrane}
\end{equation}
where we assumed again discretisation of the positions at symmetric points of the fundamental cell. We recall that $\bold{R}_2^x$ and $\bold{R}_2^y$ are the torus lattice vectors defined in \eqref{torusdefinition}. The displacement $x_2$ of the $D7_2$ stack from the origin generates a mass for the strings stretched between the brane stack and its image with respect to the orientifold plane located at the origin:
\begin{equation}
\alpha' m^2= \frac{4 \, x_2 \cdot x_2}{\alpha'}= \frac{4 \, x_2^k  x_2^l g^{\scriptscriptstyle{(2)}}_{kl}}{\alpha'}=\frac{4 \mathcal{A}_2}{{\alpha'} \text{Re}(U_2)}\left|x_2^x-iU_2 x_2^y\right|^2\equiv \frac{y(U_2)\,\mathcal{A}_2}{\alpha'}~. \label{branepositionmass}
\end{equation}
In the last equality we isolated again the complex structure modulus dependence from the K\"ahler modulus one. For more general toroidal orbifolds, the point group symmetry has to be compatible with the stabilised complex structure moduli, so that the Wilson line and brane separation quantisation already incorporates the $U_i$ dependence. 

The new lowest-lying mass states are shown in the table below. 
\vspace{0.3cm}
\begin{center}
\hspace{0 pt}
\begin{tabular}{| > {\centering\arraybackslash}m{29pt} | >{\centering\arraybackslash}m{65pt} |  >{\centering\arraybackslash}m{125pt} |  >{\centering\arraybackslash}m{115pt} | }
\hline
& $D7_1 $ & ${D7_2}$  &$ D7_3$ \\
  \hline
\rule{0pt}{15pt}$D7_1$ & $  \alpha'm^2 =0$  &\hspace{-60pt} $ \alpha' m^2 =0$  &  \hspace{-57pt} $  \alpha'm^2 = 0$ \\
  \hline
  \rule{0pt}{15pt} ${D7_2}$ & \notableentry & $ \alpha' m^2=-2|\zeta_2^{\scriptscriptstyle(3)}|+  \frac{y\mathcal{A}_2}{{\alpha'}}$  & $ \alpha' m^2 = - |\zeta_2^{\scriptscriptstyle(3)}| + \frac{\alpha' a_3^2}{\mathcal{A}_1}$ \\
  \hline
\rule{0pt}{15pt} ${D7_3}$ &  \notableentry&\notableentry  &  \hspace{-57pt} $  \alpha' m^2=0  $ \\
  \hline
\end{tabular}
\end{center}
\vspace{0.3cm}
In the small field approximation (induced by the large volume limit), the oscillator shift reads
\begin{equation}
\zeta_2^{\scriptscriptstyle(3)}=\frac 1{\pi} \text{Arctan}(2\pi\alpha' q H_2^{\scriptscriptstyle(3)})=\frac 1{\pi} \text{Arctan}\left(\frac{\alpha' k_2^{\scriptscriptstyle(3)}}{\mathcal{A}_3}\right)\approx \frac{\alpha' k_2^{\scriptscriptstyle(3)}}{\pi\mathcal{A}_3}.\label{greatVexpansion}
\end{equation}
As will be explained in details in \cref{secnewcacuum}, the K\"ahler moduli stabilisation fixes the $\mathcal{A}_i$ ratios and the tori areas are power fractions of the total volume: $\mathcal{A}_i \equiv \alpha'  r_i  \mathcal{V}^{1/3}$, with $r_1 r_2 r_3 =1$. Hence, the masses of the lowest-lying states read 
\begin{align}
&\alpha' m_{23}^2= - |\zeta_2^{\scriptscriptstyle(3)}| +\frac{ \alpha'  a_3^2}{\mathcal{A}_1} \approx  - \frac{|k_2^{\scriptscriptstyle(3)}|}{\pi r_3 \mathcal{V}^{1/3}} + \frac{a_3^2}{r_1 \mathcal{V}^{1/3}}, \\
&\alpha' m_{22}^2=-2|\zeta_2^{\scriptscriptstyle(3)}|+  \frac{y \mathcal{A}_2}{{\alpha'}}  \approx - \frac{2 |k_2^{\scriptscriptstyle(3)}|}{\pi r_3 \mathcal{V}^{1/3}} + y r_2 \mathcal{V}^{1/3}. \label{massescase1field}
\end{align}
Thus, when $\pi r_3 a_3^2 > r_1|k_2^{\scriptscriptstyle(3)}|$ the $m_{23}^2$ mass is positive for any value of the volume.  For instance, considering $a_{3x}=a_{3y}=\frac 12$ and taking $r_1=r_3$, as will be the case in the following, the condition to eliminate the tachyon in the intersection of $D7_2$ and $D7_3$ branes is 
\begin{equation}
4 \text{Re}(U_1) |k_2^{\scriptscriptstyle(3)}|<\pi |1+i U_1|^2.
\end{equation}
 For instance, in square torus this condition is reduced to $ |k_2^{\scriptscriptstyle(3)}|<\pi$ corresponding from \eqref{defH} to $n_2^{\scriptscriptstyle(3)}<\pi m_2^{\scriptscriptstyle(3)}$, $i.e$ to a flux number smaller than the wrapping number.
Concerning the second lowest-lying massive state on the $D7_2$ branes, we observe that 
\begin{equation}
m_{22}^2 \limitarrow_{\ln \mathcal{V}\rightarrow \pm \infty}  \pm \infty,
\end{equation}
hence, depending on the flux $|k_2^{\scriptscriptstyle(3)}|$ and separation $x$, $m_{22}$ turns negative when the volume falls below a specific value, $e.g.$ $\mathcal{V}_{-}$ of \eqref{Vmin}, as required for our waterfall field candidate.

\subsection{Magnetic fields on each stack}\label{model3fields}

We now consider the following configuration with magnetic fields on each stack, again denoted by a circled cross $\otimes$.
\begin{center}
\begin{tabular}{c|ccc}
& (45) & (67) & (89) \\
  \hline
 $ D7_1$ & $\cdot$ & $ \otimes$  & $\times$  \\
 $D7_2 $ &$\times$ & $\cdot$ &$ \otimes $ \\
$D7_3$ & $\otimes$ & $ \times$  & $\cdot$  \\
\end{tabular}
\end{center}
The different contributions to the annulus amplitude $\mathsf{A}_{0}$, $\mathsf{A}_{1}$ and  $\mathsf{A}_{2}$ corresponding to the neutral, single and double charged strings, read 


\begin{flalign}
8 \mathsf{A}_{0}= & \left(N_1 \bar{N}_1 W_1 \tilde{P}_2 P_3 + {N_2}\bar{N}_2 P_1 W_2 \tilde{P}_3+ N_3\bar{N}_3 \tilde{P}_1P_2W_3  \right) T_{oo}(0,0,0),& \label{annulusbis1}
\end{flalign}
\begin{flalign}
4\mathsf{A}_{1}= &-i\left(N_1{N_2}   T_{fo}(0,\zeta_1^{\scriptscriptstyle(2)}\tau,\zeta_2^{\scriptscriptstyle(3)}\tau) +\bar N_1{N_2}  T_{fo}(0,-\zeta_1^{\scriptscriptstyle(2)}\tau,\zeta_2^{\scriptscriptstyle(3)}\tau)\right)\frac{k_2^{\scriptscriptstyle(3)}\eta^3}{\vartheta_4(0)\vartheta_4(\zeta_1^{\scriptscriptstyle(2)}\tau)\vartheta_1(\zeta_2^{\scriptscriptstyle(3)}\tau)}  & \nonumber \\
 & +i\left(N_1{\bar N_2}   T_{fo}(0,\zeta_1^{\scriptscriptstyle(2)}\tau,-\zeta_2^{\scriptscriptstyle(3)}\tau) + \bar N_1{\bar N_2}  T_{fo}(0,-\zeta_1^{\scriptscriptstyle(2)}\tau,-\zeta_2^{\scriptscriptstyle(3)}\tau)\right)\frac{k_2^{\scriptscriptstyle(3)}\eta^3}{\vartheta_4(0)\vartheta_4(\zeta_1^{\scriptscriptstyle(2)}\tau)\vartheta_1(-\zeta_2^{\scriptscriptstyle(3)}\tau)}    \nonumber \\
 & -i\left(N_1{N_3}  T_{fo}(\zeta_3^{\scriptscriptstyle(1)}\tau,\zeta_1^{\scriptscriptstyle(2)}\tau,0) +N_1{\bar N_3}   T_{fo}(-\zeta_3^{\scriptscriptstyle(1)}\tau,\zeta_1^{\scriptscriptstyle(2)}\tau,0) \right)\frac{k_1^{\scriptscriptstyle(2)}\eta^3}{\vartheta_4(\zeta_3^{\scriptscriptstyle(1)}\tau)\vartheta_1(\zeta_1^{\scriptscriptstyle(2)}\tau)\vartheta_4(0)} \nonumber \\
 & +i\left( \bar N_1{N_3}  T_{fo}(\zeta_3^{\scriptscriptstyle(1)}\tau,-\zeta_1^{\scriptscriptstyle(2)}\tau,0) + \bar N_1{\bar N_3}  T_{fo}(-\zeta_3^{\scriptscriptstyle(1)}\tau,-\zeta_1^{\scriptscriptstyle(2)}\tau,0) \right)\frac{k_1^{\scriptscriptstyle(2)}\eta^3}{\vartheta_4(\zeta_3^{\scriptscriptstyle(1)}\tau)\vartheta_1(-\zeta_1^{\scriptscriptstyle(2)}\tau)\vartheta_4(0)} \nonumber \\
&- i\left(N_2 {N_3}   T_{go}(\zeta_3^{\scriptscriptstyle(1)}\tau,0,\zeta_2^{\scriptscriptstyle(3)}\tau) +N_2 {\bar N_3}   T_{go}(-\zeta_3^{\scriptscriptstyle(1)}\tau,0,\zeta_2^{\scriptscriptstyle(3)}\tau)\right)\frac{k_3^{\scriptscriptstyle(1)}\eta^3}{\vartheta_4(\zeta_3^{\scriptscriptstyle(1)}\tau)\vartheta_4(0)\vartheta_1(\zeta_2^{\scriptscriptstyle(3)}\tau)}\nonumber\\
&+i \left(\bar N_2 \bar{N}_3   T_{go}(-\zeta_3^{\scriptscriptstyle(1)}\tau,0,-\zeta_2^{\scriptscriptstyle(3)}\tau)+ \bar N_2 N_3   T_{go}(\zeta_3^{\scriptscriptstyle(1)}\tau,0,-\zeta_2^{\scriptscriptstyle(3)}\tau) \right)\frac{k_3^{\scriptscriptstyle(1)}\eta^3}{\vartheta_4(\zeta_3^{\scriptscriptstyle(1)}\tau)\vartheta_4(0)\vartheta_1(-\zeta_2^{\scriptscriptstyle(3)}\tau)}, \hspace{-5cm} \label{annulusbis2}
\end{flalign}
\begin{flalign}
8 \mathsf{A}_{2}= &-iN_1^2 W_1 P_3  T_{oo}(0,2\zeta_1^{\scriptscriptstyle(2)}\tau,0) \frac{2k_1^{\scriptscriptstyle(2)}\eta}{\vartheta_1(2\zeta_1^{\scriptscriptstyle(2)}\tau)}+i{\bar N_1}^2 W_1 P_3  T_{oo}(0,-2\zeta_1^{\scriptscriptstyle(2)}\tau,0) \frac{2k_1^{\scriptscriptstyle(2)}\eta}{\vartheta_1(-2\zeta_1^{\scriptscriptstyle(2)}\tau)} \nonumber &\\
&-i N_2^2 P_1W_2 T_{oo}(0,0,2\zeta_2^{\scriptscriptstyle(3)}\tau)  \frac{2k_2^{\scriptscriptstyle(3)}\eta}{\vartheta_1(2\zeta_2^{\scriptscriptstyle(3)}\tau)}+i{\bar N_2}^2 P_1W_2 T_{oo}(0,0,-2\zeta_2^{\scriptscriptstyle(3)}\tau)  \frac{2k_2^{\scriptscriptstyle(3)}\eta}{\vartheta_1(-2\zeta_2^{\scriptscriptstyle(3)}\tau)}& \nonumber\\
&-iN_3^2 P_2 W_3  T_{oo}(2\zeta_3^{\scriptscriptstyle(1)}\tau,0,0) \frac{2k_3^{\scriptscriptstyle(1)}\eta}{\vartheta_1(2\zeta_3^{\scriptscriptstyle(1)}\tau)}+i{\bar N_3}^2 P_2 P_3  T_{oo}(-2\zeta_3^{\scriptscriptstyle(1)}\tau,0,0) \frac{2k_3^{\scriptscriptstyle(1)}\eta}{\vartheta_1(-2\zeta_3^{\scriptscriptstyle(1)}\tau)}.  \label{annulusbis4} &
\end{flalign}
Exactly the same comments as those under \cref{annulus1,annulus2,annulus4} apply here. The M\"obius contributions have similar forms as those in \cref{moebius1,moebius2}, and are omitted here since they play no role in our arguments. They act as for the magnetised brane of \cref{section1field}, generating states in antisymmetric representations of the gauge groups. They also modify the chiral fermion multiplicity described around equation \eqref{intersectionnumber}.

The masses of the lowest-lying states of the spectrum are shown in the following table:
\vspace{0.5cm}
\begin{center}
\begin{tabular}{| > {\centering\arraybackslash}m{30pt} | >{\centering\arraybackslash}m{75pt} | >{\centering\arraybackslash}m{95pt}  |  >{\centering\arraybackslash}m{95pt} | }
\hline
& $D7_1 $ & $D7_2$ &$ {D7_3}$ \\
  \hline
\rule{0pt}{15pt}\hspace{2pt}$D7_1$ & $ \alpha' m^2 =-2|\zeta_1^{\scriptscriptstyle(2)}|$  & $ \alpha' m^2 =|\zeta_2^{\scriptscriptstyle(3)}|-|\zeta_1^{\scriptscriptstyle(2)}|$  &  $  \alpha' m^2 = |\zeta_1^{\scriptscriptstyle(2)}|-|\zeta_3^{\scriptscriptstyle(1)}|$ \\
 \hline
  \rule{0pt}{15pt} $D7_2$ & \notableentry & \hspace{-14pt}$ \alpha' m^2 =-2|\zeta_2^{\scriptscriptstyle(3)}|$  &   $ \alpha' m^2 =|\zeta_3^{\scriptscriptstyle(1)}| - |\zeta_2^{\scriptscriptstyle(3)}|$ \\
  \hline
\rule{0pt}{15pt} $D7_3$ &\notableentry  &\notableentry  &\hspace{-18pt} $ \alpha' m^2=-2|\zeta_3^{\scriptscriptstyle(1)}|$ \\
  \hline
\end{tabular}
\end{center}
\vspace{0.5cm}
We see that two different kinds of states appear: the $D7_a$--$D7_a$ (doubly charged) states, and the mixed $D7_a$--$D7_b$ ones, with $a\neq b$. The mass of the former can be uplifted as in the previous subsection and will be explained below. We can use neither Wilson lines nor brane separations to increase the mixed states masses, since these can be introduced only in directions without magnetic fields, $i.e.$ along both worldvolumes (for Wilson lines), or transverse to both stacks (for separations). In the directions along the magnetic field, zero modes of gauge potentials are gauge artifacts and thus unphysical. We must then specify the fields $H_a^{\scriptscriptstyle(i)}$ in order to eliminate the tachyons, at least at large volumes. By a simple inspection of the table above, it follows that the only way to eliminate all three potential tachyons in the $D7_a$--$D7_b$ brane intersections is to choose 
\begin{equation}
|\zeta_1^{\scriptscriptstyle(2)}|=|\zeta_2^{\scriptscriptstyle(3)}|=|\zeta_3^{\scriptscriptstyle(1)}|. \label{mixedstatecondition}
\end{equation}
The corresponding lowest-lying states then become massless.

As mentioned above, to uplift the tachyons on the $D7_a$--$D7_a$ sectors, we can introduce distance separations between branes and their images (in the direction orthogonal to their worldvolume), or Wilson lines $i.e.$ constant background gauge fields (on the unmagnetised worldvolume torus), as in \cref{section1field}. We show below a configuration keeping only one potential tachyonic state that can play the role of the waterfall field:
\vspace{0.7cm}
\begin{center}
\begin{tabular}{c|ccc}
& (45) & (67) & (89) \\
  \hline
 $ D7_1$ & $\cdot$ & $ \otimes$  & $\times$  \\
 $D7_2 $ &$\times$ & $\cdot$ &$ \otimes $ \\
$D7_3$ & $\otimes$ & $ \times$  & $\cdot$  \\
\end{tabular}
\hspace{20pt}$\underset{\text{}}{ \xrightarrow{\hspace*{2cm}} }$\hspace{20pt}
\begin{tabular}{c|lll}
& (45) & (67) & (89) \\
  \hline
 $ D7_1$ & \hspace{4pt} $\cdot$ &   \hspace{6pt}$ \otimes$  & \hspace{6pt}$\times_{A_1}$  \\
 $D7_2 $ & \hspace{5pt}$\times$ & \hspace{5pt} $\cdot_{\,\, \pm {x}_2 }$ &  \hspace{3pt} $  \otimes $ \\
$D7_3$ &  \hspace{5pt}$\otimes$ &  \hspace{6pt}$ \times_{A_3}$  &   \hspace{5pt} $\cdot$  \\
\end{tabular}
\end{center}
\vspace{0.7cm}
Using the notation of the previous subsection, we introduce (discrete) Wilson lines along the third torus $T^2_3$ for the $D7_1$ stack and along the second torus $T^2_2$ for the $D7_3$ stack, while we separate the $D7_2$ stack from its orientifold image in its transverse directions. The masses for the double charge states in the three brane stacks now become:
\begin{align}
 &\alpha' m_{11}^2= -2|\zeta_1^{\scriptscriptstyle(2)}|+  \frac{\alpha'a_1^2}{ \mathcal{A}_3} \approx  - \frac{2 \alpha' |k_1^{\scriptscriptstyle(2)}|}{\pi \mathcal{A}_2}+ \frac{\alpha' a_1^2}{ \mathcal{A}_3}  \approx - \frac{2 |k_1^{\scriptscriptstyle(2)}|}{\pi r_2 \mathcal{V}^{1/3}} + \frac{a_1^2}{ r_3 \mathcal{V}^{1/3}} ,\\
 &\alpha' m_{22}^2=-2|\zeta_2^{\scriptscriptstyle(3)}|+   \frac{y \mathcal{A}_2}{\alpha'} \approx - \frac{2 \alpha' |k_2^{\scriptscriptstyle(3)}|}{\pi \mathcal{A}_3}+   \frac{y \mathcal{A}_2}{\alpha'}= - \frac{2 |k_2^{\scriptscriptstyle(3)}|}{\pi r_3 \mathcal{V}^{1/3}} +  y r_2 \mathcal{V}^{1/3}, \label{mass22state}\\
 &\alpha' m_{33}^2= -2|\zeta_3^{\scriptscriptstyle(1)}|+  \frac{ \alpha' a_3^2}{ \mathcal{A}_2} \approx  - \frac{2\alpha' |k_3^{\scriptscriptstyle(1)}|}{\pi \mathcal{A}_1}+  \frac{\alpha' a_3^2}{ \mathcal{A}_2} \approx  - \frac{2 |k_3^{\scriptscriptstyle(1)}|}{\pi r_1 \mathcal{V}^{1/3}} + \frac{a_3^2}{ r_2\mathcal{V}^{1/3}}.
\end{align}
To obtain the second equality of each equation we used large volume expansions for $\zeta_a^{\scriptscriptstyle(j)}$ as in \cref{greatVexpansion}. The Wilson lines and brane position parameters are defined as in \cref{Wilson,positionbrane}. Similarly to the single magnetic field case of \cref{section1field}, by choosing appropriately $a_{1}$, $a_3$ and the values of the magnetic fluxes $|k_1^{\scriptscriptstyle{(2)}}|$ and $|k_3^{\scriptscriptstyle{(1)}}|$, one can eliminate the $D7_1$--$D7_1$ and  $D7_3$--$D7_3$ tachyons. For instance, as explained after \cref{massescase1field},  for $a_i=1/2$ typical for $\mathbb{Z}_2$ orbifolds, this requires flux numbers smaller than wrapping numbers. 
On the other hand, the $D7_{2}$--$D7_{2}$ state becomes tachyonic at and below a critical value of the volume that can be chosen to be around $\mathcal{V}_{-}$ (defined in \eqref{Vmin}), 
as required for the waterfall field.

\subsection{Magnetic fields on entire worldvolumes}\label{multiplefieldseachbrane}

In the previous subsection we saw that in order to eliminate the mixed-state tachyons from brane intersections we had to impose condition \eqref{mixedstatecondition}. We now relax this condition by introducing magnetic fields in all worldvolume tori as shown below:

\begin{center}
\begin{tabular}{c|ccc}
& (45) & (67) & (89) \\
  \hline
 $ D7_1$ & $\cdot$ & $ \otimes$  & $\otimes$  \\
 $D7_2 $ &$\otimes$ & $\cdot$ &$ \otimes $ \\
$D7_3$ & $\otimes$ & $ \otimes$  & $\cdot$  \\
\end{tabular}
\end{center}
The masses of the potential tachyonic states can be extracted by computing the annulus amplitude as done before and they are shown in the following table:
\begin{center}
\resizebox{\columnwidth}{!}{
\begin{tabular}{| >{\centering\arraybackslash}m{25pt} | >{\centering\arraybackslash}m{110pt} | >{\centering\arraybackslash}m{155pt}  |  >{\centering\arraybackslash}m{155pt} | }
\hline
\rule{0pt}{14pt} & $D7_1 $ & $D7_2$ &$ {D7_3}$ \\
  \hline 
 \vspace{4pt}\hspace{2pt}$D7_1$\vspace{4pt} & $ \alpha' m^2 =-2\left|\zeta_1^{\scriptscriptstyle(2)}+\zeta_1^{\scriptscriptstyle(3)}\right|$  & $  \alpha' m^2=\left|\zeta_1^{\scriptscriptstyle(3)}\pm\zeta_2^{\scriptscriptstyle(3)}\right|-\left|\zeta_1^{\scriptscriptstyle(2)}\pm\zeta_2^{\scriptscriptstyle(1)}\right| $ &  $ \alpha' m^2=\left|\zeta_1^{\scriptscriptstyle(2)}\pm\zeta_3^{\scriptscriptstyle(2)}\right|-\left|\zeta_1^{\scriptscriptstyle(3)}\pm\zeta_3^{\scriptscriptstyle(1)}\right| $\\
 \hline
\vspace{4pt}\hspace{2pt}$D7_2$\vspace{4pt}  & \notableentry & \hspace{-48pt}$ \alpha' m^2 =-2\left|\zeta_2^{\scriptscriptstyle(1)}+\zeta_2^{\scriptscriptstyle(3)}\right|$  &  $  \alpha' m^2=\left|\zeta_2^{\scriptscriptstyle(1)}\pm\zeta_3^{\scriptscriptstyle(1)}\right|-\left|\zeta_2^{\scriptscriptstyle(3)}\pm\zeta_3^{\scriptscriptstyle(2)}\right| $ \\
  \hline
\vspace{4pt}\hspace{2pt}$D7_3$\vspace{4pt}  &\notableentry  &\notableentry  & \hspace{-48pt}$\alpha' m^2 =-2\left|\zeta_3^{\scriptscriptstyle(1)}+\zeta_3^{\scriptscriptstyle(2)}\right|$ \\
  \hline
\end{tabular}
}
\end{center}
\vspace{0.3cm}
where the $\pm$ signs in the same equality have to be identical ($e.g.$ if the first $\pm$ is a $+$, the second is $+$ as well.)

The mixed states $D7_a$--$D7_b$, $a\neq b$ can be eliminated by choosing an appropriate field configuration, satisfying a system of inequalities defined by the positivity of the corresponding mass expressions in the table: 
\begin{equation}
\begin{cases}
\vspace{5pt} \left|\zeta_1^{\scriptscriptstyle(3)}\pm\zeta_2^{\scriptscriptstyle(3)}\right|-\left|\zeta_1^{\scriptscriptstyle(2)}\pm\zeta_2^{\scriptscriptstyle(1)}\right|\ge0 \\
\vspace{5pt}\left|\zeta_1^{\scriptscriptstyle(2)}\pm\zeta_3^{\scriptscriptstyle(2)}\right|-\left|\zeta_1^{\scriptscriptstyle(3)}\pm\zeta_3^{\scriptscriptstyle(1)}\right|\ge 0\\
\left|\zeta_2^{\scriptscriptstyle(1)}\pm\zeta_3^{\scriptscriptstyle(1)}\right|-\left|\zeta_2^{\scriptscriptstyle(3)}\pm\zeta_3^{\scriptscriptstyle(2)}\right|\ge 0
\end{cases}
\end{equation}
This system is solved by the following configurations
\begin{alignat}{2}
&(A- && 1) \qquad \zeta_1^{\scriptscriptstyle(3)}=\zeta_2^{\scriptscriptstyle(1)}=\zeta_3^{\scriptscriptstyle(2)}, \qquad \zeta_1^{\scriptscriptstyle(2)}= \zeta_2^{\scriptscriptstyle(3)}=\zeta_3^{\scriptscriptstyle(1)} ;  \nonumber\\
& \qquad && 2) \qquad \zeta_1^{\scriptscriptstyle(3)}=\zeta_2^{\scriptscriptstyle(1)}=-\zeta_3^{\scriptscriptstyle(2)}, \qquad \zeta_1^{\scriptscriptstyle(2)}= \zeta_2^{\scriptscriptstyle(3)}=-\zeta_3^{\scriptscriptstyle(1)} ;  \nonumber\\
& \qquad && 3) \qquad \zeta_1^{\scriptscriptstyle(3)}=-\zeta_2^{\scriptscriptstyle(1)}=\zeta_3^{\scriptscriptstyle(2)}, \qquad \zeta_1^{\scriptscriptstyle(2)}= -\zeta_2^{\scriptscriptstyle(3)}=\zeta_3^{\scriptscriptstyle(1)} ;  \nonumber\\
&\qquad && 4) \qquad \zeta_1^{\scriptscriptstyle(3)}=-\zeta_2^{\scriptscriptstyle(1)}=-\zeta_3^{\scriptscriptstyle(2)}, \qquad \zeta_1^{\scriptscriptstyle(2)}=- \zeta_2^{\scriptscriptstyle(3)}=-\zeta_3^{\scriptscriptstyle(1)} ;  \nonumber\\
&(B- &&1) \qquad \zeta_1^{\scriptscriptstyle(2)}=\zeta_1^{\scriptscriptstyle(3)}, \qquad \zeta_2^{\scriptscriptstyle(1)}=\zeta_2^{\scriptscriptstyle(3)}, \qquad  \zeta_3^{\scriptscriptstyle(1)}=\zeta_3^{\scriptscriptstyle(2)} ; \label{conditionsentireworldvolumes}\\
& &&2) \qquad \zeta_1^{\scriptscriptstyle(2)}=-\zeta_1^{\scriptscriptstyle(3)}, \qquad \zeta_2^{\scriptscriptstyle(1)}=-\zeta_2^{\scriptscriptstyle(3)}, \qquad  \zeta_3^{\scriptscriptstyle(1)}=-\zeta_3^{\scriptscriptstyle(2)} ; \nonumber 
\end{alignat}
for which all inequalities are saturated and the lowest-lying mixed states become massless.

For the solutions $(A-i)$, all the double charged states $D7_a$--$D7_a$ have identical tachyonic masses equal to $\alpha'm^2 =-2\left|\zeta_1^{\scriptscriptstyle(2)}+\zeta_1^{\scriptscriptstyle(3)}\right|$, while for solution $(B-1)$ they can have different masses. Solution $(B-2)$ is the supersymmetry preserving one, with all lowest-lying states remaining massless. 
In both $(A-i)$ and $(B-1)$ cases, the study of tachyonic states and their elimination through Wilson lines and brane separations is identical to the one of \cref{model3fields}. Nevertheless, we see that we are allowed to have more complex configurations than with only one magnetic field on each brane.

\section{Effect of waterfall field on the dS vacuum and inflation}\label{sectioneffectivetheory}

We now apply the K\"ahler moduli stabilisation mechanism described in \cref{sectionintromodel} in our model of \cref{model3fields} with matter fields living on the magnetised $D7$ branes, and study the novelty introduced by the waterfall direction. We first compute the effective field theory scalar potential for the K\"ahler moduli and the newly introduced matter fields, and then describe the new vacuum of the theory.

 The scalar potential depends on the total internal volume $\mathcal{V}=\mathcal{A}_1\mathcal{A}_2\mathcal{A}_3/{\alpha'}^3=\sqrt{\tau_1\tau_2\tau_3}$ through the F-part described in \cref{sectionintromodel}, and on the K\"ahler moduli $\tau_a$ (or equivalently on the 2-tori areas $\mathcal{A}_a$ in the present case) through the D-part. Moreover, it has a new F-part depending on the matter fields of the $D7$-branes. As we are interested in the waterfall direction, we only keep track of possible tachyonic matter field contributions to the scalar potential and put the other (massive) matter fields to zero. The canonically normalised tachyonic field, coming from the $D7_2-D7_2$  state of  \cref{model3fields}, is denoted $\varphi_-$ (and its charge conjugate $\varphi_+$) in the following. 
 For simplicity, we recall the brane configuration of  \cref{model3fields} in the following table.
 
\begin{equation}
\begin{tabular}{c|lll}
& (45) & (67) & (89) \\
  \hline
 $ D7_1$ & \hspace{4pt} $\cdot$ &   \hspace{6pt}$ \otimes$  & \hspace{6pt}$\times_{A_1}$  \\
 $D7_2 $ & \hspace{5pt}$\times$ & \hspace{5pt} $\cdot_{\,\, \pm {x}_2 }$ &  \hspace{3pt} $  \otimes $ \\
$D7_3$ &  \hspace{5pt}$\otimes$ &  \hspace{6pt}$ \times_{A_3}$  &   \hspace{5pt} $\cdot$  \\
\end{tabular}\label{config1tachyon}
\end{equation}

 For simplicity we will consider wrapping numbers $m_2^{\scriptscriptstyle(1)}=m_2^{\scriptscriptstyle(3)}=1$ and $N_2'=1$ such that the $D7_2$ gauge group is restricted to $U(1)_2$. See \cref{ChanPatonwrappings} for  the definition of  $N_2'$. The number of chiral fermions after orientifold projection is denoted by given by $n^{\scriptscriptstyle (\Omega)}_{22'}$. The tachyonic state will hence also have multiplicity $n^{\scriptscriptstyle (\Omega)}_{22'}$, corresponding to the different Landau states and related to the intersection numbers. In the following we will often refer to ``the tachyon" while describing all the degenerate tachyonic scalars together, because once the tachyon gets a non-vanishing VEV, a specific direction is fixed for all the Landau states, producing a massive field and $n^{\scriptscriptstyle (\Omega)}_{22'}-1$ Goldstone modes.

 \subsection{D-term from magnetic fields} The magnetic fields can be described in the effective theory through a D-term scalar potential 
 \begin{align}
V_D&= \sum_{a}  \frac{g^2_{U(1)_a}}{2}\left(\xi_a+\sum_n q^n_a |\varphi^n_a|^2\right)^2 + \cdots \nonumber\\
&=\sum_{a=1,3} \frac{g^2_{U(1)_a}}{2}\xi_a^2+\frac{g^2_{U(1)_2}}{2}\left(\xi_2+2 |\varphi_+|^2-2|\varphi_-|^2+\cdots\right)^2+\cdots \label{Dscalarpot}
\end{align}
 In the first line, the sum runs over  the $n$ charged scalar fields. As explained above, in the second line of \eqref{Dscalarpot} we have only kept the tachyonic field (and its charge conjugate) contributions, with charges $q_a=\pm 2$. 
  
 The Fayet-Iliopoulos parameters $\xi_a$ and gauge couplings $g^2_{U(1)_a}$ used in the D-term scalar potential depend on the K\"ahler moduli.  Indeed, from the D-term \eqref{Dscalarpot} and from the string frame expressions \eqref{defzeta}-\eqref{mass_shift}, we can write the magnetic field contribution to the mass of the matter fields in the configuration of table \eqref{config1tachyon} as 
\begin{equation}
m_{H_2}^2 \equiv 2 g^2_{U(1)_2} \xi_2= \frac{2|\zeta_2^{\scriptscriptstyle(3)}|}{\alpha'} \approx \frac{2|k_2^{\scriptscriptstyle(3)}|}{\pi\alpha'}\frac{\alpha'}{\mathcal{A}_3}\approx \frac{2|k_2^{\scriptscriptstyle(3)}|}{\pi} \frac{g_s^2}{\kappa^2 \mathcal{V}}\frac{\alpha'}{\mathcal{A}_3}. \label{masseffectivetheory}
\end{equation}
We recall that $\zeta_2^{\scriptscriptstyle(3)}$ is given in equation \eqref{defzeta} and hence the third equality holds in the small magnetic field (large volume) limit.
In order to go to the supergravity frame, we used the four dimensional Planck constant expression
\begin{equation}
\frac{1}{\kappa^2}\equiv \frac{1}{\kappa_4^2}=\frac{\tilde{\mathcal{V}}}{\kappa_{10}^2}=\frac{\tilde{\mathcal{V}}}{ \alpha'g_s^2(4\pi^2\alpha')^{3}}= \frac{\mathcal{V}}{\alpha' g_s^2}, \label{stringlength}
\end{equation}
where we restored the string units in the total volume $\tilde{\mathcal{V}}=(4\pi^2)^3{\alpha'}^3\mathcal{V}=(4\pi^2)^3\mathcal{A}_1\mathcal{A}_2\mathcal{A}_3$. 

The gauge couplings are expressed in terms of the magnetised D7 brane worldvolumes as 
\begin{equation}
\frac{1}{g^2_{U(1)_a}} = \frac{| m_a^{\scriptscriptstyle(j)}m_a^{\scriptscriptstyle(k)}|}
{g_s{\alpha'}^2}\left|\mathcal{A}_j+i\alpha'k_a^{\scriptscriptstyle(j)}\right|\left|\mathcal{A}_k+i\alpha'k_a^{\scriptscriptstyle(k)}\right|, \qquad \text{with} \quad a\neq j \neq k \neq a.   \label{couplings}
\end{equation}
In the small magnetic fields (large areas) limit, the couplings \eqref{couplings} reduce to
\begin{equation}
\frac{1}{g^2_{U(1)_a}} \approx | m_a^{\scriptscriptstyle(j)}m_a^{\scriptscriptstyle(k)}|\frac{\mathcal{A}_j\mathcal{A}_k}{g_s{\alpha'}^2}=| m_a^{\scriptscriptstyle(j)}m_a^{\scriptscriptstyle(k)}|\frac{\mathcal{V}}{g_s}\frac{\alpha'}{\mathcal{A}_a}\,, \qquad \text{with} \quad a\neq j \neq k \neq a. \label{couplingslargev}
\end{equation}
Combining equations \eqref{masseffectivetheory} to \eqref{couplings}, we deduce the expressions for the moduli dependent Fayet-Iliopoulos term
\begin{equation}
\xi_2=\frac{m_{H_2}^2}{2 g^2_{U(1)_2}} \approx  | m_2^{\scriptscriptstyle(1)}m_2^{\scriptscriptstyle(3)} |  \frac{g_s |k_2^{\scriptscriptstyle(3)}|}{\pi \kappa^2 \mathcal{V}}\frac{ \mathcal{A}_1}{\alpha'}. \label{FIfromDpot}
\end{equation}
We obtain similar expressions for $\xi_1$ and $\xi_3$ for the configuraition of \eqref{config1tachyon}, so that the D-term part of the scalar potential \eqref{Dscalarpot} reads
\begin{equation}
V_D \approx \frac{1}{\kappa^4 \mathcal{V}^2} \left(d_1 \frac{\mathcal{A}_3}{\mathcal{A}_2}+d_2 \frac{\mathcal{A}_1}{\mathcal{A}_3}+d_3\frac{\mathcal{A}_2}{\mathcal{A}_1}\right)+ m_{H_2}^2 \left(|\varphi_+|^2-|\varphi_-|^2\right) + 
2g^2_{U(1)_2} \left(|\varphi_+|^2-|\varphi_-|^2\right)^2\hspace{-3pt}, \label{potentialsugra}
\end{equation}
where we defined the K\"ahler moduli D-term parameters
\begin{equation}
 d_a\equiv \frac{g^2_{U(1)_a}}{2}\xi_a^2= \frac 12  g_s^3 | m_a^{\scriptscriptstyle(j)}m_a^{\scriptscriptstyle(k)}| \left(\frac{k_a^{\scriptscriptstyle(j)}}{\pi}\right)^2\hspace{-3pt}, \,\, {\rm with}\,\,  (a,j,k)=(\sigma(1),\sigma(2),\sigma(3)) \,\, \text{and $\sigma$ a 3-cycle.} \label{dterms}
 \end{equation}
Note again that the above $d_a$ 
correspond to the specific flux configuration of \eqref{config1tachyon}. 
 
 \subsection{F-term from brane separation} Appart from the D-term potential, the effective field theory contains a positive mass contribution for the tachyonic scalars of the model described in \cref{model3fields}. These scalars come from strings stretching between the $D7_2$ brane stack and its image, and the positive contribution to their mass is due to the distance separation between the brane and its orientifold image. It is generated by the VEV of an adjoint scalar coming from strings with both ends on the $D7_2$ stack and preserves supersymmetry, in contrast to the tachyonic contribution from the magnetic field discussed above.

More precisely, this contribution is described by a trilinear superpotential obtained by an appropriate $\mathcal{N}=1$ truncation of an $\mathcal{N}=4$ supersymmetric theory within the untwisted orbifold sector:
 \begin{equation}
 \mathcal{W}_{C_i^{7_a}}\ni \text{Tr}\left(C_1^{7_a}\left[C_2^{7_a},C_3^{7_a}\right]\right). \label{SYMcoupling}
 \end{equation}
The $C_j^{7_a}$ for $j=1,2,3$ are the three $\mathcal{N}=1$ chiral multiplets that are part of an $\mathcal{N}=4$ vector multiplet living on the $D7_a$ brane stack. 
$C_a^{7_a}$ parametrise the brane position in the transverse plane while $C_j^{7_a}$ with $j\neq a$ are the internal components of the 8d gauge fields along the two planes of the worldvolume of the $D7_a$ brane~\cite{Lust:2004fi,Camara:2004jj}. As explained above, the couplings of interest are given by equation \eqref{SYMcoupling}, with $a=2$. We can then identify the relevant superpotential in our case from\footnote{In our conventions the superpotential and all un-normalised fields are dimensionless.}:
  \begin{equation}
 \mathcal{W}_{C_i^{7_2}}= w_{ijk}C_i^{7_2}C_j^{7_2}C_k^{7_2} \ni \, c\,
 \Phi_2 \Phi_+\Phi_-.\label{superpotmattergeneric}
 \end{equation}
Here $\Phi_i$ are the un-normalised fields: $\Phi_2$ is the modulus associated with the $D7_2$ brane position $x_2$ of \cref{model3fields}, hence $C_2^{7_2}$,  while $\Phi_-$ (and $\Phi_+$) is the tachyonic matter field of interest (and its charge conjugate) assimilated to $C_1^{7_2}$ and $C_3^{7_2}$. 
When $\Phi_2$ acquires a non-vanishing VEV $\langle \Phi_2 \rangle \sim x_2$, the superpotential \eqref{superpotmattergeneric} generates a (supersymmetric) mass for the matter fields $\Phi_+$ and $\Phi_-$.

 The physical mass for the canonically normalised fields $\varphi_i$ can be computed from the physical Yukawa couplings derived from the supergravity action \cite{Kaplunovsky:1993rd,Brignole:1998dxa,Ibanez:2012zz} and expressed as 
 \begin{equation}
 \mathcal{W}_{tach}=Y_{ijk}\, \varphi_i\varphi_j\varphi_k, \qquad {\rm with} \quad Y_{ijk}={w}_{ijk} \, (\mathcal{K}_{i\bar i} \mathcal{K}_{j\bar{j}} \mathcal{K}_{k\bar{k}})^{-\frac12}e^{\frac{\kappa^2}2 \mathcal{K}} . \label{yukawa}
 \end{equation}
 $\mathcal{K}_{i\bar{i}}$ are the K\"ahler metrics of the matter fields of interest (assuming no kinetic mixing), and ${w}_{ijk}$ is the trilinear coupling of the holomorphic superpotential, which in our case is simply related to $c$ defined in \eqref{superpotmattergeneric}. In the type IIB string framework, and for the untwisted fields appearing in \eqref{SYMcoupling}, the K\"ahler metrics of the matter fields on magnetised tori read \cite{Ibanez:2012zz,Ibanez:1998rf,Lust:2004cx,Lust:2004fi,Font:2004cx}
 \begin{align}
 &\kappa^2\mathcal{K}_{C_1^{7_2}\bar{C}_1^{7_2}}
 =\frac{\pi e^{\phi_4}}{(U_1+\bar{U}_1)}\sqrt{\frac{\alpha'\mathcal{A}_1}{\mathcal{A}_2\mathcal{A}_3}}\left|\frac{m_2^{\scriptscriptstyle(3)}}{m_2^{\scriptscriptstyle(1)}}\right|\left|\frac{\mathcal{A}_3+i \alpha' k_2^{\scriptscriptstyle(3)}}{\mathcal{A}_1+i \alpha' k_2^{\scriptscriptstyle(1)}}\right|,
\\
&  \kappa^2\mathcal{K}_{C_3^{7_2}\bar{C}_3^{7_2}}
=\frac{\pi e^{\phi_4}}{(U_3+\bar{U}_3)}\sqrt{\frac{\alpha'\mathcal{A}_3}{\mathcal{A}_1\mathcal{A}_2}}\left|\frac{m_2^{\scriptscriptstyle(1)}}{m_2^{\scriptscriptstyle(3)}}\right|\left|\frac{\mathcal{A}_1+i \alpha' k_2^{\scriptscriptstyle(1)}}{\mathcal{A}_3+i \alpha' k_2^{\scriptscriptstyle(3)}}\right|,
 \\
&    \kappa^2\mathcal{K}_{C_2^{7_2}\bar{C}_2^{7_2}}
=\frac{\pi e^{\phi_4}}{{\alpha' }^2(U_2+\bar{U}_2)}\sqrt{\frac{\alpha'\mathcal{A}_2}{\mathcal{A}_1\mathcal{A}_3}}\left| m_2^{\scriptscriptstyle(1)}m_2^{\scriptscriptstyle(3)}\right|\left|{\mathcal{A}_1+i \alpha' k_2^{\scriptscriptstyle(1)}}\right|\left|{\mathcal{A}_3+i \alpha' k_2^{\scriptscriptstyle(3)}}\right|.
  \end{align}
 The $m_a^{\scriptscriptstyle(i)}$, $n_a^{\scriptscriptstyle(i)}$ integers are related to the quantised magnetic field $H_a^{\scriptscriptstyle(i)}$ with $k_a^{\scriptscriptstyle(i)}$ given in equation \eqref{defH}. We recall that in the present example we take $m_2^{\scriptscriptstyle(1)}=m_2^{\scriptscriptstyle(3)}=1$, as mentioned under \cref{config1tachyon}. The four dimensional dilaton $\phi_4$ is related to the ten dimensional one through the total volume 
 \begin{equation}
e^{\phi_4}={e^{\RV{\phi_{10}}}}{\mathcal{V}^{-\frac12}}=\frac{e^{\RV{\phi_{10}}}{\alpha'}^{3/2}}{\sqrt{\mathcal{A}_1\mathcal{A}_2\mathcal{A}_3}}. \label{defphi4}
 \end{equation}
The 10d dilaton is part of the axio-dilaton multiplet defined as
 \begin{equation}
\quad S=e^{-\RV{\phi_{10}}}+iC_0, \qquad \text{with} \quad g_s=\langle e^{\RV{\phi_{10}}}\rangle. \label{defS}
\end{equation}

In the configuration of \eqref{config1tachyon}, $H_2^{\scriptscriptstyle(3)}$ is turned on and $H_2^{\scriptscriptstyle(1)}$ vanishes. In the large volume limit,  $i.e.$ when $\alpha ' k_2^{\scriptscriptstyle(3)} \ll \mathcal{A}_3$, the magnetic flux is diluted and the K\"ahler metrics approach the unmagnetised ones. We will check later that the magnetic fields are indeed small for our purposes. In that case the K\"ahler metrics read
  \begin{align}
 &\kappa^2\mathcal{K}_{C_1^{7_2}\bar{C}_1^{7_2}} =\frac{\pi e^{\phi_4}}{(U_1+\bar{U}_1)}\sqrt{\frac{\alpha'\mathcal{A}_3}{\mathcal{A}_1\mathcal{A}_2}}\left|\frac{m_2^{\scriptscriptstyle(3)}}{m_2^{\scriptscriptstyle(1)}}\right|=\frac{1}{(U_1+\bar{U}_1)(\mathcal{T}_3+\bar{\cal T}_3)}
 ,\label{K11}\\
&  \kappa^2\mathcal{K}_{C_3^{7_2}\bar{C}_3^{7_2}}=\frac{\pi e^{\phi_4}}{(U_3+\bar{U}_3)}\sqrt{\frac{\alpha'\mathcal{A}_1}{\mathcal{A}_2\mathcal{A}_3}}\left|\frac{m_2^{\scriptscriptstyle(1)}}{m_2^{\scriptscriptstyle(3)}}\right|=\frac{1}{(U_3+\bar{U}_3)(\mathcal{T}_1+\bar{\cal T}_1)}
,\label{K33}\\
&    \kappa^2\mathcal{K}_{C_2^{7_2}\bar{C}_2^{7_2}}=\frac{\pi e^{\phi_4}}{(U_2+\bar{U}_2)}\sqrt{\frac{\mathcal{A}_1\mathcal{A}_2\mathcal{A}_3}{{\alpha'}^3}} \,\left| m_2^{\scriptscriptstyle(1)}m_2^{\scriptscriptstyle(3)}\right|=
\frac1{(S+\bar{S})(U_2+\bar{U}_2)}\label{K22}\,,
  \end{align}
 where we used that in the toroidal case the $\mathcal{T}_i$ moduli are expressed in terms of the tori areas through
\begin{equation}
\mathcal{T}_i=\frac{e^{-\RV{\phi_{10}}}\mathcal{A}_j\mathcal{A}_k}{{\alpha'}^2}+ ia_i\, , \quad i\neq j \neq k \neq i.\label{deftau}
\end{equation}
In the equalities of \cref{K11,K22,K33} we also explicitly took $m_2^{\scriptscriptstyle(1)}=m_2^{\scriptscriptstyle(3)}=1$. These K\"ahler metrics follow from a K\"ahler potential of the usual form
\begin{align}
\kappa^2 \mathcal{K}=&-\ln\left[(S+\bar{S})(U_2+\bar{U}_2)- |C_2^{7_2}|^2\right] \nonumber\\
&- \ln\left[ (\mathcal{T}_2+\bar{\mathcal{T}}_2) \prod_{i,j=1,3}\left(  |\epsilon_{i2j}| (\mathcal{T}_i+\bar{\mathcal{T}}_i)(U_j+\bar{U}_j) - |\epsilon_{i2j}| |C_j^{7_2}|^2+\cdots\right) \right]. \label{Kahlerpotmatter}
\end{align}
In the last line, there is an implicit summation on the $j$ index, and $\epsilon_{i2j}$ is the standard fully antisymmetric symbol. In the above K\"ahler potential we did not include the quantum corrections of equation \eqref{CorrectedKahler} which are subdominant here. 

From equations \eqref{K11} to \eqref{K22} we see that the physical Yukawa couplings \eqref{yukawa} read
\begin{align}
Y_{ijk}&=\kappa^3 {w}_{ijk} \left(\frac{1}{ (S+\bar{S})( \mathcal{T}_1+\bar{\mathcal{T}}_1)( \mathcal{T}_3+\bar{\mathcal{T}}_3) \prod_l (U_l+\bar{U}_l)}\right)^{-\frac12}\left((S+\bar{S})\prod_l (\mathcal{T}_l+\bar{\mathcal{T}}_l)(U_l+\bar{U}_l)\right)^{-\frac12} \nonumber \\
 &=\kappa^3 w_{ijk} \frac1{\sqrt{\mathcal{T}_2+\bar{\mathcal{T}}_2\vphantom{3^{3}}}}=\kappa^3 w_{ijk}\,  {g_s^{\scriptscriptstyle{1/2}}}\sqrt{\frac{\mathcal{A}_2}{\alpha'\mathcal{V}}}\,.\label{yukawaphysical}
\end{align}
We have made use of the definitions \eqref{defS}, \eqref{defphi4} and \eqref{deftau} to express the various moduli in terms of the physical quantities. From \eqref{yukawaphysical} we can extract the internal volume dependence of the canonically normalised tachyonic fields superpotential \eqref{superpotmattergeneric}
 \begin{equation}
 \mathcal{W}_{tach}= {g_s^{\scriptscriptstyle{1/2}}}{\kappa^3} \sqrt{\frac{\mathcal{A}_2}{\alpha'\mathcal{V}}} \varphi_2 \varphi_+ \varphi_- ,\label{superpotvarphi}
 \end{equation} 
which generates a F-term scalar potential.  

\paragraph{Mass term} When $\varphi_2$ gets a non-vanishing VEV $\langle \varphi_2 \rangle\neq0$, the F-term gives a mass to the tachyonic fields
 \begin{equation}
V_F\ni \kappa^{-4}\sum_i \left| \frac{\partial \mathcal{W}_{tach}}{\kappa \partial{\varphi}_i} \right|^2= \frac{g_s}{\left|m_2^{\scriptscriptstyle(1)}m_2^{\scriptscriptstyle(3)}\right|}|\langle \varphi_2 \rangle |^2 \frac{\mathcal{A}_2}{\alpha'\mathcal{V}} \left(|\varphi_+|^2+|\varphi_-|^2\right)\equiv m_{x_2}^2\left(|\varphi_+|^2+|\varphi_-|^2\right). \label{Fpotmatter}
\end{equation}
 In the above equation, we defined $m_{x_2}$ as the physical mass coming from the brane position $x_2$. 

From equation \eqref{K22} we read the $\Phi_2$ K\"ahler metric and deduce the expression for the canonically normalised field 
\begin{equation}
\varphi_2=\frac{\kappa^{-1}}{\sqrt{\vphantom{\tilde{U}}(U_2+\bar{U}_2)(S+\bar{S})}}\Phi_2=\frac{\kappa^{-1} g_s^{\scriptscriptstyle{1/2}}}{\sqrt{\vphantom{\tilde{U}}U_2+\bar{U}_2}}\Phi_2. \label{canonicalphi0}
\end{equation} 
We recall that $\Phi_2$ is the dimensionless complexifyed scalar modulus related to the brane position on $T^2_2$, given by 
\begin{equation}
\Phi_2=x_2^x-iU_2 x_2^y. \label{phi0}
\end{equation}
Hence from equations \eqref{Fpotmatter} to \eqref{phi0} we deduce that 
\begin{equation}
m_{x_2}^2=|\langle \varphi_2\rangle|^2 \, {g_s}\frac{\mathcal{A}_2}{\alpha'\mathcal{V}}=\frac{g_s^2}{\kappa^2\mathcal{V}} \frac{\mathcal{A}_2}{\alpha'|U_2+\bar{U}_2|}\left|x_2^x-iU_2 x_2^y\right|^2\equiv y(U_2) \frac{g_s^2}{\kappa^2\mathcal{V}}\frac{\mathcal{A}_2}{\alpha'} \label{mx2}.
\end{equation}
Replacing ${\kappa^2\mathcal{V}}/g_s^2$ by ${\alpha'}$ through \eqref{stringlength}, we find back the string mass formula \eqref{branepositionmass} derived in \cref{model3fields}, except for irrelevant powers of $2$ which come from the fact that in the current part we derived the mass term without explicitly applying the orientifold and orbifold projections. In the following we use the last form of \eqref{mx2}.

\paragraph{Quartic term} In order to analyse the phase transition of the waterfall field, we need to keep track of the quartic terms in addition to the mass terms. For the D-term scalar potential the quartic contributions were already included in the expansion \eqref{Dscalarpot}. 
The full F-term scalar potential can be computed through the supergravity formula using the total superpotential $\mathcal{W}=\mathcal{W}_0+\mathcal{W}_{C_i^{7_2}}$ containing the flux-dependent constant described above \cref{noscaleV} and the ${C_i^{7_2}}$ dependent part of \cref{SYMcoupling,superpotmattergeneric}, together with the total K\"ahler potential  including the  $C_i^{7_2}$ dependence of \eqref{Kahlerpotmatter} and the quantum corrections of \cref{CorrectedKahler}. From this F-term we can extract the quartic contribution of the waterfall field.

Nevertheless, the leading corrections in $g_s$ are easily obtained by expanding the K\"ahler potential \eqref{Kahlerpotmatter} with respect to the tachyonic field $\varphi_-$ (or rather its non-canonically normalised ``parent" $C_1^{7_2}$ or $C_3^{7_2}$), thus neglecting the $C_2^{7_2}$ dependence in the logarithm of the first line together with the one-loop quantum corrections. The leading quartic contribution for the tachyonic scalar field potential then simply reads
\begin{equation}
V_F \ni\frac{g_s^2}{(U_2+\bar{U}_2)} \frac{\mathcal{A}_2}{\alpha'\mathcal{V}}|\langle \Phi_2\rangle|^2 |\varphi_-|^4= y(U_2) \frac{g_s^2}{\mathcal{V}}\frac{\mathcal{A}_2}{\alpha'}  |\varphi_-|^4 = \kappa^2{m_{x_2}^2} |\varphi_-|^4. \label{quartic}
\end{equation}
Thus, it turns out that the leading quartic contribution
comes entirely from the expansion of the $e^{\kappa^2\mathcal{K}}$ factor in the supergravity formula. The dependence on the moduli of this term is identical to the one of the mass term as it comes from the $\mathcal{K}^{i\bar{j}}{\cal D}_i\mathcal{W}{\cal D}_{\bar{j}}\overline{\mathcal{W}}$ part of the F-term scalar potential, with the derivative taken with respect to the $\varphi_+$ field.

\subsection{New vacuum} \label{secnewcacuum}
Summing the D-term and F-term contributions \eqref{Dscalarpot}, \eqref{Fpotmatter} and \eqref{quartic} for the matter fields with the F-term scalar potential for the volume modulus\RV{, which is shown in \cref{Vlargelimit}}, we obtain the effective scalar potential to minimise in order to obtain the physical vacuum. It reads
\begin{align}
V(\mathcal{A}_i, \varphi_{\pm})&= V_F(\mathcal{V})+V_F(\mathcal{A}_i,\varphi_{\pm})+V_D(\mathcal{A}_i,\varphi_{\pm})+\cdots \nonumber\\
&=V_F(\mathcal{V})+ m_{x_2}^2 \left(|\varphi_+|^2+|\varphi_-|^2\right)+{\kappa^2} {m_{x_2}^2} |\varphi_-|^4+ \cdots \nonumber \\
&\hspace{1cm}+\sum_{b=1,3} \frac{g^2_{U(1)_b}}{2}\xi_b^2+\frac{g^2_{U(1)_2}}{2}\left(\xi_2+2|\varphi_+|^2-2|\varphi_-|^2\right)^2 + \cdots \label{scalarpotstring}
\end{align}

\paragraph{K\"ahler moduli minimisation} As motivated after \cref{gammaxi}, we first minimise the scalar potential with respect to ratios of the internal areas moduli $\mathcal{A}_i$, letting free the total volume $\mathcal{V}$ and neglecting for the moment the matter fields. This is similar to what was done in \cite{Antoniadis:2020stf}, with nevertheless a slightly different expression for the D-term before the minimisation. This is why we perform again the minimisation in our precise model.
 Defining the ratios 
\begin{equation} 
u\equiv\frac{\mathcal{A}_3}{\mathcal{A}_2}, \qquad v\equiv\frac{\mathcal{A}_1}{\mathcal{A}_3}, \qquad \frac{1}{uv}=\frac{\mathcal{A}_2}{\mathcal{A}_1}, \label{uvratios}
\end{equation}
the D-term part of the scalar potential \eqref{potentialsugra} reads
\begin{equation}
V_D(\mathcal{A}_i)=V_D(\mathcal{V},u,v)=\frac{1}{\kappa^4\mathcal{V}^2}\left(d_1 u + d_2 v + \frac{d_3}{uv}\right), \label{VDratios}
\end{equation}
where the $d_i$ parameters are defined in equation \eqref{dterms}. $V_D$ is minimised by
\begin{equation}
u_0=\left(\frac{d_2d_3}{d_1^2}\right)^{\frac13}\hspace{-4pt}, \qquad v_0=\left(\frac{d_1d_3}{d_2^2}\right)^{\frac 13}\hspace{-4pt}, 
\end{equation}
which gives the following tori moduli
\begin{equation}
\mathcal{A}_1=\alpha' \left(\frac{d_3}{d_2}\right)^{\frac 13} \mathcal{V}^{\frac 13}, \quad \mathcal{A}_2= \alpha' \left(\frac{d_1}{d_3}\right)^{\frac 13}\mathcal{V}^{\frac 13}, \quad \mathcal{A}_3=\alpha' \left(\frac{d_2}{d_1}\right)^{\frac 13} \mathcal{V}^{\frac 13},
\label{modulistabresult}
\end{equation}
while its expression at the minimum becomes:
\begin{equation}
V_D(\mathcal{V})=V_D(\mathcal{V},u_0,v_0)=\frac{3(d_1d_2d_3)^{\frac13}}{\kappa^4\mathcal{V}^2}\equiv \frac{d}{\kappa^4\mathcal{V}^2}.
\end{equation}
In the last equality we defined 
\begin{equation}
d\equiv3(d_1d_2d_3)^{\frac13}=\frac32 g_s^3 \left|\frac{m_1^{\scriptscriptstyle(3)}m_2^{\scriptscriptstyle(1)}m_3^{\scriptscriptstyle(2)}}{m_1^{\scriptscriptstyle(2)}m_2^{\scriptscriptstyle(3)}m_3^{\scriptscriptstyle(1)}}\right|^{\frac13} \left(\frac{n_1^{\scriptscriptstyle(2)}n_2^{\scriptscriptstyle(3)}n_3^{\scriptscriptstyle(1)}}{\pi^3}\right)^{\frac23}, \label{totaldtermparameter}
\end{equation}
 giving back the D-term contribution shown in equation \eqref{Vlargelimit}, 
 but with a specific value of $d$ related to the parameters of our model, \RV{as explained hereafter}.
 
From \eqref{masseffectivetheory}, \eqref{couplings}, \eqref{dterms} and  \eqref{mx2}, one finds that the masses and couplings for the matter fields take the form
\begin{align}
&m^2_{H_2}= {2\sqrt{2}}{\left|{m_2^{\scriptscriptstyle(1)}m_2^{\scriptscriptstyle(3)}}\right|^{-\frac12}} \,\frac{\sqrt{g_s}}{\kappa^2}\frac{ \left(d_1^2d_2\right)^{\frac16}}{\mathcal{V}^{\frac 43}}, \hspace{30pt} g^2_{U(1)_2}=\frac{g_s}{\mathcal{V}^{\frac 23}}  \left(\frac{d_1}{d_3}\right)^{\frac13}\left|m_2^{\scriptscriptstyle(1)}m_2^{\scriptscriptstyle(3)}\right|^{-1}, \nonumber \\
&m_{x_2}^2 = \frac{g_s^2}{\kappa^2} \,y(U_2) \left(\frac{d_1}{d_3}\right)^{\frac13} \frac 1{\mathcal{V}^{\frac23}}~\cdot \label{parametersstabilised}
\end{align}
	For the configuration of \cref{model3fields}, the condition $|\zeta_1^{\scriptscriptstyle(2)}|=|\zeta_2^{\scriptscriptstyle(3)}|=|\zeta_3^{\scriptscriptstyle(1)}|$ in \eqref{mixedstatecondition} is necessary to eliminate the tachyons from different brane intersections. Together with the moduli stabilisation condition \eqref{modulistabresult}, this gives the following relations that the fluxes must satisfy
\begin{equation}
n_2^{\scriptscriptstyle(3)}=\left|\frac{m_3^{\scriptscriptstyle(1)}m_3^{\scriptscriptstyle(2)}}{m_2^{\scriptscriptstyle(1)}m_1^{\scriptscriptstyle(2)}}\right| \, n_1^{\scriptscriptstyle(2)} , \qquad n_3^{\scriptscriptstyle(1)}=\left|\frac{m_3^{\scriptscriptstyle(1)}m_1^{\scriptscriptstyle(3)}}{m_2^{\scriptscriptstyle(1)}m_2^{\scriptscriptstyle(3)}}\right|\, n_1^{\scriptscriptstyle(2)} , \label{fluxessimplemodel}
\end{equation} 
leading to the following expression for the D-term parameter introduced in \cref{totaldtermparameter}:
\begin{equation} d=\frac 32 \, g_s^3 \, \left|\frac{m_3^{\scriptscriptstyle(1)}m_3^{\scriptscriptstyle(2)}m_1^{\scriptscriptstyle(3)}}{m_2^{\scriptscriptstyle(1)}m_2^{\scriptscriptstyle(3)}m_1^{\scriptscriptstyle(2)}} \right|\left(\frac{n_1^{\scriptscriptstyle(2)}}{\pi}\right)^2= \frac 32 \, g_s^3 \left(\frac k\pi\right)^2, \qquad {\text{with}} \qquad k^2\equiv ({n_1^{\scriptscriptstyle(2)}})^2 \left|\frac{m_3^{\scriptscriptstyle(1)}m_3^{\scriptscriptstyle(2)}m_1^{\scriptscriptstyle(3)}}{m_2^{\scriptscriptstyle(1)}m_2^{\scriptscriptstyle(3)}m_1^{\scriptscriptstyle(2)}}\right|. \label{defk}
	\end{equation} 
We recall that the wrapping numbers $m_a^{\scriptscriptstyle(j)}$ are also subject, together with the brane multiplicities $N_a$ (or $N_a'$  introduced in \cref{ChanPatonwrappings}), to tadpole cancellation conditions\footnote{In the absence of 3-form fluxes the tadpole cancellation conditions for the $m_a^{\scriptscriptstyle(j)}$ simply read $N_a' m_a^{\scriptscriptstyle(j)} m_a^{\scriptscriptstyle(k)} -16 =0$, $a\neq j \neq k\neq a$,  for each $D7_a$ brane stack.}. We also recall that we have chosen $m_2^{\scriptscriptstyle(1)}=m_2^{\scriptscriptstyle(3)}=1$, even if we kept generality in \cref{totaldtermparameter,parametersstabilised,fluxessimplemodel,defk}. Note that the model of \cref{multiplefieldseachbrane} with more general flux configurations leads to less constrained flux parameters $n_a^{\scriptscriptstyle(j)}$ than those of \cref{fluxessimplemodel}.

\RV{Before moving to the study of the global minimum of the scalar potential, we come back to the point discussed under \eqref{gammaxi}. As explained there, the ratios of the internal tori areas correspond to moduli orthogonal to the total internal volume $\mathcal{V}$. One can thus fix them at their VEVs, while keeping $\mathcal{V}$ free, as long as their masses are larger than the volume modulus mass. In order to compare the different masses, one must consider canonically normalised fields, which in our setup can be introduced through the following basis
 \begin{align}
& \phi=\sqrt{\frac 23} \ln \mathcal{V}\\
& U=\frac 12 \ln\left(\frac{\tau_1}{\tau_2}\right)=\frac 12 \ln\left(\frac{\mathcal{T}_1+\bar{\mathcal{T}}_1}{\mathcal{T}_2+\bar{\mathcal{T}}_2}\right)=\frac 12 \ln\left(\frac{\mathcal{A}_2}{\mathcal{A}_1}\right)=\frac 12 \ln\left(\frac 1{uv}\right)\\
& V=\frac 1{2\sqrt 3} \ln\left(\frac {\tau_1 \tau_2}{\tau_3^2}\right)=\frac 1{2\sqrt 3} \ln\left(\frac {(\mathcal{T}_1+\bar{\mathcal{T}}_1) (\mathcal{T}_2+\bar{\mathcal{T}}_2)}{(\mathcal{T}_3+\bar{\mathcal{T}}_3)^2}\right)=\frac 1{2\sqrt 3} \ln\left(\frac{\mathcal{A}_3^2}{\mathcal{A}_1 \mathcal{A}_2}\right)=\frac 1{2\sqrt{3}} \ln\left(\frac uv\right).
\end{align}
The D-term scalar potential \eqref{VDratios} can be expressed in this basis and the masses $m^2_{\phi}$, $m^2_U$, $m^2_V$ for $\phi$, $U$ and $V$ are extracted from the second derivatives of the K\"ahler moduli dependent part of $V_F+V_D$. These masses read:
\begin{align}
&m^2_U=m^2_V=\frac{2d}{\kappa^2}e^{-\sqrt{6}\phi}, \\
&m^2_{\phi}=\frac{6d}{\kappa^2}e^{-\sqrt{6}\phi} - \frac 94 \frac C{\kappa^2} e^{-3\sqrt{\frac 32}\phi}\left(3\sqrt{6} \phi -28 + 6q\right),
\end{align}
and we recall that $q$, $C$ were introduced in \cref{modelparameters}. As summarized in \cref{sectionhybridinflation}, the study of \cite{Antoniadis:2020stf} shows that inflation occurs for a certain value of the parameter $x$, related to $d$ through \cref{modelparameters,xdefinition}, and  takes place for $\phi \in [\phi_-;\phi_+]$. We checked that in this region we indeed have $m^2_{\phi}\ll m^2_U,m^2_V$ (by at least a factor of $\sim 25$), so that the minimisation procedure is consistent.}

\paragraph{Global minimum and waterfall direction} After stabilisation of the transverse moduli ratios \RV{($u$, $v$ or $U$, $V$)}, the left-over parameters of the total scalar potential \eqref{scalarpotstring} can be replaced using equations \eqref{parametersstabilised}. In the simple case under consideration, with fluxes as in \eqref{fluxessimplemodel}, the mass and coupling of \cref{parametersstabilised} read
\begin{align}
&m^2_{H_2}= {2}\frac{g_s^2}{\kappa^2 \mathcal{V}^{\frac 43}}\frac{ k}{\pi} \left|\frac 1{\prod_{a\neq j}m_a^{\scriptscriptstyle(j)}}\right|^{\frac 16}, \hspace{30pt} g^2_{U(1)_2}=\frac{g_s}{\mathcal{V}^{\frac 23}}  \left|\frac 1{\prod_{a\neq j}m_a^{\scriptscriptstyle(j)}}\right|^{\frac 13}, \nonumber \\
&m_{x_2}^2 = \frac{g_s^2}{\kappa^2\mathcal{V}^{\frac23}} \,y(U_2) \left|\frac{{m_2^{\scriptscriptstyle(1)}}^2{m_2^{\scriptscriptstyle(3)}}^2}{m_1^{\scriptscriptstyle(2)}m_1^{\scriptscriptstyle(3)}m_3^{\scriptscriptstyle{(1)}}m_3^{\scriptscriptstyle(2)}}\right|^{\frac13} . \label{parameterssimplefluxes}
\end{align}
Neglecting the massive $\varphi_+$ field and expressing the volume modulus dependent contribution $V_F(\mathcal{V})+ V_D(\mathcal{V})$ through \eqref{Vlargelimit}, the scalar potential \eqref{scalarpotstring} is written as 
\begin{align}
V(\mathcal{V}, \varphi_-)
&=\frac{C}{\kappa^4} \left( -\frac{\ln \mathcal{V}-4+q}{{\mathcal V}^3}-\frac{3\sigma}{2\mathcal{V}^2}\right)+\frac 12 m_Y^2(\mathcal{V})|\varphi_-|^2+\frac{ \lambda (\mathcal{V})}4 |\varphi_-|^4,\label{scalarpotVl}
\end{align}
taking the same form as eq.~\eqref{Vsfs}, with $\varphi_-$  playing the role of the waterfall field $Y$. Its mass and coupling read 
\begin{align}
&\frac 12 m_Y^2(\mathcal{V})= (m_{x_2}^2-m_{H_2}^2)= \frac {g_s^2\, y(U_2)}{\kappa^2\mathcal{V}^{\frac 23}} \left|\frac1{m_1^{\scriptscriptstyle(2)}m_1^{\scriptscriptstyle(3)}m_3^{\scriptscriptstyle{(1)}}m_3^{\scriptscriptstyle(2)}}\right|^{\frac13}\left(1-\left(\frac{\mathcal{V}_{c2}}{\mathcal{V}}\right)^{\frac23}\right), \label{masswaterfall} \\
& \lambda({\cal V})= 4 \left(2{g_{U(1)_2}^2}+\kappa^2{m_{x_2}^2}\right) = \frac{4g_s}{\mathcal{V}^{\frac 23}} \left|\frac 1{\prod_{a\neq j}m_a^{\scriptscriptstyle(j)}}\right|^{\frac 13} \left( \vphantom{\frac11} 2 + g_s \, y(U_2) \right). \label{couplingwaterfall} 
\end{align}
In \cref{masswaterfall} we explicitly took  $m_2^{\scriptscriptstyle(1)}=m_2^{\scriptscriptstyle(3)}=1$ and defined the critical volume $\mathcal{V}_{c2}$ at which $\varphi_-$ becomes tachyonic, $i.e.$ for $\mathcal{V}<\mathcal{V}_{c2}$, 
\begin{equation}
\mathcal{V}_{c2}\equiv \left(\frac{2k }{\pi y(U_2)}\right)^{\frac32}\left|{m_1^{\scriptscriptstyle(2)}m_1^{\scriptscriptstyle(3)}{m_3^{\scriptscriptstyle(1)}}{m_3^{\scriptscriptstyle(2)}}}\right|^{\frac14}. 
\label{Vtachyon}
\end{equation}

As expected, $\mathcal{V}_{c2}$ depends on the fluxes through $k$ defined in \cref{defk} and on the $D7_2$ brane position through $y(U_2)$ defined through \eqref{mx2}. 
We also remark from  \cref{couplingwaterfall} that the main contribution to the quartic coupling $\lambda$ comes from the D-term part of the potential, since the F-term contribution is suppressed by a power of $g_s$.

When the mass term $m_Y^2$ of \cref{masswaterfall} becomes negative, the waterfall field $Y$ (our tachyonic field $\varphi_-$) rolls down its potential to the new vacuum at $\langle \varphi_-\rangle=\langle Y\rangle=\pm v_2$. From \cref{vacuumbrokenphase,scalarpotVl} we see that the  value of the potential at this new vacuum is 
\begin{align}
V(\mathcal{V},v_2)&=V_F(\mathcal{V})+V_D(\mathcal{V})- \frac{m_Y^4}{4\lambda}({\cal V}) \nonumber\\[5pt]
&= \frac{C}{\kappa^4} \left( -\frac{\ln \mathcal{V}-4+q}{{\mathcal V}^3}-\frac{3\sigma}{2\mathcal{V}^2}\right) - \frac {C_2}{\kappa^4\mathcal{V}^{\frac23}}\left(1-\left(\frac{\mathcal{V}_{c2}}{\mathcal{V}}\right)^{\frac23}\right)^2. 
\label{newvacuumformula}
\end{align}
 We recall here the expressions of various parameters introduced before
\begin{equation}
q= \frac \xi{2 \gamma}, \quad \sigma = \frac{2d}{9 {\mathcal{W}_0}^2\gamma}, \quad C= -3{\mathcal{W}_0}^2\gamma>0, \quad y(U_2)=  \frac{\left|x_2^x-iU_2 x_2^y\right|^2}{|U_2+\bar{U}_2|}, \quad d=\frac 32 g_s^3 \left(\frac{k}{\pi}\right)^2, \label{modelparametersrecall}
\end{equation}
and define $C_2$, the coefficient of the tachyonic contribution to the vacuum energy through
\begin{equation}
C_2\equiv -\frac{g_s^3y^2(U_2)}{4\left( \vphantom{\frac11} 2 + g_s \, y(U_2)\right)}\left|\frac{1}{m_1^{\scriptscriptstyle(2)}m_1^{\scriptscriptstyle(3)}{m_3^{\scriptscriptstyle(1)}}{m_3^{\scriptscriptstyle(2)}}}\right|^{\frac13}.
\end{equation}
We observe that $C_2$ and $\mathcal{V}_{c2}$ are not independent, their are related by  
 \begin{equation}
 C_2=\beta_2 \frac{d}{3 \mathcal{V}_{c2}^{\frac43}}, \qquad
 \beta_2\equiv\frac{2}{2+g_s y(U_2)} \in [0,1], \label{CY}
 \end{equation}
where the  parameter $\beta_2$ expresses the relative contributions to the quartic coupling from the F-term versus the D-term. From \eqref{mx2} we see that as $y(U_2)>0$, $\beta_2$ lies between $0$ and $1$. For $\beta_2=1$ the D-term dominates whereas for $\beta_2=0$ the F-term dominates. It is clear from \eqref{CY} that the tachyonic contribution becomes maximal (in absolute value) for $\beta_2=1$, when the quartic coupling is dominated by the D-term. 

 Let us discuss now the physics of the waterfall direction. As explained in \cref{sectionhybridinflation}, the waterfall field can generate the desired scenario for the end of inflation. It has to become tachyonic when the volume modulus $\mathcal{V}$ (identified to the inflaton $\phi$ through \cref{inflaton}) reaches the bottom of its potential $V_F(\mathcal{V})+V_D(\mathcal{V})$. This situation corresponds to $\mathcal{V}_{c2} \approx \mathcal{V}_-$, where $\mathcal{V}_-$ is the value of the volume at its minimum, expressed by \eqref{Vmin}. From \cref{Vtachyon} we see that the value $\mathcal{V}_{c2}$ depends on $y(U_2)$, $k$ and a ratio of flux and wrapping numbers, hence it is easy to choose $\mathcal{V}_{c2}$ near $\mathcal{V}_-$. 
  
We see from the scalar potential expression \eqref{newvacuumformula} that once $C$ and $d$ are determined by the inflationary phase, and $ \mathcal{V}_{c2}$ fixed to $\mathcal{V}_-$, the coefficient $C_2$ is the only parameter to tune the minimum. From the relation of \cref{CY} we see that in fact, only $\beta_2$ can be used for fixed $C$ and $d$. As $\beta_2$ depends only on the product $g_ s y(U_2)$, we express $\mathcal{V}_{c2}$ in terms of $d$, $g_s$ and $g_s y(U_2)$ using \cref{Vtachyon,defk}. It reads:
 \begin{equation}
 \mathcal{V}_{c2}=\frac{4\sqrt{2}}{3}\frac{\sqrt{\pi}}{(y(U_2)g_s)^{\frac32}} g_s^{-\frac32} d \left|\frac{m_1^{\scriptscriptstyle(2)}}{n_1^{\scriptscriptstyle(2)}}\right|^{\frac12} \label{Vcgy}
 \end{equation}
 Hence, 
 in principle, we can first fix the product $g_ s y(U_2)$ to have the desired $\beta_2$  and tune the minimum, then
 choose the values of $g_s$ and of the ratio of the flux/wrapping numbers on the second torus to tune the critical volume. 
  
\begin{figure}[h!]
\begin{minipage}{0.45\textwidth}
   \hspace{-0.2cm}\includegraphics[scale=0.35]{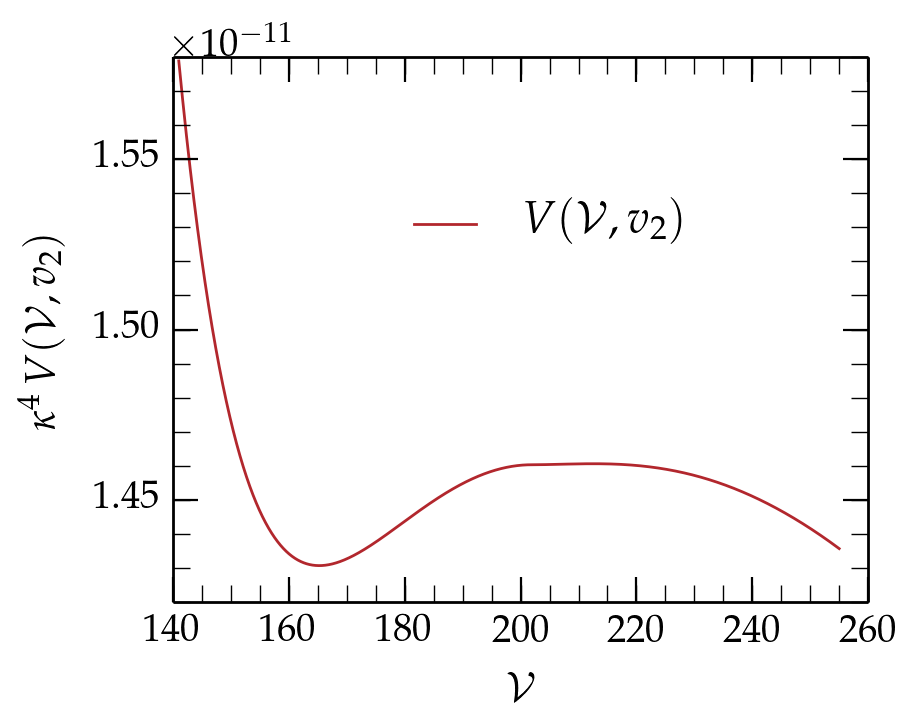}      
  \end{minipage}
    \hspace{0.3cm}
  \begin{minipage}{0.45\textwidth}
   \hspace{0cm}\includegraphics[scale=0.35]{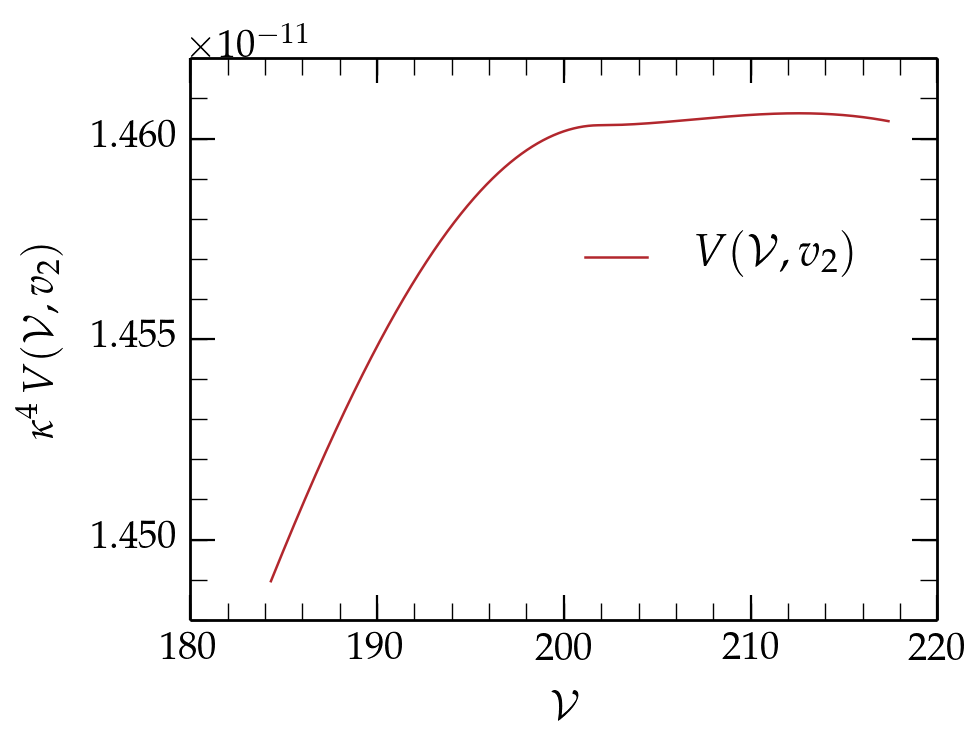}      
  \end{minipage}
  \caption{Value $V(\mathcal{V},v_2)$ of the global minimum of the scalar potential as a function of the internal volume, for the parameters of \eqref{parametersnum}.}
\label{scalarpotwaterfall}
 \end{figure}

\paragraph{Example of numerical values}  We give now an explicit example of parameters supporting the above discussion. In the inflationary scenario \cite{Antoniadis:2020stf} discussed in \cref{sectionhybridinflation}, the values of $x$ and $C$ are fixed by observational constraints to
\begin{equation}
 x\approx 3.3 \times 10^{-4}, \qquad C=e^{-3q}\times 7.81 \times 10^{-4}\equiv e^{-3q} C_0.  \label{C}
  \end{equation}
From \cref{modelparameters,Vmin,xdefinition} we extract for $q=0$ the values of $\mathcal{V}_-$, the minimum of the modulus part of the potential, and the $d$ magnetic flux parameter
\begin{equation}
\mathcal{V}_-\approx201.9,\qquad d\approx5.65 \times 10^{-6}.
\end{equation}
 We compute numerically the global minimum of the potential given in \eqref{newvacuumformula} and see that in order to have an almost vanishing value at the minimum, we need to tune the tachyonic coefficient to $C_2\approx5.136 \times 10^{-9}$, which through equation \eqref{CY} would impose the value
\begin{equation}
\beta^{\Lambda=0}= 5.136 \times 10^{-9} \, \frac{3{\mathcal{V}_-}^{\frac43}}{d}\approx3.228. \label{tachampvanishingvacuum}
\end{equation}
Nevertheless, as $\beta_2\in[0,1]$, we see that we cannot tune the vacuum energy to zero in the simple model of the current section. We come back to this point in detail in the next section. From \cref{CY} we see that the largest value $\beta_2\approx1$ is obtained for small $g_s y(U_2)$ and taking for instance $g_s y(U_2)\approx 10^{-2}$ in equation \eqref{Vcgy} we obtain
\begin{equation}
 \mathcal{V}_{c2}\approx\mathcal{V}_-\approx201.9=1.89 \times 10^{-2} g_s^{-\frac32} \left|\frac{m_1^{\scriptscriptstyle(2)}}{n_1^{\scriptscriptstyle(2)}}\right|^{\frac12}, 
 \end{equation}
which has to be satisfied together with the relations on $d$ and $\beta_2$ given by \cref{defk,CY}
\begin{equation}
\beta_2=\frac{2}{2+g_sy(U_2)}\approx1, \qquad d=\frac 32 \, g_s^3 \, \left|\frac{m_3^{\scriptscriptstyle(1)}m_3^{\scriptscriptstyle(2)}m_1^{\scriptscriptstyle(3)}}{m_1^{\scriptscriptstyle(2)}} \right|\left(\frac{n_1^{\scriptscriptstyle(2)}}{\pi}\right)^2\approx5.65\times10^{-6}.
\end{equation}
We recall that we consider the case $m_2^{\scriptscriptstyle(1)}=m_2^{\scriptscriptstyle(3)}=1$.
The following parameters 
\begin{align}
&g_s=2.596\times10^{-3}, \qquad n_1^{\scriptscriptstyle(2)}=1, \qquad m_1^{\scriptscriptstyle(2)}= 2,\nonumber \\
&m_1^{\scriptscriptstyle(3)}=10, \quad  m_3^{\scriptscriptstyle(1)}=17,\qquad m_3^{\scriptscriptstyle(2)}=25, \qquad y(U_2)=3.85, \label{parametersnum}
\end{align}
give the desired values for $d$, $\mathcal{V}_{c2}$ and $\beta_2\approx1$. Of course, there is an infinite set of other choices of parameters giving the same values. We show in the left panel of Figure \ref{scalarpotwaterfall}  the value of the potential at the global minimum (including the waterfall), located at $\mathcal{V}_0\approx160$. However, as explained above, this value is not vanishing. The next section tackles this point in details.

On the other hand, the right panel of Figure \ref{scalarpotwaterfall} shows that the tachyonic field gives indeed a ``warterfall" direction. Falling in this direction leads to an increase of the slow-roll parameters marking the end of the inflationary phase. A precise computation of the slow-roll parameters along the inflaton trajectory is necessary  to extract the extra number of e-folds until the end of inflation and compare it to the case without waterfall field \cite{Antoniadis:2020stf}. This number depends on $\mathcal{V}_--\mathcal{V}_{c2}$ and is model dependent.

\paragraph{Validity of our approximations} Before the end of this section, we stress the fact that with the parameters \eqref{parametersnum}, the volume modulus is large and the flux numbers $k_a^{\scriptscriptstyle (i)}$ small, so that the large volume approximations of $e.g.$ \cref{masseffectivetheory,couplingslargev} hold. 
We also want to check that the vacuum expectation value $v_2$ of the waterfall field stays small (in $\kappa$ units), so that the quartic expansion of \eqref{Dscalarpot} holds. From \cref{newmin,masswaterfall} this VEV is expressed as
\begin{equation}
 \langle \varphi_- \rangle=\langle Y \rangle  = \pm v_2 = \pm \frac{|m_Y|}{\sqrt{\lambda}} = \pm \frac{1}{\sqrt{2} \kappa}\sqrt{\frac{g_sy(U_2)}{2+g_s y(U_2)}} \left|1-\left(\frac{\mathcal{V}_{c2}}{\mathcal{V}}\right)^{\frac23}\right|^{\frac12}.\label{valueVEV}
\end{equation}
It follows that  $\langle \varphi_- \rangle$ is entirely determined by $g_s y(U_2)$ and $\mathcal{V}_{c2}$. With the parameters of \eqref{parametersnum} and the volume modulus $\mathcal{V}_{c2}\geqslant\mathcal{V}_0\gtrsim160$, which is the range of Figure \ref{scalarpotwaterfall}, one finds a VEV $v_2$ satisfying $0\leq \kappa |v_2| \simlt  0.4\sqrt{g} \approx 0.02 $. The quartic expansion \eqref{Dscalarpot} is thus indeed sufficient. 

\section{Lowering the global minimum} \label{tuningminimum}

We have seen in \cref{secnewcacuum} that in our type IIB framework with three orthogonal $D7$ branes, a waterfall field can be implemented through a doubly charged state stretching between $e.g.$ the $D7_2$ brane and its orientifold image. The mass of such a state depends on the internal volume (our inflaton) and we showed that under a certain critical volume this state becomes tachyonic, generating a waterfall direction typical of hybrid inflation models described  in \cref{sectionhybridinflation}. 

The first motivation for the introduction of such a waterfall direction was that it is responsible for the end of inflation. The second motivation was that the waterfall field, through its negative contribution to the scalar potential, lowers the value of the global minimum and can in principle tune the cosmological constant to the almost vanishing value observed today. Nevertheless, as we explained near the end of \cref{secnewcacuum}, due to the relation \eqref{CY} we are not able to choose independently the position $\mathcal{V}_{c2}$ and the depth of the waterfall related to $C_2$. The remaining freedom in the choice of the waterfall depth lies in the $\beta_2$ coefficient, whose value $\beta^{\Lambda=0}\approx 3.228$, needed to tune the vacuum energy to zero, cannot be reached in our example where $\beta_2\leq1$. In this section we investigate how to modify the model in order to bypass the constraint imposed by \cref{CY} and lower the global minimum.

We first verify in \cref{fieldtheorydescription} that in the field theoretical description, where we can choose freely the mass and coupling parameters while keeping their volume dependence, the tuning of the global minimum is indeed possible. In \cref{oneoneachbrane} we come back to the simple case studied in the previous section. We show that the natural tentative to tune the vacuum through the use of the parameter $q$, not constrained by the inflationary phase, does not work. We also study if  the contributions of the $\gamma$, $\xi$ quantum corrections to the tachyonic mass and coupling, gives extra freedom and helps to evade relation \eqref{CY}. We show that, as these quantum corrections stay small, they don't play an important role. 
We hence examine in \cref{severaltachyons} if adding more tachyons, coming from the two other $D7$-brane stacks, allows to tune the vacuum energy to zero. We find that even if these additional tachyons lower indeed the global minimum, their contribution still determined by $d$, constrained by the inflationary phase, is not sufficient to tune the vacuum energy to zero. Nevertheless in \cref{sectionfourthstack}, we show that adding a forth magnetised stack, parallel to an already present one, adds additional tachyonic contributions to the scalar potential, allowing to tune the vacuum energy.

\subsection{Field theoretical description}\label{fieldtheorydescription}

We first look at the possibility to tune the vacuum energy of the model with arbitrary parameters, $i.e.$ in the field theoretical description. We thus take arbitrary values for the mass and quartic parameters of the F-terms and D-terms, but keep the volume dependences as in the string theory setup of the previous sections. The scalar potential is written as

\begin{align}
V(\mathcal{V},\varphi_-)&= \frac{C}{\kappa^4} \left( -\frac{\ln \mathcal{V}-4+q}{{\mathcal V}^3}-\frac{3\sigma}{2\mathcal{V}^2}\right) + \frac12 \left(\vphantom{4^{4^4}}m^2_F(\mathcal{V})-m^2_D(\mathcal{V})\right) \varphi_-^2 + \frac 14 \left(\lambda_F( \mathcal{V})+ \lambda_D(\mathcal{V})\vphantom{4^{4^4}}\right) \varphi_-^4 \nonumber\\
&= \frac{C}{\kappa^4} \left( -\frac{\ln \mathcal{V}-4+q}{{\mathcal V}^3}-\frac{3\sigma}{2\mathcal{V}^2}\right) + \frac 12 \left(\vphantom{4^{4^4}} \frac{\mu^{0}_F}{\mathcal{V}^{\frac23}}-\frac{\mu^0_D}{\mathcal{V}^{\frac43}}\right) \varphi_-^2 + \frac 14 \frac{(\lambda_F^0+ \lambda_D^0)}{\mathcal{V}^{\frac23}}\varphi_-^4\,\,.
\end{align}
As described in the previous sections on a particular example, for  $\mathcal{V}<\mathcal{V}_{c2}$ the matter field becomes tachyonic and gets a non-vanishing VEV $v\neq0$. The scalar potential gets a contribution $-{m^4}/{4\lambda}({\cal V})$ when $\varphi_-$ sits at its VEV and the dependence of the global minimum in the volume then reads
\begin{align}
V(\mathcal{V},v)&=\frac{C}{\kappa^4} \left( -\frac{\ln \mathcal{V}-4+q}{{\mathcal V}^3}-\frac{3\sigma}{2\mathcal{V}^2}\right)-\frac{{\mu^0_F}^2}{4(\lambda^0_F+\lambda^0_D)}\frac 1{\mathcal{V}^{\frac23}}\left(1-\left(\frac{\mathcal{V}_{c2}}{\mathcal{V}}\right)^{\frac23}\right)^2,\label{Vefffieldtheory}
\end{align}
with $\mathcal{V}_{c2}^{2/3}=\mu^0_D/\mu^0_F$. The D-term parameter $\mu^0_D$ is related to the flux parameter $d$ and thus to $x$, relevant during the inflationary phase. We see from \cref{Vefffieldtheory} that in the field theoretical description, one can tune $\mu^0_D$,  $\mathcal{V}_{c}=(\mu^0_D/\mu^0_F)^{3/2}$ and the coefficient $C_2={{\mu^0_F}^2}/{4(\lambda^0_F+\lambda^0_D)}$ independently. This was not the case in the simple configuration described in \cref{secnewcacuum} due to relation \eqref{CY} between $C_2$, $d$ and $\mathcal{V}_{c2}$, which translates the fact that in our string theory setup, the $\mu^0_D,\mu^0_F, \lambda^0_D$ and $\lambda^0_F$ parameters cannot be chosen independently.

In the next subsections we will investigate if more complex configurations can allow the tuning of the scalar potential at the global minimum within our string theory setup.
 
\subsection{Simple case studied previously}\label{oneoneachbrane}

\paragraph{Dependence on the $q$ parameter} We now come back on the discussion on the tuning of the global minimum in the configuration discussed since the beginning of \cref{sectioneffectivetheory}, $i.e.$ with the flux configuration \eqref{config1tachyon}. The value of the scalar potential at the global minimum was expressed in \eqref{newvacuumformula}. We first examine if the use of the parameter $q$ could liberate the constraint on the waterfall depth $C_2$ (which is related to $\mathcal{V}_{c2}\approx\mathcal{V}_-$), by shifting $\mathcal{V}_-$ arbitrarily.
%
From \cref{modelparameters,xdefinition,Vmin} we express the following parameters dependence in $q$ and $x$:
\begin{align}
&\mathcal{V}_-=e^{-q} \exp\left(\frac{13}{3}-W_0\left(-e^{-x-1}\right)\right), \quad \sigma=-e^{q-\frac{16}3-x},\hspace{0.2cm}  d=-\frac 32 C \sigma, \quad C\equiv -3{\mathcal{W}_0}^2\gamma.  \label{sigma_d1Vminus}
\end{align}
We recall again that the parameters $x$ and $C$ determine the inflationary phase \cite{Antoniadis:2020stf} and are fixed from the observations to
\begin{equation}
 x\approx 3.3 \times 10^{-4}, \qquad C=e^{-3q}\times 7.81 \times 10^{-4}\equiv e^{-3q} C_0.  \label{C}
  \end{equation}
From the definition \eqref{sigma_d1Vminus} and constraints \eqref{C}, it seems that the $q$ parameter could indeed help to tune the vacuum energy, by shifting the value of $\mathcal{V}_-$, and thus the tachyonic contribution's coefficient $C_2$ defined in \cref{CY}. Applying the constraint on the waterfall position $\mathcal{V}_{c2}\approx\mathcal{V}_-$, the $C_2$ dependence on $q$ reads
  \begin{align}
 C_2&=\beta_2 \frac{d}{3 \mathcal{V}_{c}^{\frac43}}\approx\beta_2 \frac{d}{3 \mathcal{V}_{-}^{\frac43}}=\frac{-\beta_2 C\sigma}{2 \mathcal{V}_{-}^{\frac43}}= \beta_2 \frac{C_0}2 e^{-\frac 23 q} \times \frac{e^{-\frac{16}3-x}}{\left(\exp\left(\frac{13}{3}-W_0\left(-e^{-x-1}\right)\right)\right)^{\frac 43}}. \label{CYVminus}
 \end{align} 
Replacing $C_2$ from \eqref{CYVminus}, it follows that the scalar potential of \eqref{newvacuumformula} is nevertheless scale invariant with respect to $q$. It can indeed be expressed as
  \begin{align}
V(\mathbb{V},v)=\frac{C_0}{\kappa^4} \frac{4-\ln \mathbb{V}}{\mathbb{V}^3}+\frac{d_0}{\kappa^4\mathbb{V}^2}\left(1-\frac{\beta_2}3\left(1-\left(\frac{\mathbb{V}}{\mathbb{V}_-}\right)^{\frac23}\right)^2\right),\label{Veffinvariantq}
\end{align}
in terms of the $q$-dependent variables
\begin{equation}
 \mathbb{V}=e^{q}\, \mathcal{V}, \quad \mathbb{V}_-= e^q \, \mathcal{V}_-= \exp\left(\frac{13}{3}-W_0\left(-e^{-x-1}\right)\right), \quad d_0\equiv\frac32 C_0e^{-\frac{16}{3}-x}=e^{2q}d\,, \label{qinvparam}
 \end{equation}
 which absorb the explicit $q$-dependence of $V$.
 Hence, $V(\mathbb{V},v)$ only depends on $x$, $C_0$, both fixed by the inflationary phase, and $\beta_2$. 
As mentioned already, it is clear that the greater $\beta_2$ is, the lower the global minimum is. Hence the value $\beta_2=1$ gives the lowest minimum, which is then totally fixed by $x$ and $C_0$.

We conclude that in the simple case studied in the previous section, the value of the vacuum at the global minimum is totally fixed by the constraints on the inflationary phase and the waterfall scenario implementation, and that neither the $q$ nor $\beta_2$ parameters can help to lower it.

\paragraph{Influence of $\gamma$ corrections to the squared mass and quartic term}
 
 In the previous sections we neglected the contributions to the F-term squared mass and quartic terms for $\varphi_-$ coming from $\gamma$ and $\xi$ factors. We now examine if these corrections could add supplementary freedom allowing us to choose independently the mass and quartic coupling of the tachyonic field. As explained in the field theory description of \cref{fieldtheorydescription}, in this way one could tune the vacuum energy.
 
The aforementioned corrections can be read from the F-term supergravity formula through the expansion in the  $\varphi_-$ (or $C_2^{7_2}$) variable of the K\"ahler potential. The first corrections (in the $g_s$ and $\gamma$ expansion) to the mass and quartic contributions read
\begin{align}
\frac 12 m^2_{\gamma}&=\kappa^{-2} \frac{6 g_s \mathcal{W}_0^2 \gamma}{(U_1+\bar{U}_1)^2(U_2+\bar{U}_2)(U_3+\bar{U}_3)^2 } \frac{ \ln \mathcal{V}-2+q}{\mathcal{V}^3}
=-2\kappa^{-2}C_0 \frac{g_s}{f(U_i)} \frac{ \ln \mathbb{V}-2}{\mathbb{V}^3},\label{masscorrection}\\
\frac 14 \lambda_{\gamma}&=  \frac{2 g_s \mathcal{W}_0^2\gamma}{(U_1+\bar{U}_1)^2(U_2+\bar{U}_2)(U_3+\bar{U}_3)^2 } \frac{ 5\ln \mathcal{V}-9+5q}{\mathcal{V}^3}
=-\frac 23 C_0 \frac{g_s}{f(U_i)} \frac{ 5\ln \mathbb{V}-9}{\mathbb{V}^3}, \label{coupcorrection}
\end{align}
with $f(U_i)=(U_1+\bar{U}_1)^2(U_2+\bar{U}_2)(U_3+\bar{U}_3)^2$ and $\mathbb{V}$ introduced in \cref{qinvparam}. Recall that $-C>0$, so that these parameters are indeed positive. The mass and quartic terms of \cref{masswaterfall,couplingwaterfall} associated with these additional contributions now read:
\begin{align}
\frac 12 {m'_Y}^2(\mathcal{V})&= (m_{x_2}^2-m_{H_2}^2+\frac 12 m_{\gamma}^2)\nonumber\\
&= \frac {g_s^2\, x_2(U_2)}{\kappa^2\mathcal{V}^{\frac 23}} \left(\frac{1}{m_1^{\scriptscriptstyle(2)}m_1^{\scriptscriptstyle(3)}m_3^{\scriptscriptstyle{(1)}}m_3^{\scriptscriptstyle(2)}}\right)^{\frac13}\hspace{-0.2pt}\left(1-\left(\frac{\mathcal{V}_{c2}}{\mathcal{V}}\right)^{\frac23}\right)-\frac{2 Cg_s}{f(U_i)} \frac{ \ln \mathcal{V}-2+q}{\kappa^{2}\mathcal{V}^3}, \label{correctedmass} \\
 \frac14 \lambda'({\cal V})&= \left(2{g_{U(1)_2}^2}+\kappa^2{m_{x_2}^2}+\frac 14 \lambda_{\gamma}^2\right) \nonumber\\
 &= \frac{g_s}{\mathcal{V}^{\frac 23}} \left(\frac 1{\prod_{a\neq j}m_a^{\scriptscriptstyle(j)}}\right)^{\frac 13} \left( \vphantom{\frac11} 2 + g_s \, y(U_2) \right)-\frac 23 C \frac{g_s}{f(U_i)} \frac{ 5\ln \mathcal{V}-9+5q}{\mathcal{V}^3}. \label{correctedcoupling} 
\end{align}
The new critical volume cannot be computed analytically now. Nevertheless the $\gamma$ correction is suppressed by a factor $\mathcal{V}^{\scriptscriptstyle 7/3}$ and stays small for the values considered previously at large volume. Indeed, from \cref{correctedmass,correctedcoupling} we see that the coefficients in front of the previous contributions and $\gamma$ corrections scale as $g_s^2 y(U_2)$ against $g_sC$ for the mass, and $g_s$ against $Cg_s$ for the quartic coupling $\lambda$, so that it is not possible to balance the huge volume suppression $\mathcal{V}^{\scriptscriptstyle 7/3}$ of the $\gamma$ corrections.

We also remark that the corrections of \cref{masscorrection,coupcorrection} are effectively independent of the $q$ parameter since they only depend on $\mathbb{V}$, as the other contributions.

\subsection{Additional tachyons from other D7-brane stacks}\label{severaltachyons}


We now study the possibility of having several tachyons similar to the one described previously. We start with the addition of a second tachyon, generating a second waterfall direction. As the position of the second waterfall is only constrained to be at volumes $\mathcal{V}<\mathcal{V}_{c2}\approx\mathcal{V}_-$, we expect to have more freedom on the height of this second waterfall scalar potential contribution. We consider the following configuration:
\begin{equation}
\begin{tabular}{c|lll}
& (45) & (67) & (89) \\
  \hline
 $ D7_1$ & \hspace{4pt} $\cdot$ &   \hspace{6pt}$ \otimes$  & \hspace{6pt}$\times_{A_1}$  \\
 $D7_2 $ & \hspace{5pt}$\times$ & \hspace{5pt} $\cdot_{\,\, \pm {x}_2 }$ &  \hspace{3pt} $  \otimes $ \\
$D7_3$ &  \hspace{5pt}$\otimes$ &  \hspace{6pt}$ \times$  &   \hspace{5pt} $\cdot_{\pm x_3}$  \\
\end{tabular}
\end{equation}
The $D7_3$ brane tachyon is not eliminated by a Wilson line anymore. We introduce a position $x_3$ for the brane on the third torus $T^2_3$, eliminating the tachyon at large volumes, exactly as the one from the $D7_2$ brane. The mass of the string state is indeed of the form
\begin{equation}
\alpha' m_{33}=-\frac{2 |k_3^{\scriptscriptstyle{(1)}}|\alpha'}{\pi \mathcal{A}_1} + \frac{z(U_3) A_3}{\alpha'},
\end{equation}
where the function $z(U_3)$ plays a role similar to $y(U_2)$ in the previous sections and is directly related to the brane position $x_3$. As for the tachyon studied previously, we describe the new effective theory of the second tachyon $\psi_-$ through its masses $m_{x_3}$ and $m_{H_3}$, generated respectively by an F-term and a D-term, and the corresponding quartic couplings. Their expressions are similar to those of \cref{masswaterfall,couplingwaterfall} for the $D7_2$--$D7_2$ tachyon $\varphi_-$, replacing the fluxes and tori areas by the respective ones for the $D7_3$--$D7_3$ state. The corresponding parameters for this $D7_3$--$D7_3$ state are denoted with a $3$ subscript. For instance, $\mathcal{V}_{c3}$ is the critical volume of this second tachyon, corresponding to the position of the second waterfall.
For $\mathcal{V}_{c3}<\mathcal{V}_{c2}$, the study of the first phase transition does not change with respect to the single tachyon configuration. Indeed for $\mathcal{V}_{c3}<\mathcal{V}\leq\mathcal{V}_{c2}$, the second tachyon sits at its vanishing VEV $\langle\psi_-\rangle=0$ and does not contribute to the potential. Then, when $\mathcal{V}\leq\mathcal{V}_{c3}$ the second tachyonic field gets a non-vanishing VEV $v_3\neq0$ and its contribution to the scalar potential reads
\begin{equation}
V(\mathcal{V},v_3)=-\frac{m_Z^4}{\lambda_Z}(\mathcal{V})= - \frac {C_3}{\kappa^4\mathcal{V}^{\frac23}}\left(1-\left(\frac{\mathcal{V}_{c3}}{\mathcal{V}}\right)^{\frac23}\right)^2,
\end{equation}
with
  \begin{equation}
 \hspace{-0.2cm} \mathcal{V}_{c3}\equiv \left(\frac{2k }{\pi z(U_3)}\right)^{\frac32}\left|\frac{m_1^{\scriptscriptstyle(2)}m_1^{\scriptscriptstyle(3)}{m_2^{\scriptscriptstyle(1)}}{m_2^{\scriptscriptstyle(3)}}}{{m_3^{\scriptscriptstyle{(1)}}}^5{m_3^{\scriptscriptstyle(2)}}^5}\right|^{\frac14}\hspace{-0.1cm},
\hspace{0.3cm} C_3= \beta_3 \frac{d}{3 \mathcal{V}_{c3}^{\frac43}}, \label{C3}\hspace{0.3cm} \beta_3=\frac{2 }{2+g_s z(U_3)| m_3^{\scriptscriptstyle(1)}m_3^{\scriptscriptstyle(2)}|} \in [0,1]. \hspace{-0.1cm} 
 \end{equation}
 For $\mathcal{V}<\mathcal{V}_{c3} \leq \mathcal{V}_{c2}$, the dependence of the global minimum of the scalar potential hence reads
 \begin{align}
V(\mathcal{V},v_2,v_3)&=\frac{C}{\kappa^4} \left( -\frac{\ln \mathcal{V}-4+q}{{\mathcal V}^3}-\frac{3\sigma}{2\mathcal{V}^2}\right) -\sum_{a=2,3} \frac {C_a}{\kappa^4\mathcal{V}^{\frac23}}\left(1-\left(\frac{\mathcal{V}_{ca}}{\mathcal{V}}\right)^{\frac23}\right)^2.\label{twotachyonsscalar}
\end{align}

A short comment is in order on the way the global minimum is determined when several tachyons appear. The mass and coupling of $\varphi_-$ expressed in \cref{masswaterfall,couplingwaterfall} and the similar ones for $\psi_-$ are the ``bare" ones. As the `first' tachyon gets a VEV before the second one, contributions should appear due to interaction terms. These interaction terms come from the supergravity formula for the scalar potential through the expansion of the K\"ahler potential. 
Namely, corrections to the F-term mass and quartic coupling of $\psi_-$ due to the VEV of the $\varphi_-$ field are of the form:
 \begin{align}
&m_{x_3, corrections}^2 \sim \kappa^2 (m_{x_3}^2 + m_{x_2}^2) \langle \varphi_-\rangle ^2 + \kappa^4 m_{x_3}^2  \langle \varphi_-\rangle ^4 + \cdots \\
&\lambda_{corrections} \sim \kappa^2   m_{x_3, corrections}^2 = \kappa^4 (m_{x_3}^2 + m_{x_2}^2) \langle \varphi_-\rangle ^2 +  \kappa^4 m_{x_3}^2  \langle \varphi_-\rangle ^4+\cdots
\end{align}
 As long as $\langle \varphi_-\rangle $ stays small (compared to $\kappa$) these corrections are negligible in front of the ``bare" parameters and only shift the values of $C_3$ or $\mathcal{V}_{c3}$ by a small amount. Conversely, once $\psi_-$ gets a non-vanishing VEV, corrections to the first tachyon parameters also appear but are negligible and only shift lightly the values of $C_2$ or $\mathcal{V}_{c2}$.
 
\begin{figure}[h]
\center
 \includegraphics[scale=0.35]{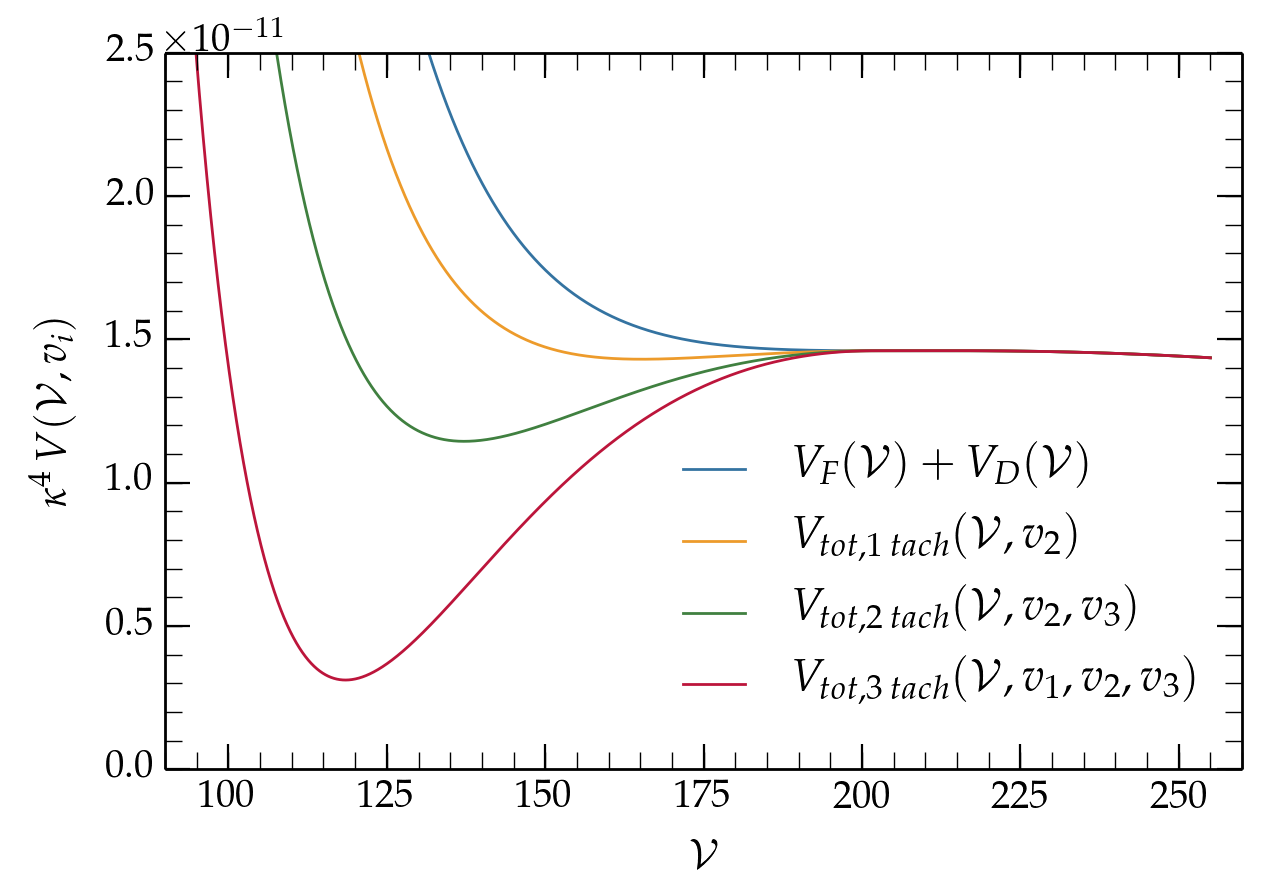}      
   \caption{Value of the global minimum of the effective scalar potential as a function of $\mathcal{V}$, without ({\it blue}), with one ({\it orange}), two ({\it green}) or three ({\it red}) tachyons. The parameters are such that $\mathcal{V}_{c2}=\mathcal{V}_-$, $\mathcal{V}_{c3}=0.99 \mathcal{V}_-$ and $\mathcal{V}_{c1}=0.98 \mathcal{V}_-$. }
   \label{Figure3tachyons}
    \end{figure}

We now turn back to the study of the global minimum. We see through \eqref{C3} that the amplitude $C_3$ of the tachyonic contribution and its critical volume $\mathcal{V}_{c3}$ are directly related. To get a large tachyonic contribution, we need to increase $C_3$, implying a smaller critical volume $\mathcal{V}_{c3}$.  Nevertheless, at small volumes the moduli part (the first contribution) of the scalar potential \eqref{twotachyonsscalar} dominates because it increases as $1/\mathcal{V}^3$, against $1/\mathcal{V}^{2}$ for the tachyonic contributions. Hence if $\mathcal{V}_{c3}$ is small, the tachyonic contribution only appears at small volumes and cannot compensate the moduli part.
In fact, it turns out that the largest contribution to the scalar potential from the second tachyon is for $\mathcal{V}_{c3}\approx \mathcal{V}_{c2}$ and hence $C_3 \approx C_2$. We see from Figure \ref{Figure3tachyons} that the second tachyon ({\it green curves}) contribution indeed lowers the value of the global minimum but is not sufficient to tune the vacuum energy to zero.  

We are thus naturally led to consider adding a third tachyon on the last brane $D7_1$. The treatment is identical to the one for the first two and its contribution is described by a critical volume $\mathcal{V}_{c1}$ related to the corresponding coefficient $C_1$. When $\mathcal{V}<\mathcal{V}_{c1}\leq \mathcal{V}_{c3} \leq \mathcal{V}_{c2}$, all three tachyons sit at their respective non-vanishing VEV. The value of the global minimum of the scalar potential is then as in \cref{twotachyonsscalar} but with a sum over the three tachyons:  \begin{align}
V(\mathcal{V},v_1,v_2,v_3)&=\frac{C}{\kappa^4} \left( -\frac{\ln \mathcal{V}-4+q}{{\mathcal V}^3}-\frac{3\sigma}{2\mathcal{V}^2}\right) -\sum_{a=1,2,3} \frac {C_a}{\kappa^4\mathcal{V}^{\frac23}}\left(1-\left(\frac{\mathcal{V}_{ca}}{\mathcal{V}}\right)^{\frac23}\right)^2.\label{threetachyonsscalar}
\end{align}

   From Figure \ref{Figure3tachyons} we see that the third tachyon is not sufficient yet to lower the global minimum to zero. In fact, this is understandable by the fact that when $\mathcal{V}_{c1}\approx\mathcal{V}_{c2}\approx\mathcal{V}_{c3}$, the value of the global minimum expressed as in \eqref{twotachyonsscalar} (but with the sum on $a=1,2,3$  tachyons) is almost similar to the one with only one tachyon, but with an effective tachyonic contribution coefficient equal to $C_1+C_2+C_3$ instead of $C_2$. As the $C_i$ are also related to the $\mathcal{V}_{ci}$ the only parameter to tune is $\beta_1+\beta_2+\beta_3\leq3$, which is always smaller than the desired value $\beta_2^{\Lambda=0}\approx3.228$ introduced in \eqref{tachampvanishingvacuum}. One can also wonder if having magnetic fluxes on the entire worldvolumes would allow to relax the relation between the $\mathcal{V}_{ci}$ and the $C_i$ in order to go above this bound, but we show in \Cref{entireworldvolumetachyons} that a configuration as in  \cref{multiplefieldseachbrane}  does not help.

It is now clear, as can be understood from the above discussion, that the addition of a fourth tachyon would allow for an effective $\beta=\sum_{i \, tachyons} \beta_i$ that could be higher that the value $\beta^{\Lambda=0}\approx3.228$, allowing to tune the vacuum energy to zero. In the next subsection we implement this idea in an example with a fourth $D7$-brane stack, parallel to one of the stacks previously studied. 
 
 \subsection{Adding a fourth magnetised stack}\label{sectionfourthstack}
As explained in the previous subsection, a fourth tachyon seems necessary to tune the vacuum energy of the minimum. One way to achieve this is by adding a fourth $D7$ brane stack, parallel to one of the one already present, say $D7_2$. We thus consider the following configuration
 \begin{equation}
\begin{tabular}{c|lll}
& (45) & (67) & (89) \\
  \hline
 $ D7_1$ & \hspace{4pt} $\cdot_{\pm x_1}$ &   \hspace{6pt}$ \otimes$  & \hspace{6pt}$\times$  \\
 $D7_{2a} $ & \hspace{5pt}$\times$ & \hspace{5pt} $\cdot_{\,\, \pm {x}_{2a} }$ &  \hspace{3pt} $  \otimes $ \\
  $D7_{2b} $ & \hspace{5pt}$\times$ & \hspace{5pt} $\cdot_{\,\, \pm {x}_{2b} }$ &  \hspace{3pt} $  \otimes $ \\
$D7_3$ &  \hspace{5pt}$\otimes$ &  \hspace{6pt}$ \times$  &   \hspace{5pt} $\cdot_{\pm x_3}$  \\
\end{tabular}
\end{equation}
The two $D7_{2i}$ branes can be studied exactly as before. The $D7_{2i}$--$D7_1$, $D7_{2i}$--$D7_3$, $D7_{2i}$--$D7_{2i}$ states are hence identical to the ones studied in \cref{model3fields}. The necessary condition to eliminate the mixed-state tachyons is similar to \eqref{mixedstatecondition}:
\begin{equation}
|\zeta_1^{\scriptscriptstyle(2)}|=|\zeta_{2a}^{\scriptscriptstyle(3)}|=\zeta_{2b}^{\scriptscriptstyle(3)}|=|\zeta_3^{\scriptscriptstyle(1)}|. \label{mixedstateconditionbis}
\end{equation}
 The new ingredient comes from the $D7_{2a}$--$D7_{2b}$ states. The magnetic fields produce the following mass for the lowest-lying states
   \begin{equation}
 \alpha'm^2=-|\zeta_{2a}^{\scriptscriptstyle(3)}|-|\zeta_{2a}^{\scriptscriptstyle(3)}|=-2|\zeta_{2a}^{\scriptscriptstyle(3)}|,
\end{equation}
where in the last equality we used equation \eqref{mixedstateconditionbis}.
The $D7_{2a}$--$D7_{2b}$ states also receive contributions from their relative distance, $i.e.$ from the separation in the second torus $T^2_2$ due to the different brane localisations $x_{2a}$ and $x_{2b}$. We recall that \begin{equation}
{x}_{2i} \equiv x_{2i}^x \, \bold{R}_{2x} + x_{2i}^y \, \bold{R}_{2y} \quad \text{with} \quad x_{2i}^x,x_{2i}^y \in \mathbb{Q}, \quad i=a,b. \label{position2branes}
\end{equation}
The mass contribution is then similar to the one of \eqref{branepositionmass}, with $x_2$ replaced by $x_{2ab}=x_{2a}-x_{2b}$. It reads
\begin{equation}
\alpha' m^2= \frac{x_{2ab} \cdot x_{2ab}}{\alpha'}= \frac{x_{2ab}^k  x_{2ab}^l g^{\scriptscriptstyle{(2)}}_{kl}}{\alpha'}=\frac{4 \mathcal{A}_2}{{\alpha'} \text{Re}(U_2)}\left|x_{2ab}^x-iU_2 x_{2ab}^y\right|^2\equiv \frac{y_{ab}(U_2)\,\mathcal{A}_2}{\alpha'}. \label{2branespositionmass}
\end{equation}
The total $D7_{2a}$--$D7_{2b}$ lowest-lying state mass then is
\begin{equation}
 \alpha' m_{22}^2=-2|\zeta_{2a}^{\scriptscriptstyle(3)}|+   \frac{y_{ab} \mathcal{A}_2}{\alpha'} \approx - \frac{2 \alpha' |k_{2a}^{\scriptscriptstyle(3)}|}{\pi \mathcal{A}_3}+   \frac{y_{ab} \mathcal{A}_2}{\alpha'}. \label{mass2a2bstate}
 \end{equation}

In the effective theory, the new mass contributions come from a D- and F-term, as in the previous cases. The second brane orthogonal to the $T_2^2$ torus give additional contributions to the $D$-term scalar potential obtained from the previous formula \eqref{Dscalarpot}, where we recall that the sum runs over the different $U(1)$ factors:
 \begin{align}
V_D&= \sum_{a}  \frac{g^2_{U(1)_a}}{2}\left(\xi_a+\sum_n q^n_a |\varphi^n_a|^2\right)^2 + \cdots \nonumber\\
&=  \frac{g^2_{U(1)_{2a}}}{2}\left(\xi_{2a}-2 |\varphi_{2a\, -}|^2-|\varphi_{2ab\,-}|^2 +\cdots\right)^2 +\frac{g^2_{U(1)_{2b}}}{2}\left(\xi_{2b}-2|\varphi_{2b\,-}|^2-|\varphi_{2ab\,-}|^2+\cdots\right)^2  \nonumber\\
&\hspace{0.5cm}+ \sum_{a=1,3}  \frac{g^2_{U(1)_a}}{2}\left(\xi_a-2 |\varphi_{a\,-}|^2+\cdots\right)^2 + \cdots \label{Dscalarpot2branes}
\end{align}
with the FI terms $\xi_i$ expressed from the fluxes as in \cref{FIfromDpot}. The additional $D7_{2b}$ brane adds a contribution to the $d_2$ term, defined in \eqref{dterms}, which now reads
\begin{equation}
d_2=\frac{g^2_{U(1)_{2a}}}{2}\xi_{2a}^2+\frac{g^2_{U(1)_{2b}}}{2}\xi_{2b}^2=g^2_{U(1)_{2a}}\xi_{2a}^2. \label{newd2}
\end{equation}
In the last equality we used the flux condition \eqref{mixedstateconditionbis} and the fact that for unit wrapping numbers $g^2_{U(1)_{2a}}=g^2_{U(1)_{2b}}$ since the two stacks are parallel, as can be seen from equation \eqref{couplingslargev}.
The D-term contributions to the masses and quartic couplings of the $\varphi_{2a,\, -}$, $\varphi_{2b,\, -}$ and $\varphi_{2ab,\, -}$ fields can be expressed by expanding the scalar potential \eqref{Dscalarpot2branes}. The masses have the same expressions while there is a factor of 2 difference between the quartic couplings of the doubly charged states and the bi-charged $D7_{2a}$--$D7_{2b}$ state.

The F-term contributions to the mass and quartic couplings can be derived as in the previous subsections, see \cref{mx2} and around, and read
\begin{align}
&m_{x_{2i}}^2=y_i(U_2) \frac{g_s^2}{\kappa^2\mathcal{V}}\frac{\mathcal{A}_2}{\alpha'}, \quad i=a,b, \qquad m_{x_{2ab}}^2=y_{ab}(U_2) \frac{g_s^2}{\kappa^2\mathcal{V}}\frac{\mathcal{A}_2}{\alpha'}, \label{mx2branes} \nonumber \\
&\lambda_{x_{2i}}=4 \kappa^2 m_{x_{2i}}^2, \qquad \lambda_{x_{2ab}} \sim 4 \kappa^2 m_{x_{2ab}}^2,
\end{align}
where we recall that $y_{ab}$ was defined in \cref{2branespositionmass} using $x_{2ab}=x_{2a}-x_{2b}$ and $y_i(U_2)$ are of course defined with respect to the respective brane positions $x_{2i}$.  There are some subtleties for the low-energy derivation of the mass and quartic couplings for the $D7_{2a}$ -- $D7_{2b}$ tachyon, because it does not appear in the same way as the $D7_{2i}$ -- $D7_{2i}$ tachyons in the K\"ahler potential and has a different superpotential expression. Nevertheless, as expected from the string mass formula, we obtain the dependences as in \eqref{mx2branes}.

The minimisation procedure follows as in the case with the three tachyons of \cref{severaltachyons}. In the present case, there are four tachyons coming from the doubly charged states between each stack and its image, and a fifth one from the $D7_{2a}$--$D7_{2b}$ sector. The value of the scalar potential at the minimum hence  reads
 \begin{align}
V(\mathcal{V},v_i)&=\frac{C}{\kappa^4} \left( -\frac{\ln \mathcal{V}-4+q}{{\mathcal V}^3}-\frac{3\sigma}{2\mathcal{V}^2}\right) - \sum_{i=1}^5 \frac {C_i}{\kappa^4\mathcal{V}^{\frac23}}\left(1-\left(\frac{\mathcal{V}_{ci}}{\mathcal{V}}\right)^{\frac23}\right)^2, \label{five_tachyons}
\end{align}
where the sum runs over the five tachyons mentioned above, hence $i=1,2a,2b,2ab,3$. The critical volumes and tachyonic contribution amplitudes can be computed as before and read
 \begin{align}
& {\mathcal{V}_{c1}}^{\frac23}\equiv \frac {2 |k_1^{\scriptscriptstyle{(2)}}|}{w(U_1)\pi} \left(\frac{d_2}{d_1}\right)^{\frac 13}\hspace{-4 pt},\quad  {\mathcal{V}_{c2i}}^{\frac23}\equiv \frac {2 |k_{2a}^{\scriptscriptstyle{(3)}}|}{y_i(U_2) \pi} \left(\frac{d_3}{d_2}\right)^{\frac 13}\hspace{-4 pt}, \,\, i=a,b,ab, \quad  {\mathcal{V}_{c3}}^{\frac23}\equiv \frac {2 |k_3^{\scriptscriptstyle{(1)}}|}{z(U_3) \pi} \left(\frac{d_1}{d_3}\right)^{\frac 13}\hspace{-4 pt}, \nonumber \\[5pt]
  &C_i= \beta_i \frac{d}{3 \mathcal{V}_{ci}^{\frac 43}}, \quad i=1,2ab, 3, \qquad  \qquad C_{j}= \frac 12 \beta_j\frac{d}{3 \mathcal{V}_{cj}^{\frac 43}}, \quad j=2a,2b, \label{CaVca5tachyons} \\
  & \beta_i=\frac{2 }{2+g_s f_i},\quad i=1,2a,2b,3 \quad {\rm with} \quad (f_1,f_{2a},f_{2b},f_3)=(w,y_a,y_b,z), \quad \beta_{2ab}=\frac{1 }{1+g_s y_{ab}}.\nonumber
 \end{align}
There is a small subtlety coming from the addition of a second parallel brane $D7_{2b}$, which modifies the $d_2$ parameter as in \cref{newd2}, and is responsible for the factor $\frac 12$ in $C_{2a},C_{2b}$. This factor is not present in $C_{2ab}$, because of the factor $2$ between the D-term quartic couplings mentioned under \cref{newd2}.
As in the case with three tachyons discussed under \eqref{threetachyonsscalar}, we look at the maximum value of the tachyonic amplitude, reached for almost equal $\mathcal{V}_{c,i}\approx\mathcal{V}_{-}$ and saturated value for $\beta_i=1$:
\begin{equation}
C_1+C_{2a}+C_{2b}+C_{2ab}+C_3 \approx 4 \frac{d}{3\mathcal{V}_{-}^{\frac43}}>\beta^{\Lambda=0} \frac{d}{3\mathcal{V}_{-}^{\frac43}}. \label{maxbeta5tachyons}
\end{equation}
Hence the value $\beta^{\Lambda=0}$ introduced in \cref{tachampvanishingvacuum} can be reached with the current configuration, $i.e.$ through the addition of the fourth brane $D7_{2b}$, and the value of the global minimum of the potential can be tuned to almost zero. With the saturated bound of equation \eqref{maxbeta5tachyons}, the sum of the tachyonic contributions to the global minimum is greater than the moduli contribution and an AdS vacuum is obtained. There are several options to tune the global minimum: one can either lower the $\beta_i$ parameters or choose smaller tachyonic critical volumes (except for the first waterfall field responsible for the end of inflation). 

\begin{figure}[h!]
\begin{minipage}{0.45\textwidth}
   \hspace{-0.2cm}\includegraphics[scale=0.33]{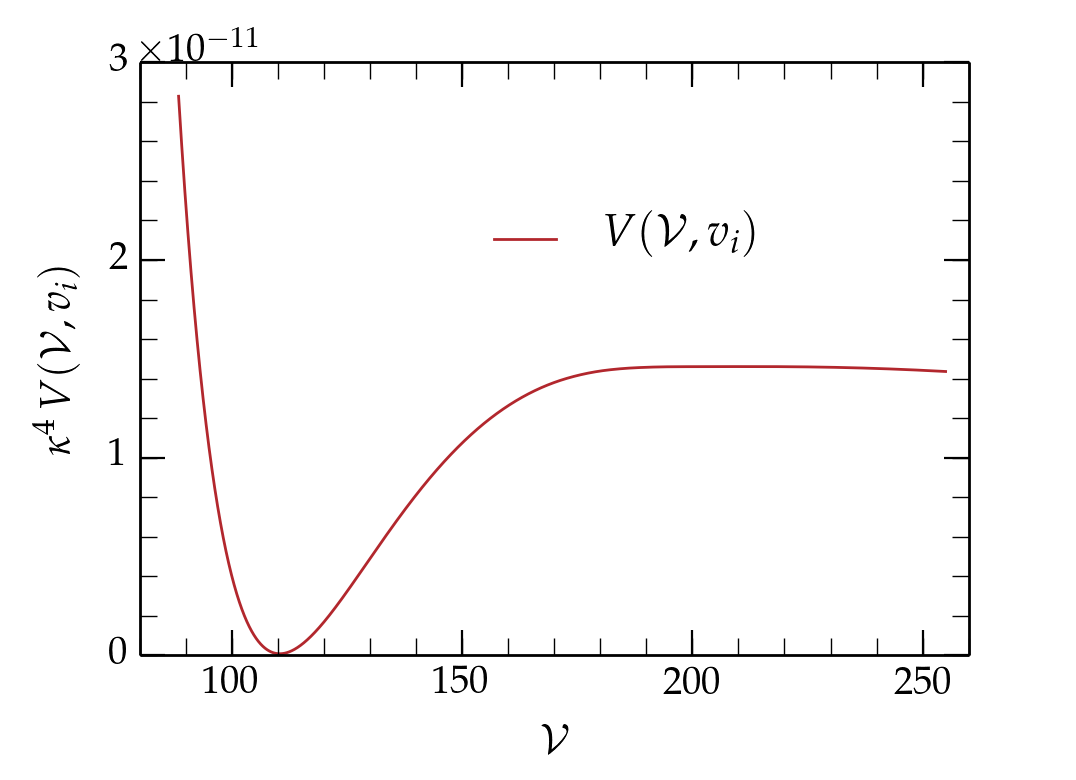}      
  \end{minipage}
    \hspace{0.3cm}
  \begin{minipage}{0.45\textwidth}
   \hspace{0cm}\includegraphics[scale=0.34]{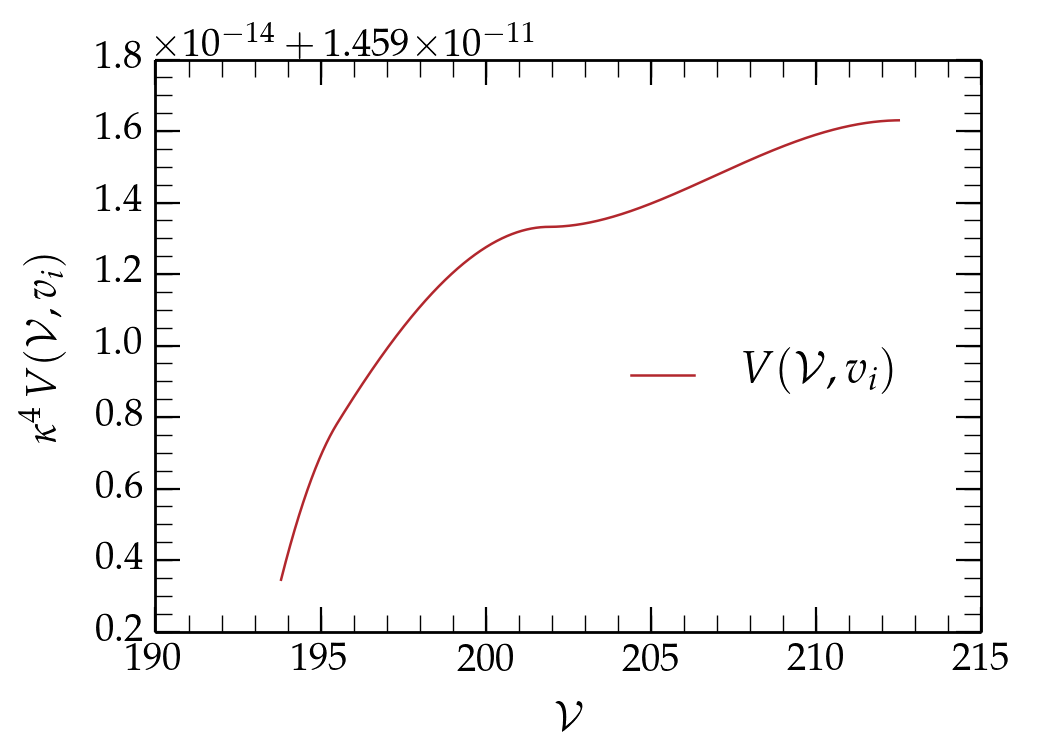}      
  \end{minipage}
  \caption{Value $V(\mathcal{V},v_i)$ of the global minimum of the scalar potential as a function of the internal volume, for the parameters \eqref{parameters5tachyons}. We show  {\it (left panel)} the almost vanishing value of the global minimum and focus {\it (right panel)} on the waterfall zone near $\mathcal{V}_-\approx 201.9$.}
\label{scalarpotwaterfall5tachyons}
 \end{figure}

Taking for simplicity only unit wrapping numbers, the choice of parameters
\begin{align}
&g_s= 8.025\times10^{-3} \qquad n_1^{\scriptscriptstyle(2)}=12, \qquad m_i^{\scriptscriptstyle{(j)}}=1,  \qquad x(U_1)=0.185,\nonumber\\
& z(U_3)=0.189, \quad y_a(U_2) =0.0881, \quad y_b(U_2)=0.098, \quad y_{ab}(U_2)=0.09, \label{parameters5tachyons}
\end{align}
gives the following values for $d$, the critical volumes and the $\beta_i$ coefficients:
\begin{align}
&d=5.65 \times10^{-6}, \quad \beta_i\approx1, \,\, i=1,2a,2b,2ab,3, \quad \mathcal{V}_{c2a}=201.9\approx\mathcal{V}_-, \\
&\mathcal{V}_{c1}=187.6,  \quad \mathcal{V}_{c2b}=172.1, \quad \mathcal{V}_{c2ab}=195.5, \quad \mathcal{V}_{c3}=181.7. \nonumber
\end{align}
Figure \ref{scalarpotwaterfall5tachyons} shows the value of the global minimum of the scalar potential as a function of the internal volume for the parameters of \eqref{parameters5tachyons}. We see that with this choice the cosmological constant can indeed be tuned to an almost vanishing positive value.
 
 \section{Conclusion}
In this work we have  shown that the essential  principles of hybrid inflation can be naturally implemented in the context of a mechanism of moduli stabilisation providing a metastable de Sitter vacuum within type  IIB  flux compactifications, using only perturbative quantum corrections. 
Identifying the inflaton field with the internal volume modulus, it is found that slow-roll inflation can be readily implemented, however, the minimum of the potential corresponds to a false vacuum with an unacceptably large cosmological constant, of the order of the inflation scale. This is where hybrid inflation can come into rescue by introducing an extra steep `waterfall' direction from a saddle point near the previous minimum, down to a new vacuum that can accommodate the present amount of dark energy.

In this paper, this issue is naturally resolved within our framework by considering the case where the role of the waterfall field is realised by an appropriate open string excitation located  on the  three $D7$-brane stacks. Indeed charged states on the branes and their intersections receive a tachyonic contribution due to the coupling of the magnetic field with the internal spin and a positive (supersymmetric) contribution when Wilson lines along the branes worldvolume are turned on, or branes separation in the transverse directions.
We have shown that for appropriate magnetic fluxes and brane separations, the minimum of the potential occurs at a critical value of the internal volume where  charged  tachyonic states appear playing the role of the waterfall field.  Inflation is realised with 60 e-folds accumulated  as the inflaton rolls towards the minimum, while the variation of the inflaton field  is small compared to the Planck scale as in small field  inflation models, consistently with the validity of the effective field theory and swampland distance conjecture. Moreover, the magnetic fluxes generate  the  appropriate coupling with
the waterfall fields which is necessary to realise the transition to the true vacuum.  In conclusion, the  main features of the
proposed framework described above  tally with the general principles of hybrid inflation, establishing a  firm  ground for the  implementation of this scenario in string theory.

\section*{Acknowledgments}
Work partially performed by I.A. as International professor of the Francqui Foundation, Belgium.

\appendix
\section{Theta functions and $so(2)$ characters} \label{AppendixA}

The Jacobi theta functions are introduced as 

\begin{align}
\vartheta\left[^{\alpha}_{\beta}\right](z|\tau)&=\sum_n q^{\frac12 (n+\alpha)^2}e^{2\pi i (n+\alpha)(z+\beta)} \nonumber \\
&=e^{2\pi i \alpha (z+\beta)q^{\frac{\alpha^2}2}}\prod_{n=1}^{\infty}(1-q^n) (1+q^{n+\alpha-\frac 12}e^{2\pi i (z+\beta)})(1+q^{n-\alpha-\frac 12}e^{-2\pi i (z+\beta)}).
\end{align}
From them we define the four theta functions
\begin{align}
&\vartheta_1(z|\tau)\equiv \vartheta\left[^{1/2}_{1/2}\right](z|\tau), \qquad& \vartheta_2(z|\tau)\equiv \vartheta\left[^{1/2}_{\,\,\,0}\right](z|\tau),  \nonumber\\
& \vartheta_3(z|\tau)\equiv\vartheta\left[^{0}_{0}\right](z|\tau),  &\vartheta_4(z|\tau)\equiv\vartheta\left[^{\,\,\,0}_{1/2}\right](z|\tau),
\end{align}
used in the superstring amplitudes we consider in this work. The four level-one $so(2)$ characters read 
\begin{align}
&O_2(z)=\frac{\vartheta_3(z|\tau)+\vartheta_4(z|\tau)}{2\eta}, \quad V_2(z)=\frac{\vartheta_3(z|\tau)-\vartheta_4(z|\tau)}{2\eta}, \nonumber \\
&S_2(z)=\frac{\vartheta_2(z|\tau)-\vartheta_1(z|\tau)}{2\eta}, \quad C_2(z)=\frac{\vartheta_2(z|\tau)+\vartheta_1(z|\tau)}{2\eta}, \label{characters}
\end{align}
with the Dedekind function defined by
\begin{equation}
\eta(\tau)=q^{\frac1{24}}\prod_{n=1}^{\infty}(1-q^n)~.
\end{equation}
The space-time characters used in $\mathbb{Z}_2\times\mathbb{Z}_2$ toroidal orbifolds are constructed from the $so(2)$ characters \eqref{characters} and read :
\begin{align}
\tau_{oo}&=V_2O_2O_2O_2+O_2V_2V_2V_2-S_2S_2S_2S_2-C_2C_2C_2C_2 \nonumber\\
 \tau_{og}&=O_2V_2O_2O_2+V_2O_2V_2V_2-C_2C_2S_2S_2-S_2S_2C_2C_2 \nonumber\\
 \tau_{oh}&=O_2O_2O_2V_2+V_2V_2V_2O_2-C_2S_2S_2C_2-S_2C_2C_2S_2 \nonumber\\
\tau_{of}&=O_2O_2V_2O_2+V_2V_2O_2V_2-C_2S_2C_2S_2-S_2C_2S_2C_2 \nonumber\\
\nonumber\\
\tau_{go}&=V_2O_2S_2C_2+O_2V_2C_2S_2-S_2S_2V_2O_2-C_2C_2O_2V_2 \nonumber\\
 \tau_{gg}&=O_2V_2S_2C_2+V_2O_2C_2S_2-S_2S_2O_2V_2-C_2C_2V_2O_2 \nonumber\\
 \tau_{gh}&=O_2O_2S_2S_2+V_2V_2C_2C_2-C_2S_2V_2V_2-S_2C_2O_2O_2 \nonumber\\
\tau_{gf}&=O_2O_2C_2C_2+V_2V_2S_2S_2-S_2C_2V_2V_2-C_2S_2O_2O_2 \nonumber\\
\label{Z2characters}\\
\tau_{ho}&=V_2S_2C_2O_2+O_2C_2S_2V_2-C_2O_2V_2C_2-S_2V_2O_2S_2 \nonumber\\
 \tau_{hg}&=O_2C_2C_2O_2+V_2S_2S_2V_2-C_2O_2O_2S_2-S_2V_2V_2C_2 \nonumber\\
 \tau_{hh}&=O_2S_2C_2V_2+V_2C_2S_2O_2-S_2O_2V_2S_2-C_2V_2O_2C_2 \nonumber\\
\tau_{hf}&=O_2S_2S_2O_2+V_2C_2C_2V_2-C_2V_2V_2S_2-S_2O_2O_2C_2 \nonumber\\
\nonumber\\
\tau_{fo}&=V_2S_2O_2C_2+O_2C_2V_2S_2-S_2V_2S_2O_2-C_2O_2C_2V_2 \nonumber\\
 \tau_{fg}&=O_2C_2O_2C_2+V_2S_2V_2S_2-C_2O_2S_2O_2-S_2V_2C_2V_2 \nonumber\\
 \tau_{fh}&=O_2S_2O_2S_2+V_2C_2V_2C_2-C_2V_2S_2V_2-S_2O_2C_2O_2 \nonumber\\
\tau_{ff}&=O_2S_2V_2C_2+V_2C_2O_2S_2-C_2V_2C_2O_2-S_2O_2S_2V_2 \nonumber
\end{align}
The $T_{kj}$ characters used in the $T^6/\mathbb{Z}_2\times \mathbb{Z}_2$  model of \cref{section3} are \cite{Angelantonj:2002ct,Larosa:2003mz}
\begin{align}
T_{ko}=\tau_{ko}+\tau_{kg}+\tau_{kh}+\tau_{kf}, \quad T_{kg}=\tau_{ko}+\tau_{kg}-\tau_{kh}-\tau_{kf}, \nonumber \\
T_{kh}=\tau_{ko}-\tau_{kg}+\tau_{kh}-\tau_{kf}, \quad T_{kf}=\tau_{ko}-\tau_{kg}-\tau_{kh}+\tau_{kf},  
\end{align}
for $k=o,f,h,g$.

\section{Momenta and windings sums} \label{Appendixsums}
In absence of B-field background, the $T^2_i$ torus momenta, lying on the dual lattice $\Lambda_i^*$ defined under equation \eqref{torusdefinition}, read
\begin{equation}
\bold{p}_i=m_k \bold{R}_{i}^{*k}, \quad m_k\in \mathbb{Z}.
\end{equation}
Defining the $T^2_i$ torus windings, lying on the lattice $\Lambda_i$, through
\begin{equation}
\bold{L}_i=n^l\bold{R}_{il} \quad n^l \in \mathbb{Z},
\end{equation}
we introduce left and right momenta
\begin{equation}
\bold{p}_{iL,R}=\left(m_k \pm g^{(i)}_{kl}n^l \right)\bold{R}_i^{*k}.
\end{equation}
 The $T^2_i$ torus partition function is then defined by 
 \begin{equation}
 \Lambda_i=\sum_{m,n}\frac{q^{\frac{\alpha'}{4}\bold{p}_{iL}\cdot\bold{p}_{iL}}\,\bar{q}^{\frac{\alpha'}{4}\bold{p}_{iR}\cdot\bold{p}_{iR}}}{|\eta(\tau)|^4}~.
 \end{equation}
 The Klein-bottle windings and momenta sums read
 \begin{equation}
  W_i=\sum_{n}\frac{q^{\frac{1}{2\alpha'}\bold{L}_{i}\cdot\bold{L}_{i}}}{\eta(2i{\rm Im}\tau)^2}~, \qquad   P_i=\sum_{n}\frac{e^{-2\pi \ell \alpha'\bold{p}_{i}\cdot\bold{p}_{i}}}{\eta(i\ell)^2}~,
 \end{equation} 
 and the annulus ones are
  \begin{equation}
  W_i=\sum_{n}\frac{e^{-2\pi \ell \frac{1}{4\alpha'} \bold{L}_{i}\cdot\bold{L}_{i}}}{\eta(i\ell)^2}~, \qquad   P_i=\sum_{m}\frac{q^{\frac{\alpha'}2\bold{p}_{i}\cdot\bold{p}_{i}}}{\eta(i{\rm Im}\tau/2)^2}~,
 \end{equation} 
 with $\ell$ being the modulus of the double cover of either the Klein bottle or the annulus \cite{Angelantonj:2002ct}.
 
  \section{Tachyons from magnetic fields on the entire D7-branes worldvolumes}\label{entireworldvolumetachyons}

In this appendix we study the tachyons generated by a configuration with three $D7$-brane stacks with magnetic fields on the entire worldvolumes. This is motivated because we saw in \cref{severaltachyons} that \cref{CY,C3} fix the relation between the critical volumes and the amplitudes through the flux parameter $d$. This parameter plays a crucial role in the inflationary phase and is fixed by observations. In the simple flux configuration of \cref{severaltachyons}, all fluxes were taken equal, hence $d_1=d_2=d_3$. One may wonder if allowing for different $d_a$ would relax relations between the tachyonic contribution scalings and the critical volumes, by introducing $d_a$ in the relations similar to \cref{CY,C3}. 

According to the study of section \ref{multiplefieldseachbrane}, it is possible to have different doubly charged states masses (and hence different $d_a$) by putting magnetic fields on the entire brane worldvolumes, as shown in the following table.
\begin{center}
\begin{tabular}{c|lll}
& (45) & (67) & (89) \\
  \hline
 $ D7_1$ & \hspace{4pt} $\cdot_{\pm x_1}$ &   \hspace{6pt}$ \otimes$  & \hspace{6pt}$\otimes$  \\
 $D7_2 $ & \hspace{5pt}$\otimes$ & \hspace{5pt} $\cdot_{\,\, \pm {x}_2 }$ &  \hspace{3pt} $  \otimes $ \\
$D7_3$ &  \hspace{5pt}$\otimes$ &  \hspace{6pt}$ \otimes$  &   \hspace{5pt} $\cdot_{\pm x_3}$  \\
\end{tabular}
\end{center}
\vspace{0.2cm}

We recall that the magnetic fields are subject to conditions $(A-i)$ or $(B-i)$ of equation \eqref{conditionsentireworldvolumes} to eliminate the mixed states tachyons. In order to have the possibility for different (non-vanishing) $d_a$, we choose the configuration of fluxes in condition $(B-1)$ of equation \eqref{conditionsentireworldvolumes} that we recall here for simplicity
\begin{equation}
(B- 1)  \qquad \zeta_1^{\scriptscriptstyle(2)}=\zeta_1^{\scriptscriptstyle(3)}, \qquad \zeta_2^{\scriptscriptstyle(1)}=\zeta_2^{\scriptscriptstyle(3)}, \qquad  \zeta_3^{\scriptscriptstyle(1)}=\zeta_3^{\scriptscriptstyle(2)}. \label{repeatconditioinfluxes}
\end{equation}

An important point is that when magnetic fields are plugged on the entire worldvolumes, one cannot use Wilson lines $A_i$ anymore to eliminate the tachyons from the doubly charged states. The only way is to use brane separations $x_i$, which indeed eliminate tachyons at large volumes but lead to tachyons under a certain critical volume. This was phenomenon was described in details in the previous subsections. In the present case, we thus have to consider one tachyon for each doubly charged $D7_i$--$D7_i$ state. 
As before, the tachyonic masses contributions generated by the magnetic fluxes and brane separation at the string level read
\begin{align}
&\alpha'm^2_{11}=-2 \left|\zeta_1^{\scriptscriptstyle{(2)}}+\zeta_1^{\scriptscriptstyle{(3)}}\right|+ \frac{w \mathcal{A}_1}{\alpha'}\underset{(B-1)}{=}-4 \left|\zeta_1^{\scriptscriptstyle{(2)}}\right|+ \frac{w \mathcal{A}_1}{\alpha'} \approx -\frac{4 |k_1^{\scriptscriptstyle{(2)}}|\alpha'}{\pi \mathcal{A}_2}+ \frac{w \mathcal{A}_1}{\alpha'}, \nonumber \\
&\alpha'm^2_{22}=-2 \left|\zeta_2^{\scriptscriptstyle{(1)}}+\zeta_2^{\scriptscriptstyle{(3)}}\right|+ \frac{y \mathcal{A}_2}{\alpha'}\underset{(B-1)}{=}-4 \left|\zeta_2^{\scriptscriptstyle{(1)}}\right|+ \frac{y \mathcal{A}_2}{\alpha'} \approx -\frac{4 |k_2^{\scriptscriptstyle{(1)}}|\alpha'}{\pi \mathcal{A}_1}+ \frac{y \mathcal{A}_2}{\alpha'},  \\
&\alpha'm^2_{33}=-2 \left|\zeta_3^{\scriptscriptstyle{(1)}}+\zeta_3^{\scriptscriptstyle{(2)}}\right|+ \frac{z \mathcal{A}_3}{\alpha'}\underset{(B-1)}{=}-4 \left|\zeta_3^{\scriptscriptstyle{(1)}}\right|+ \frac{z \mathcal{A}_3}{\alpha'} \approx -\frac{4 |k_3^{\scriptscriptstyle{(1)}}|\alpha'}{\pi \mathcal{A}_1}+ \frac{z \mathcal{A}_3}{\alpha'}. \nonumber
\end{align}
 In the low energy effective theory this corresponds to $d_a$ parameters of the form 
 \begin{equation}
 d_a=\frac 12g_s^3 |m_a^{\scriptscriptstyle{(j)}}m_a^{\scriptscriptstyle{(k)}}|\left(\frac{2k_a^{\scriptscriptstyle{(j)}}}{\pi}\right)^2,\quad a \neq j \neq k \neq a.
 \end{equation}
 Remember that the moduli stabilisation conditions depend on these $d_a$ and are given by \eqref{modulistabresult}. Together with \eqref{repeatconditioinfluxes} these conditions allow to express $e.g.$ $n_1^{\scriptscriptstyle{(3)}}$, $n_2^{\scriptscriptstyle{(1)}}$ and $n_3^{\scriptscriptstyle{(2)}}$ with respect to $n_1^{\scriptscriptstyle{(2)}}$, $n_2^{\scriptscriptstyle{(3)}}$, $n_3^{\scriptscriptstyle{(1)}}$ and the $m_a^{\scriptscriptstyle{(j)}}$, hence leaving only three independent flux numbers together with the wrapping numbers.

After some  straightforward manipulations we check that when the volume is inferior to all the critical volumes, $i.e.$ when for any value of $a$, $\mathcal{V}<V_{c,a}$, the scalar potential reads
 \begin{align}
V(\mathcal{V},v_1,v_2,v_3)&=\frac{C}{\kappa^4} \left( -\frac{\ln \mathcal{V}-4+q}{{\mathcal V}^3}-\frac{3\sigma}{2\mathcal{V}^2}\right) - \sum_{a=1}^3 \frac {C_a}{\kappa^4\mathcal{V}^{\frac23}}\left(1-\left(\frac{\mathcal{V}_{ca}}{\mathcal{V}}\right)^{\frac23}\right)^2 \label{threetachyons_allwv}
\end{align}
with again
 \begin{align}
& {\mathcal{V}_{c1}}^{\frac23}\equiv \frac {4 |k_1^{\scriptscriptstyle{(2)}}|}{w(U_1)\pi} \left(\frac{d_2}{d_1}\right)^{\frac 13},\quad  {\mathcal{V}_{c2}}^{\frac23}\equiv \frac {4 |k_2^{\scriptscriptstyle{(1)}}|}{y(U_2) \pi} \left(\frac{d_2}{d_1}\right)^{\frac 13}, \quad  {\mathcal{V}_{c3}}^{\frac23}\equiv \frac {4 |k_3^{\scriptscriptstyle{(1)}}|}{z(U_3) \pi} \left(\frac{d_1}{d_3}\right)^{\frac 13}, \nonumber \\
  &C_a= \beta_a \frac{d}{3 \mathcal{V}_{ca}},\qquad \label{CaVca} \beta_a=\frac{2 }{2+g_s f_a(U_a)| m_a^{\scriptscriptstyle(j)}m_a^{\scriptscriptstyle(k)}|} \in [0,1], \quad (f_1,f_2,f_3)=(w,y,z).
 \end{align}
 Hence we see that even with different $d_a$ as in the current configuration, the relations \eqref{CaVca} between the critical volumes and the amplitudes of the tachyonic contributions only imply $d=3(d_1d_2d_3)^{\frac 13}$, as in the simpler case with only one magnetic field per brane. The potential is thus identical to the one with three tachyons ({\it  red curve}) of Figure \ref{Figure3tachyons}.
 

\begin{thebibliography}{10}

\bibitem{Kachru:2003aw}
S.~Kachru, R.~Kallosh, A.~D. Linde and S.~P. Trivedi, \emph{{De Sitter vacua in
  string theory}},
  \href{http://dx.doi.org/10.1103/PhysRevD.68.046005}{\emph{Phys. Rev. D} {\bf
  68} (2003) 046005}, [\href{https://arxiv.org/abs/hep-th/0301240}{{\tt
  hep-th/0301240}}].

\bibitem{Conlon:2005ki}
J.~P. Conlon, F.~Quevedo and K.~Suruliz, \emph{{Large-volume flux
  compactifications: Moduli spectrum and D3/D7 soft supersymmetry breaking}},
  \href{http://dx.doi.org/10.1088/1126-6708/2005/08/007}{\emph{JHEP} {\bf 08}
  (2005) 007}, [\href{https://arxiv.org/abs/hep-th/0505076}{{\tt
  hep-th/0505076}}].

\bibitem{Antoniadis:2018hqy}
I.~Antoniadis, Y.~Chen and G.~K. Leontaris, \emph{{Perturbative moduli
  stabilisation in type IIB/F-theory framework}},
  \href{http://dx.doi.org/10.1140/epjc/s10052-018-6248-4}{\emph{Eur. Phys. J.
  C} {\bf 78} (2018) 766}, [\href{https://arxiv.org/abs/1803.08941}{{\tt
  1803.08941}}].

\bibitem{Antoniadis:2019rkh}
I.~Antoniadis, Y.~Chen and G.~K. Leontaris, \emph{{Logarithmic loop
  corrections, moduli stabilisation and de Sitter vacua in string theory}},
  \href{http://dx.doi.org/10.1007/JHEP01(2020)149}{\emph{JHEP} {\bf 01} (2020)
  149}, [\href{https://arxiv.org/abs/1909.10525}{{\tt 1909.10525}}].

\bibitem{Antoniadis:2002tr}
I.~Antoniadis, R.~Minasian and P.~Vanhove, \emph{{Noncompact Calabi-Yau
  manifolds and localized gravity}},
  \href{http://dx.doi.org/10.1016/S0550-3213(02)00974-4}{\emph{Nucl. Phys. B}
  {\bf 648} (2003) 69--93}, [\href{https://arxiv.org/abs/hep-th/0209030}{{\tt
  hep-th/0209030}}].

\bibitem{Antoniadis:1998ax}
I.~Antoniadis and C.~Bachas, \emph{{Branes and the gauge hierarchy}},
  \href{http://dx.doi.org/10.1016/S0370-2693(99)00102-1}{\emph{Phys. Lett. B}
  {\bf 450} (1999) 83--91}, [\href{https://arxiv.org/abs/hep-th/9812093}{{\tt
  hep-th/9812093}}].

\bibitem{Cremmer:1983bf}
E.~Cremmer, S.~Ferrara, C.~Kounnas and D.~V. Nanopoulos, \emph{{Naturally
  Vanishing Cosmological Constant in N=1 Supergravity}},
  \href{http://dx.doi.org/10.1016/0370-2693(83)90106-5}{\emph{Phys. Lett. B}
  {\bf 133} (1983) 61}.

\bibitem{Ellis:1983ei}
J.~R. Ellis, C.~Kounnas and D.~V. Nanopoulos, \emph{{Phenomenological SU(1,1)
  Supergravity}},
  \href{http://dx.doi.org/10.1016/0550-3213(84)90054-3}{\emph{Nucl. Phys. B}
  {\bf 241} (1984) 406--428}.

\bibitem{Becker:2002nn}
K.~Becker, M.~Becker, M.~Haack and J.~Louis, \emph{{Supersymmetry breaking and
  alpha-prime corrections to flux induced potentials}},
  \href{http://dx.doi.org/10.1088/1126-6708/2002/06/060}{\emph{JHEP} {\bf 06}
  (2002) 060}, [\href{https://arxiv.org/abs/hep-th/0204254}{{\tt
  hep-th/0204254}}].

\bibitem{Burgess:2003ic}
C.~P. Burgess, R.~Kallosh and F.~Quevedo, \emph{{De Sitter string vacua from
  supersymmetric D terms}},
  \href{http://dx.doi.org/10.1088/1126-6708/2003/10/056}{\emph{JHEP} {\bf 10}
  (2003) 056}, [\href{https://arxiv.org/abs/hep-th/0309187}{{\tt
  hep-th/0309187}}].

\bibitem{Antoniadis:2020stf}
I.~Antoniadis, O.~Lacombe and G.~K. Leontaris, \emph{{Inflation near a
  metastable de Sitter vacuum from moduli stabilisation}},
  \href{http://dx.doi.org/10.1140/epjc/s10052-020-08581-9}{\emph{Eur. Phys. J.
  C} {\bf 80} (2020) 1014}, [\href{https://arxiv.org/abs/2007.10362}{{\tt
  2007.10362}}].

\bibitem{Linde:1993cn}
A.~D. Linde, \emph{{Hybrid inflation}},
  \href{http://dx.doi.org/10.1103/PhysRevD.49.748}{\emph{Phys. Rev. D} {\bf 49}
  (1994) 748--754}, [\href{https://arxiv.org/abs/astro-ph/9307002}{{\tt
  astro-ph/9307002}}].

\bibitem{Frey:2002hf}
A.~R. Frey and J.~Polchinski, \emph{{N=3 warped compactifications}},
  \href{http://dx.doi.org/10.1103/PhysRevD.65.126009}{\emph{Phys. Rev. D} {\bf
  65} (2002) 126009}, [\href{https://arxiv.org/abs/hep-th/0201029}{{\tt
  hep-th/0201029}}].

\bibitem{Kachru:2002he}
S.~Kachru, M.~B. Schulz and S.~Trivedi, \emph{{Moduli stabilization from fluxes
  in a simple IIB orientifold}},
  \href{http://dx.doi.org/10.1088/1126-6708/2003/10/007}{\emph{JHEP} {\bf 10}
  (2003) 007}, [\href{https://arxiv.org/abs/hep-th/0201028}{{\tt
  hep-th/0201028}}].

\bibitem{Gukov:1999ya}
S.~Gukov, C.~Vafa and E.~Witten, \emph{{CFT's from Calabi-Yau four folds}},
  \href{http://dx.doi.org/10.1016/S0550-3213(00)00373-4}{\emph{Nucl. Phys. B}
  {\bf 584} (2000) 69--108}, [\href{https://arxiv.org/abs/hep-th/9906070}{{\tt
  hep-th/9906070}}].

\bibitem{Grisaru:1986kw}
M.~T. Grisaru, A.~E.~M. van~de Ven and D.~Zanon, \emph{{Four Loop Divergences
  for the N=1 Supersymmetric Nonlinear Sigma Model in Two-Dimensions}},
  \href{http://dx.doi.org/10.1016/0550-3213(86)90449-9}{\emph{Nucl. Phys. B}
  {\bf 277} (1986) 409--428}.

\bibitem{Antoniadis:1997eg}
I.~Antoniadis, S.~Ferrara, R.~Minasian and K.~S. Narain, \emph{{R**4 couplings
  in M and type II theories on Calabi-Yau spaces}},
  \href{http://dx.doi.org/10.1016/S0550-3213(97)00572-5}{\emph{Nucl. Phys. B}
  {\bf 507} (1997) 571--588}, [\href{https://arxiv.org/abs/hep-th/9707013}{{\tt
  hep-th/9707013}}].

\bibitem{Antoniadis:2003sw}
I.~Antoniadis, R.~Minasian, S.~Theisen and P.~Vanhove, \emph{{String loop
  corrections to the universal hypermultiplet}},
  \href{http://dx.doi.org/10.1088/0264-9381/20/23/009}{\emph{Class. Quant.
  Grav.} {\bf 20} (2003) 5079--5102},
  [\href{https://arxiv.org/abs/hep-th/0307268}{{\tt hep-th/0307268}}].

\bibitem{Larosa:2003mz}
M.~Larosa and G.~Pradisi, \emph{{Magnetized four-dimensional Z(2) x Z(2)
  orientifolds}},
  \href{http://dx.doi.org/10.1016/S0550-3213(03)00551-0}{\emph{Nucl. Phys. B}
  {\bf 667} (2003) 261--309}, [\href{https://arxiv.org/abs/hep-th/0305224}{{\tt
  hep-th/0305224}}].

\bibitem{Aldazabal:1998mr}
G.~Aldazabal, A.~Font, L.~E. Ib{\`a}\~nez and G.~Violero, \emph{{D = 4, N=1,
  type IIB orientifolds}},
  \href{http://dx.doi.org/10.1016/S0550-3213(98)00666-X}{\emph{Nucl. Phys. B}
  {\bf 536} (1998) 29--68}, [\href{https://arxiv.org/abs/hep-th/9804026}{{\tt
  hep-th/9804026}}].

\bibitem{Bianchi:1991eu}
M.~Bianchi, G.~Pradisi and A.~Sagnotti, \emph{{Toroidal compactification and
  symmetry breaking in open string theories}},
  \href{http://dx.doi.org/10.1016/0550-3213(92)90129-Y}{\emph{Nucl. Phys. B}
  {\bf 376} (1992) 365--386}.

\bibitem{Bianchi:1990yu}
M.~Bianchi and A.~Sagnotti, \emph{{On the systematics of open string
  theories}}, \href{http://dx.doi.org/10.1016/0370-2693(90)91894-H}{\emph{Phys.
  Lett. B} {\bf 247} (1990) 517--524}.

\bibitem{Angelantonj:2002ct}
C.~Angelantonj and A.~Sagnotti, \emph{{Open strings}},
  \href{http://dx.doi.org/10.1016/S0370-1573(02)00273-9}{\emph{Phys. Rept.}
  {\bf 371} (2002) 1--150}, [\href{https://arxiv.org/abs/hep-th/0204089}{{\tt
  hep-th/0204089}}].

\bibitem{Abouelsaood:1986gd}
A.~Abouelsaood, C.~G. Callan, Jr., C.~R. Nappi and S.~A. Yost, \emph{{Open
  Strings in Background Gauge Fields}},
  \href{http://dx.doi.org/10.1016/0550-3213(87)90164-7}{\emph{Nucl. Phys. B}
  {\bf 280} (1987) 599--624}.

\bibitem{Bachas:1995ik}
C.~Bachas, \emph{{A Way to break supersymmetry}},
  \href{https://arxiv.org/abs/hep-th/9503030}{{\tt hep-th/9503030}}.

\bibitem{Angelantonj:2000hi}
C.~Angelantonj, I.~Antoniadis, E.~Dudas and A.~Sagnotti, \emph{{Type I strings
  on magnetized orbifolds and brane transmutation}},
  \href{http://dx.doi.org/10.1016/S0370-2693(00)00907-2}{\emph{Phys. Lett. B}
  {\bf 489} (2000) 223--232}, [\href{https://arxiv.org/abs/hep-th/0007090}{{\tt
  hep-th/0007090}}].

\bibitem{Lust:2004fi}
D.~L\"ust, S.~Reffert and S.~Stieberger, \emph{{Flux-induced soft supersymmetry
  breaking in chiral type IIB orientifolds with D3 / D7-branes}},
  \href{http://dx.doi.org/10.1016/j.nuclphysb.2004.11.030}{\emph{Nucl. Phys. B}
  {\bf 706} (2005) 3--52}, [\href{https://arxiv.org/abs/hep-th/0406092}{{\tt
  hep-th/0406092}}].

\bibitem{Camara:2004jj}
P.~G. Camara, L.~E. Ib{\`a}\~nez and A.~M. Uranga, \emph{{Flux-induced
  SUSY-breaking soft terms on D7-D3 brane systems}},
  \href{http://dx.doi.org/10.1016/j.nuclphysb.2004.11.035}{\emph{Nucl. Phys. B}
  {\bf 708} (2005) 268--316}, [\href{https://arxiv.org/abs/hep-th/0408036}{{\tt
  hep-th/0408036}}].

\bibitem{Kaplunovsky:1993rd}
V.~S. Kaplunovsky and J.~Louis, \emph{{Model independent analysis of soft terms
  in effective supergravity and in string theory}},
  \href{http://dx.doi.org/10.1016/0370-2693(93)90078-V}{\emph{Phys. Lett. B}
  {\bf 306} (1993) 269--275}, [\href{https://arxiv.org/abs/hep-th/9303040}{{\tt
  hep-th/9303040}}].

\bibitem{Brignole:1998dxa}
A.~Brignole, L.~E. Ib{\`a}\~nez and C.~Mu\~noz, \emph{{Soft supersymmetry
  breaking terms from supergravity and superstring models}},
  \href{http://dx.doi.org/10.1142/9789812839657_0003}{\emph{Adv. Ser. Direct.
  High Energy Phys.} {\bf 18} (1998) 125--148},
  [\href{https://arxiv.org/abs/hep-ph/9707209}{{\tt hep-ph/9707209}}].

\bibitem{Ibanez:2012zz}
L.~E. Ib{\`a}\~nez and A.~M. Uranga, \emph{{String theory and particle physics:
  An introduction to string phenomenology}}.
\newblock Cambridge University Press, 2, 2012.

\bibitem{Ibanez:1998rf}
L.~E. Ib{\`a}\~nez, C.~Munoz and S.~Rigolin, \emph{{Aspect of type I string
  phenomenology}},
  \href{http://dx.doi.org/10.1016/S0550-3213(99)00264-3}{\emph{Nucl. Phys. B}
  {\bf 553} (1999) 43--80}, [\href{https://arxiv.org/abs/hep-ph/9812397}{{\tt
  hep-ph/9812397}}].

\bibitem{Lust:2004cx}
D.~L\"ust, P.~Mayr, R.~Richter and S.~Stieberger, \emph{{Scattering of gauge,
  matter, and moduli fields from intersecting branes}},
  \href{http://dx.doi.org/10.1016/j.nuclphysb.2004.06.052}{\emph{Nucl. Phys. B}
  {\bf 696} (2004) 205--250}, [\href{https://arxiv.org/abs/hep-th/0404134}{{\tt
  hep-th/0404134}}].

\bibitem{Font:2004cx}
A.~Font and L.~E. Ib{\`a}\~nez, \emph{{SUSY-breaking soft terms in a MSSM
  magnetized D7-brane model}},
  \href{http://dx.doi.org/10.1088/1126-6708/2005/03/040}{\emph{JHEP} {\bf 03}
  (2005) 040}, [\href{https://arxiv.org/abs/hep-th/0412150}{{\tt
  hep-th/0412150}}].

\end{thebibliography}

\end{document}